\documentclass[aip,amsmath,amssymb,reprint]{revtex4-1}

\usepackage{graphicx}
\usepackage{bm}
\usepackage[utf8]{inputenc}
\usepackage[T1]{fontenc}
\usepackage{mathptmx}
\usepackage{subcaption}

\usepackage{chemformula}
\usepackage{siunitx} 
\usepackage[normalem]{ulem} 
\usepackage{marginnote}

\newcommand{\f}[2]{f^{#1}_{#2}}
\newcommand{\Vinfo}[4]{\left\langle #1#2 \right| \left| #3#4 \right\rangle }

\newcommand{\V}[4]{V^{#1#2}_{#3#4}}

\newcommand{\Rr}{\mathcal{R}}

\newcommand{\F}[2]{\mathcal{F}^{#1}_{#2}}

\newcommand{\Fb}[2]{\bar{F}^{#1}_{#2}}

\newcommand{\W}[4]{\mathcal{W}^{#1#2}_{#3#4}}

\newcommand{\Wb}[4]{\bar{W}^{#1#2}_{#3#4}}

\newcommand{\taup}[4]{\tau^{#1#2}_{#3#4}}

\newcommand{\ts}[2]{t^{#1}_{#2}}
\newcommand{\td}[4]{t^{#1#2}_{#3#4}}

\newcommand{\ro}{r_{0}}
\newcommand{\rs}[2]{r^{#1}_{#2}}
\newcommand{\rd}[4]{r^{#1#2}_{#3#4}}

\newcommand{\Pm}[2]{P_-(#1#2)}

\newcommand{\undemi}{\frac{1}{2}}
\newcommand{\unquart}{\frac{1}{4}}

\newcommand{\adag}[1]{\hat{a}_#1^\dag}
\newcommand{\apasdag}[1]{\hat{a}_#1}

\usepackage{booktabs} 
\usepackage{float} 

\newcommand{\Dirac}{DIRAC}
\newcommand{\pEOM}{P-EOM}
\newcommand{\EOMMBPT}{EOM-MBPT(2)}
\newcommand{\pEOMMBPT}{P-EOM-MBPT(2)}
\newcommand{\MBPT}{PT(2)}
\newcommand{\pMBPT}{P-PT(2)}

\newcommand{\ip}{IP}
\newcommand{\pip}{P-IP}
\newcommand{\MBPTip}{PT(2)-IP}
\newcommand{\pMBPTip}{P-PT(2)-IP}
\newcommand{\IPA}{$ \text{IP}_{\text{App}} $}
\newcommand{\ee}{EE}
\newcommand{\pee}{P-EE}
\newcommand{\MBPTee}{PT(2)-EE}
\newcommand{\pMBPTee}{P-PT(2)-EE}

\newcommand{\ea}{EA}
\newcommand{\pea}{P-EA}
\newcommand{\MBPTea}{PT(2)-EA}
\newcommand{\pMBPTea}{P-PT(2)-EA}

\newcommand{\CoutOV}[2]{$n^{#1}_{\text{o}} n^{#2}_{\text{v}}$}
\newcommand{\CoutComp}[1]{$\mathcal{O}\left(N^{#1}\right)$}

\newcommand{\fullip}{full-IP}

\newcommand{\davtz}{dyall-av3$ \zeta $}

\newcommand{\XOm}[1]{\ch{#1O^-}}

\newcommand{\HamilDC}{$ {}^{\text{2}}\text{DC}^{\text{M}} $}
\newcommand{\HamilDCG}{$ {}^{\text{2}}\text{DCG}^{\text{M}} $}

\newcommand{\Supplemental}{{Supplementary Information}}

\newcommand{\trenteetun}{$X_{3/2} $}
\newcommand{\trentedeux}{$A_{3/2} $}

\newcommand{\douze}{$A_{1/2} $}

\newcommand{\inftycorr}{$[\infty \SI{}{\hartree}]$}
\newcommand{\corrlim}[2]{$\left[{#1};\ {#2} \right]$\SI{}{\hartree}}

\begin{document}

\preprint{AIP/123-QED}

\title{The performance of approximate equation of motion coupled cluster for valence and core states of heavy element systems}

\author{Loic Halbert}
\email{loic.halbert@univ-
	lille.fr}
\author{André Severo Pereira Gomes}%
\altaffiliation{Author to whom correspondence should be addressed:}  
 \email{andre.gomes@univ-
 	lille.fr}
\affiliation{ 
Université de Lille, CNRS, UMR 8523—PhLAM—Physique des Lasers 
Atomes et Molécules,
F-59000 Lille, France
}

\begin{abstract}

The equation of motion coupled cluster singles and doubles model (EOM-CCSD) is an accurate, black-box correlated electronic structure approach to investigate electronically excited states and electron attachment or detachment processes. It has also served as a basis for developing less computationally expensive approximate models such as partitioned EOM-CCSD (P-EOM-CCSD), the second-order many-body perturbation theory EOM (\EOMMBPT), and their combination (\pEOMMBPT) [S.\ Gwaltney et al., {Chem.\ Phys.\ Lett.} \textbf{248}, 189-198 (1996)]. In this work we outline an implementation of these approximations for four-component based Hamiltonians and investigate their accuracy relative to EOM-CCSD for valence excitations, valence and core ionizations and electron attachment, and this for a number of systems of atmospheric or astrophysical interest containing elements across the periodic table. We have found that across the different systems and electronic states of different nature considered, 
partition EOM-CCSD yields results with the largest deviations from the reference, whereas second-order based approaches tend show a generally better agreement with EOM-CCSD. We trace this behavior to the imbalance brought about by the removal of excited state relaxation in the partition approaches, with respect to degree of electron correlation recovered.
\end{abstract}

\maketitle

\section{Introduction}

Developments in electronic structure methods of recent decades~\cite{Loos2020,Bokarev2019,Izsk2019} have allowed theory to play a more important role in helping interpret increasingly complex experiments. One case arises in connection to the recent developments on coherent light sources such as those generated in synchrotrons~\cite{Alov2005,Fadley2010,Doucet_2011,Couprie2014} or X-ray free-electron lasers (XFEL)~\cite{Bergmann_2017_rsc_XFEL,Young2018}, which have enabled significant improvements in resolution when exploring high-energy processes involving electronic excitations, such as in X-ray absorption spectroscopy (XAS). But theory can also be of help in lower-energy regimes, be in photoionization experiments~\cite{Gunzer2018} in UV/visible energy range or in the determination of electron affinities~\cite{RienstraKiracofe2002,Richard2016,Chakraborty2020}.

Among the approaches that are generally capable of treating core and valence excited states, as well as states representing electron attachment and detachment processes, we find methods such as complete-active space second-order perturbation theory (CASPT2)~\cite{Grell2015,Lundberg2019} and multireference CI (MRCI)~\cite{Maganas2019}, which thanks to their flexibility can describe systems which are multiconfigurational already for their ground states. Multi-reference coupled cluster approaches such as the state-specific (SS-MRCC) approach of~\citet{Brabec2012}, the state-universal (UGA-SUMRCC) approach of~\citet{Sen2013} and the valence-universal (IH-FSCC) approach of~\citet{Dutta2014fs} have also been successfully applied to investigate core spectra, though their use has been more limited than that of MRCI or CASPT2. And more recently, a mulireference extension of the algebraic diagrammatic construction (ADC) family of methods~\cite{Dreuw2014,Wenzel2014} by~\citet{Mazin2023} has been proposed to address core spectra. 

For systems which the ground-state can be well-represented by a single configuration an array of other methods are also available, such as time-dependent DFT (TD-DFT)~\cite{Besley2020,Besley2021,Brena2006,Fouda2017,DeSantis2022}, the ADC family of methods~\cite{Dreuw2014,Wenzel2014}, as well as the family of methods based on coupled cluster theory, such as linear-response (LR-CC)~\cite{Koch1990,Coriani2015} and equation of motion (EOM)~\cite{Bartlett2007,Coriani2012,Coriani2015,Sadybekov2017,Vidal2019,Peng2015,Park2019,Matthews2020} coupled cluster. EOM-CC is closely connected to the valence-universal multireference Fock-space coupled cluster (FSCC) approach (see~\citet{Musial2008b} for a detailed analysis of the two formalisms) and will yield results which are indistingushable from FSCC for ionization energies~\cite{Shee2018} (if the FSCC model spaces are flexible enough), though for excited states it tends to overestimate the FSCC energies~\cite{Musial2008a,Musial2008b,Real2009}.

It should be noted that for certain situations such as in the calculation of core and valence ionization energies, methods based on the calculation of energy differences (such $\Delta$HF~\cite{Bagus1965,Bagus1999,NavesdeBrito1991}, $\Delta$MP2~\cite{Shim2011,South2016}, $\Delta$DFT~\cite{Besley2009,Fouda2017,PueyoBellafont2015,Takahata2012}, $\Delta$CC~\cite{Watts1990,Zheng2019}) can be very useful, since they are very effective in accounting for orbital relaxation (upon ionization) and electron correlation, but at the expense of requiring the solution of the mean-field problem for open-shells that may be difficult to converge, not to mention the need to find broken symmetry solutions when equivalent centers are present.

Thus, when single-reference approaches are applicable, a good balance between ease of use and accuracy for both valence and core processes is found in coupled cluster based approaches such as EOM-CC (or equivalently LR-CC) based on the coupled cluster singles doubles (CCSD) model. That said, in view of the still significant computational cost of CCSD it is interesting to explore to which extent approximations to EOM-CCSD/LR-CCSD can still provide accurate results while reducing computational cost.

With respect to EOM-CCSD, two main classes of approximations that show promise for valence and core state but that nevertheless have not been extensively explored are partition EOM-CCSD (P-EOM-CCSD~\cite{Nooijen1995_EA,Goings2014,Dutta2014,Dutta2018}), which approximates the doubles-doubles block of matrix representation of the similarity transformed Hamiltonian, and a second-order approximation to EOM-CC (\EOMMBPT)\cite{Nooijen1995,Stanton1995,Gwaltney1996,Nooijen1997} in which, the matrix elements of similarity transformed Hamiltonian are approximated to second order. 

In the case of LR-CC, approximations have been introduced as part of the CCn family of methods (which includes CC3~\cite{Christiansen1995b,Koch1997}, which approximates the treatment of triple excitations as done in the CCSDT method), and in analogy to the EOM-CCSD based approximations above one has the CC2~\cite{Christiansen1995a} method, in which the doubles amplitude equation is approximated, as is the doubles-doubles block of the CC Jacobian ($\mathbf{A}_{DD}$). Since its inception CC2 has become one of the \textit{de facto} standard approaches for exploring molecular properties with low-order computational scaling, and a basis for further approximated methods~\cite{Izsk2019}.

A first comparison of the performance of the approximate EOM methods above and CC2 has been carried out by~\citet{Goings2014}, and it was
found that CC2 showed better performance for valence excited states than P-\EOMMBPT{} whereas the reverse was true for Rydberg states. A thorough analysis of the connection between EOM-CCSD approximations and CC2 was subsequently presented by~\citet{Tajti2016}, which showed among other things that the poorer performance of CC2 for Rydberg states was connected to an imbalance in error cancellation associated with the diagonal approximation in $\mathbf{A}_{DD}$. The performance of EOM-based approaches for valence and core ionization energies has also been investigated by~\citet{Dutta2018}, and it was found that the \EOMMBPT{} approaches followed rather well the EOM-CCSD results, when compared to a EOM-IP-CCSDT reference.

To date and to the best of our knowledge, explorations of these different approaches in general (and of P-EOM-CCSD and (P-)\EOMMBPT{} in particular) have been carried out for non-relativistic Hamiltonians. However, it is now widely recognized that in order to arrive at qualitative and quantitative agreement with experiment, relativistic effects\cite{Saue1997,Pyykko1979,pyykko_relativistic_1988,Pyykk2011} must be taken into account for valence processes of molecules containing elements from the middle to the bottom of the periodic table, and for core processes of molecules containing relatively light elements such as chloride~\cite{Opoku2022}. 

The aim of this paper is therefore to extend our work on relativistic EOM-CCSD for valence~\cite{Shee2018} and core~\cite{Halbert2021} states, presenting a pilot implementation of P-EOM-CCSD and (P-)\EOMMBPT{} in the \Dirac{} code, and to assess their performance with respect to EOM-CCSD for benchmark systems containing heavy and superheavy elements--with a particular emphasis on halogenated species, in view of their importance to atmospheric chemistry and physics~\cite{SaizLopez2011,Steinhauser2014}. With that, we provide a first comparison across the periodic table between EOM-CCSD and its approximations in strictly identical conditions (same basis sets, correlation space and Hamiltonians) for both valence and core excitation and ionization energies, as well as electron attachment energies.

Based on previous benchmarks~\cite{Goings2014,Tajti2016,Dutta2018} we consider the behavior of EOM-based approaches across the periodic table should already give us some qualitative idea of how other approximated methods should behave. That said, gauging the performance of CC2 for heavy elements remains of interest, since to the best of our knowledge there is no data on this topic in the literature. A focus of subsequent works will be on bridging this gap, by implementing the approximate EOM, CC2 and other low-scaling approaches in the newly developed EOM and response theory code~\cite{Yuan2023a} as part of the ExaCorr module~\cite{Pototschnig2021} of \Dirac{}~\cite{DiracSaue2020}.

This paper is organized as follows : after a brief review of the formalism for P-EOM-CCSD, \EOMMBPT{} and \pEOMMBPT, we compare their performance to that of EOM-CCSD for (a) valence ionizations of I$_3^-$ and halogen monoxide ions XO$^-$ (X: Cl, Br, I, At); (b)  core ionizations of HCl, HBr and I$^{-}$; (c) electron affinities of I$_3^-$, CH$_2$I$_2$ and CH$_2$IBr; and (d) excitation energies of I$_3^-$ and CH$_2$I$_2$, followed by our conclusions and perspectives for future work.

\section{Theory\label{title:section_Theoretical_part}}

\subsection{Equation of motion coupled cluster}

The Coupled-Cluster wave function $|\Phi_{\mathrm{CC}} \rangle$ is defined as~\onlinecite{Bartlett2007}~:
	\begin{align}
	| \Phi_{\mathrm{CC}} \rangle = e^{\hat{T}} |\Phi_0 \rangle,
	\label{eq:CC}
	\end{align}
	where $\left| \Phi_0 \right\rangle $ denotes the Hartree-Fock determinant for the ground-state and $\hat{T}$ the cluster operator, which here shall be restricted to single and double excitations,
    \begin{equation}
        \hat{T}=\hat{T}_1 +\hat{T}_2 = \sum\limits_{ia}{t_i^a\{\hat{a}_a^\dag \hat{a}_i}\} + \dfrac{1}{4}\sum\limits_{aibj}t_{ij}^{ab}\{\hat{a}_a^\dag\hat{a}_b^\dag
	\hat{a}_j\hat{a}_i\}
    \end{equation}
    with $ \hat{a}^{\dagger} $ and $\hat{a} $ denoting respectively creation and annihilation operators and $ \ts ai $  and $ \td abij $ the corresponding amplitudes to be determined. Here, and in the following : $a,b,c,.. $ will indicate {particle lines}, $ i,j,k,... $ {hole lines}, and $ p,q,r,s,... $ either holes or	particles~\cite{Crawford2007}.
	
	In order to define the EOM-CCSD method we start from the normal-ordered Hamiltonian
	\begin{align}
	\hat{H}_\mathrm{N} &= \hat{H} - \langle\Phi_0 | \hat{H} |  \Phi_0 \rangle \nonumber \\
    &=\sum\limits_{pq} \left\langle p|f|q\right\rangle \{ 
	\hat{a}^\dag _p 
	\hat{a}_q\}+\frac{1}{4}\sum\limits_{pqrs} \left\langle 
	pq||rs\right\rangle  
	\{ 
	\hat{a}^\dagger_p 
	\hat{a}^\dag_q \hat{a}_s \hat{a}_r\} \nonumber\\
    &=\sum\limits_{pq} f_{pq} \{ 
	\hat{a}^\dag _p 
	\hat{a}_q\}+\frac{1}{4}\sum\limits_{pqrs} g_{pqrs}  
	\{ 
	\hat{a}^\dagger_p 
	\hat{a}^\dag_q \hat{a}_s \hat{a}_r\} 
	\end{align}
	where $f_{pq}$ and $g_{pqrs}$ represent the matrix/tensor representations of the Fock operator and 2-electrons integral respectively; and second, the similarity transformed Hamiltonian $\bar{H}$,
    \begin{equation}    
    \bar{H}=e^{-\hat{T}}\hat{H}_{\text{N}}e^{\hat{T}}.
    \end{equation}
    
	With these, and choosing a parametrization for an electronic state $x$ other than the ground state on the basis of a reference coupled cluster wavefunction,
    \begin{equation}
    |\Phi_x \rangle =\Rr_{x} |\Phi_{\mathrm{CC}}\rangle = \Rr_{x} 	e^{\hat{T}}   |\Phi_0 	\rangle   
    \end{equation}
    the EOM problem for electronically excited, electron attachment and electron detachment states is given by an eigenvalue equation
 	\begin{align}
	\bar{H}\mathbf{R}&=\omega \mathbf{R},  
	\label{eq:REigenstate}
	\end{align}
 that is solved by standard iterative procedures (see~\cite{Shee2018,Halbert2021} and references therein). 

    The operator $\Rr$, for excited states (\ee{}) is given by the linear expansion
    \begin{equation}
   \Rr^{EE}_{x} = \ro + {^{x}}\rs ai \adag a \apasdag i+\unquart {^{x}}\rd abij \adag a \adag b \apasdag j \apasdag i \ ;
    \end{equation}
	with $\ro = 1$ for the ground-state and $\ro = 0$ otherwise; for electron detachment (ionization energies,  \ip{}) a particle line disapears, yielding
	\begin{align}
	\Rr^{IP}_{{x}} = {^{x}}\rs {}i  \apasdag i+\undemi {^{x}}\rd a{}ij \adag a 
	\apasdag i \apasdag j \ , \label{R-ip-definition}
	\end{align}
    whereas for electron attachment (electron affinities, \ea{}) a hole line disapears, yielding
	\begin{align}
	\Rr^{EA}_{{x}} = {^{x}}\rs a{}  \adag a+\undemi {^{x}}\rd ab{}i \adag a \apasdag i
	\adag b  \ ,
	\end{align}
	However, in all of these cases we can identify the same block structure in the matrix representation of $\bar{H}$, 
    \begin{equation}
	\left(\begin{array}{cc}
	\textcolor{black}{\bar{H}_{SS}}& \bar{H}_{SD}\\ \bar{H}_{DS} & 
	\bar{H}_{DD}
	\end{array} \right)\\%
	\label{eq:EOM_matrix}
    \end{equation}
    consisting of matrix elements between singly excited (1h1p), attached (1p) or detached (1h) configuration ($\bar{H}_{SS}$); between doubly excited (2h2p), or singly attachment or detachment configurations accompanied by relaxation (2p1h/2h1p) ($\bar{H}_{DD}$); and between these two manifolds ($\bar{H}_{SD}$ and $\bar{H}_{DS}$).
 	
	To obtain the ionization potentials of the core electrons, we use the Core-Valence-Separation (CVS) technique ~\cite{cederbaum_many-body_1980,Dreuw2014,Coriani2015}, in which the eigenvector vector solution $ \mathbf{R}_k $ is limited in size by the fact of taking into account only the molecular orbitals (MO) lower than a certain value in energy arbitrarily defining the  valence/core limit. For algorithm we use a projection operator $ \mathcal{P} $ 	working on the trial vector $\sigma $ ~\cite{Shee2018} which acts on a molecular orbital $ i $ to a virtual  molecular  orbital $ a $, only if $ i $ belongs to the core space (labelled "$I$").
	
	\begin{equation}
	\left\{\begin{aligned} \mathcal{P}_{I}
 \sigma_{i}^{a} &=0 & \text { if } i 
	\neq I 
	\\ \mathcal{P}_{I}
 \sigma_{i j}^{a b} &=0 & \text { if } i \neq I \text { 
	or } 
	j \neq 
	I \end{aligned}\right.
	\end{equation}
	\begin{equation}
	\mathcal{P}_{I}
 \left(\textbf{H} \mathcal{P}_{I}
	\mathbf{R}_{k}\right)=\omega_{k} \mathcal{P}_{I}
 \mathbf{R}_{k} \label{eq:CVS}
	\end{equation}
	
	This technique had been tested  on several system and gives accurate and reproducible results~\cite{Halbert2021}.
	
	\subsection{Approximated methods}
	
	As mentioned before, we shall consider two main families of approximated methods here. The first is the partitioned EOM-CCSD (P-EOM-CCSD). Its starting point is the EOM equation, in which a given eigenvector $ R_k $ with eigenvalue $ \omega_k $ is separated into two parts; a part of interest 'a' and a part 'b', orthogonal to	'a', 
	\begin{equation}
	\begin{array}{c@{\ }c}
	&\left(\begin{array}{cc}
	\bar{H}_{SS}& \bar{H}_{SD}\\ \bar{H}_{DS} & \bar{H}_{DD}
	\end{array} \right)\\
	\end{array}
	\begin{array}{c@{\ }c}
	&\left(\begin{array}{c}
	R_{k_{a}}\\ R_{k_{b}}
	\end{array} \right)\\
	\end{array}
	= \omega_k \begin{array}{c@{\ }c}
	&\left(\begin{array}{c}
	R_{k_{a}}\\ R_{k_{b}}
	\end{array} \right)\\
	\end{array}
	\end{equation}
	  This expression can be rewritten as 
	\begin{equation}
	\left\{\begin{aligned} \bar{H}_{SS}.R_{k_{a}}+\bar{H}_{SD}.R_{k_{b}} 
	&=\omega_k 
	R_{k_{a}}\\ \bar{H}_{DS}.R_{k_{a}}+\bar{H}_{DD}.R_{k_{b}} 
	&=\omega_k 
	R_{k_{b}} \end{aligned}\right.
	\end{equation}
	and from it one can then define an effective Hamiltonian and an associated eigenvalue problem
	\begin{align}
	\bar{H}_{SS}(\omega_k){}_{\text{eff}}&=\bar{H}_{SS}+\bar{H}_{SD}\left(\omega_k.\mathbf{1}
	-\bar{H}_{DD} 
	\right)^{-1}\bar{H}_{DS}\nonumber\\
	\bar{H}_{SS}(\omega_k){}_{\text{eff}}&=\omega_k R_{k_{a}}
	\end{align}

	Carrying out a development limited to the zeroth order ~\cite{Lwdin1963,lawley_ab_1987,Geertsen1989} 	$ \bar{H}_{DD}$ is replaced by zero-order approximation $ H^{\mathbf{[0]}}_{DD}$, which is diagonal for (semi-)canonical Hartree-Fock orbitals ~\cite{Gwaltney1996}. With that, the EOM-CCSD matrix now becomes
 	\begin{equation}
	\left(\begin{array}{cc}
	\bar{H}_{SS}& \bar{H}_{SD}^{\textcolor{white}{[X]}}\\ \bar{H}_{DS} & 
	\bar{H}^{\mathbf{[0]}}_{DD}
	\end{array} \right)
     \end{equation}
    Working equations for $\sigma$-vectors and intermediates are given in appendix~\ref{title:appendix_1}.
	
	In the \EOMMBPT{} method~\cite{Stanton1995}, we devising a second-order approximation to the molecular Hamiltonian,
	\begin{equation}
	\bar{H}^{[2]}=\left\langle 0\left|\bar{H}^{[2]}\right| 
	0\right\rangle+\F pq {}^{[2]} \left\lbrace \adag p \apasdag 
	q\right\rbrace  
	+\frac{1}{4} \W pqrs{}^{[2]}  \left\lbrace \adag p \adag q \apasdag s 
	\apasdag 
	r\right\rbrace  +...
	\end{equation}
	The first term is just the energy of the reference state at the second order of perturbation. $ \F pq {}^{[2]} $ and $ \W pqrs {}^{[2]}  $ are matrix element, some {Effective Hamiltonians}, but in fact intermediates (see~\citet{Gauss1995}), developed at the second order.
	
    As an exemple, $ \F vv $ depends on $ \taup aemn=\td aemn+ \ts am \ts en - \ts em \ts 
	an $, but now : 
	\begin{align}
	_{}\F vv = \F ab = & \f ab - \ts am \f mb +  \ts fm \Vinfo 	mafb 
	\nonumber \\
	&- \undemi  
	\begin{Bmatrix} \text{Complete : } & \taup aemn  \Vinfo mnbe \\ \text{MBPT(2) : }& \td 	
	aemn   \Vinfo mnbe  
	\end{Bmatrix} 
	\end{align}
	the limitation to the second order now imposes amplitudes being developed at first order
	\begin{align}
	{\ts ai }^{[1]}=\dfrac{\f ai}{\varepsilon_i -\varepsilon_a}, \quad {\td abij 
	}^{[1]}=\dfrac{\Vinfo 
		abij}{\varepsilon_i +\varepsilon_j -\varepsilon_a-\varepsilon_b} ,
    \label{eq:amplitude_t_MP2}
	\end{align}
	and this obtained from MP2, so that the EOM matrix now has the form (see~\citet{Stanton1995})
    \begin{equation}
     \left(\begin{array}{cc}
	 \bar{H}^{[2]}_{SS}& \bar{H}^{[2]}_{SD}\\ \bar{H}^{[2]}_{DS} & 
	 \bar{H}^{[2]}_{DD}
	 \end{array} \right).
    \end{equation}
    Working equations for $\sigma$-vectors and intermediates are given in appendix~\ref{title:appendix_1}. Finaly, it is possible to combine the \EOMMBPT{} and P-EOM approaches in the so-called \pEOMMBPT{} scheme, whereby the $ \bar{H}^{\mathbf{[2]}}_{DD}$ block is further approximated to $ \bar{H}^{\mathbf{[0]}}_{DD}$, with $ \ts ai $ and $ \td abij $ calculated as in 	eq.~\ref{eq:amplitude_t_MP2}, or in matrix form
 \begin{equation}
     \left(\begin{array}{cc}
	 \bar{H}^{[2]}_{SS}& \bar{H}^{[2]}_{SD}\\ \bar{H}^{[2]}_{DS} & 
	 \bar{H}^{\mathbf{[0]}}_{DD}
	 \end{array} \right).
 \end{equation}

\subsection{Computational scaling} 

The computational cost of EOM-CCSD scales as \CoutComp{6} (\CoutOV{2}{4}, \CoutOV{3}{3} and \CoutOV{4}{2} for EE, IP and EA respectively, with $n_\text{o},n_\text{v}$ denoting the number of correlated occupied and virtual spinors). For the iterative determination of eigenvalues, assembling $ H_{DD} $ shows a \CoutOV{2}{3} scaling for EOM-IP and \CoutOV{1}{4} for EOM-EA; in the construction of intermediates  the cost is \CoutComp{6}, but that is done outside any iterative steps. The \EOMMBPT{} method will show the same scaling as above, though saving arises from replacing the costly iterative \CoutComp{6} determination of the ground-state CC amplitudes by the \CoutComp{5} non-iterative determination of MP2 amplitudes. 

In the case of P-EOM-CCSD, the cost in the iterative determination of eigenvalues is reduced to \CoutComp{5}~\cite{Gwaltney1996} for EE (\CoutOV{2}{3}). For IP and EA the cost of assembling $ H_{DD} $ and $ H_{DS} $ is respectively \CoutComp{4} and \CoutComp{5}, though with the use of intermediates the cost during the iterative procedure falls to \CoutComp{4}. Finally, for 
\pEOMMBPT{} we again avoid the determination of the CC amplitudes, and for EE the formation of the  $ H_{DS} $ and the iterative determination of eigenvalues both have \CoutComp{5} cost.

\section{Computational details}
	
	All calculations were carried out with a development version of the DIRAC package~\cite{DIRAC19,DiracSaue2020}. The implementations were carried out in the RELCCSD module, which allows for exploitation of point group symmetry, including linear symmetry, in EOM-CC calculations~\cite{Shee2018,Halbert2021}.
	
	We employ the same geometries used in prior studies : thus, for XO$^-$, CH$_2$I$_2$, CH$_2$IBr and I$_3^-$ these were taken from the work of~\citet{Shee2018}, whereas for HX from the study by~\citet{Visscher1996}. We employ the \davtz~\cite{Dyall2006} basis sets for the heavy elements, and Dunning's aug-cc-pVTZ~\cite{Kendall1992} for oxygen, carbon and hydrogen. All basis sets are used uncontracted. 

	Unless otherwise noted, in our calculations we employ the eXact 2-Component molecular mean-field (X2Cmmf) approach~\cite{Sikkema2009}, based either on the Dirac-Coulomb (\HamilDC) or Dirac-Coulomb-Gaunt (\HamilDCG) Hamiltonians. For the calculations based on the \HamilDCG, two-electron integrals over small-component basis sets $ (\text{SS}|\text{SS}) $ are approximated by a point charge model~\cite{Visscher1997}, and a Gaussian distribution is used to model the finite size of the nuclei~\cite{Visscher1997_nuclearpotential}.

    Since the size of EOM-\{\ip; \ea; \ee\} matrices are generally very large due to the (2h2p/2h1p/2p1h) configurations, eigenvalues and eigenvectors are obtained using the generalization of the Davidson iterative algorithm for non-symmetric matrices~\cite{Davidson1975,Hirao1982,Shee2018,Halbert2021}.
     
    When all the spinors have been correlated during the calculations we used the notation $[\infty \SI{}{\hartree}]$, otherwise we follow the notation : $\left[\text{lower limit };\ \right.$ $\left. \text{upper limit}\right]$\SI{}{\hartree} to indicate, via spinor energy lower and upper bounds, the extent of the spinor space  retained for the transformation from the atomic (AO) to the molecular spinor (MO) basis.

This work focuses on different halogenated systems. For the valence \ip{}s (1h) the studied systems are I$_3^-$ \corrlim{-3}{12} and XO$^-$ with $X \in \left\lbrace \text{Cl};\ \text{Br};\ \text{I};\ \text{At};\ \text{Ts} \right\rbrace$ \inftycorr{}. In all cases, the first four solutions are calculated. Then, for the {potential energy curves (PECs)} of the monoxides (excluding tennessine) we employed  \corrlim{-10}{100}, and spectroscopic constants are obtained by a polynomial fit of the absolute electronic state energies for internuclear distances comprised between \SI{1.4}{\angstrom} and \SI{2.4}{\angstrom}. For core \ip s, the systems studied are HCl, HBr and I$^{-}$, and in these cases we correlate all electrons and all virtuals (\inftycorr{}).%

For \ea{} (1p), we considered I$_3^{-}$, for which we calculated four electronic states, and the halomethanes CH$_2$I$_2$ and CH$_2$IBr, for which we calculated six electronic states. The correlated electrons are within the limits \corrlim{-3}{6} for CH$_2$I$_2$ and \corrlim{-4}{6} for CH$_2$IBr.%

For \ee{} (1h1p), we present here the results obtained for the systems I$_3^-$ and CH$_2$I$_2$. The correlated electrons are within the limits \corrlim{-3}{12} for I$_3^-$ and \corrlim{-3}{6} for CH$_2$I$_2$.%

 All graphs have been prepared with Matplotlib~\cite{Hunter:2007} python library. The dataset associated with this manuscript is available at the Zenodo repository~\cite{halbert2023-dataset}.
	
\section{Results}

Before beginnning the discussion of our results, we introduce some shorthand notations that will be used throughout (for EE and EA a strictly analogous shorthand notations will be used) : first, \ip{} will refer to EOM-IP-CCSD calculations whereas \pip{}  will refer to P-EOM-IP-CCSD calculations, and \MBPTip{} and \pMBPTip{} will refer to the second-order perturbation theory analogs. Second, we will generally present the difference between EOM-IP-CCSD and the others appraches by $\Delta E_{\ip}$. 

In addition to these, we will provide a measure of whether electronic states are dominated by singly ionized (SI) main configurations through the notation ``\%SI'' accompanied by percentage ranges; for example, \%SI in $[95\%;98\%]$ will denote the value of the square of the largest $r_i$ coefficients in Eq.~\ref{R-ip-definition} falls under 0.95 and 0.98 for the states under consideration. If this value is close to 100\%, a given state can be considered as monoconfigurational, and lower values will generally indicate an increased multiconfigurational character for a particular state. 

Finally, we note that in what follows we will mostly focus on the comparison between theoretical approaches. The experimental data for the systems under consideration is available in the \Supplemental.
	
\subsection{Ionization energies\label{title:IP}}

\subsubsection{Valence ionizations, equilibrium structures}

We begin our discussion on valence ionization energies with the triiodide I$_3^-$ species. This hypervalent anion is interesting due to being one of the relatively rare molecular anions which are stable, and for the multiple pathways it offers in its photodissociation. The species  has been studied theoretically \cite{Gomes2010,Wang2014,Shee2018} and experimentally \cite{Zhu2001,Choi2000}. Our results are given in table~\ref{tab:I3m1_IP_Gaunt_tab}.

	\begin{table} 
		\fontsize{8}{6}\selectfont
		\centering
		\caption{Ionization energies (\ip{} in \SI{}{\electronvolt}) for I$_3^-$  employing the \HamilDCG{} Hamiltonian, and $\Delta E_{\ip}$ values for the different approximate methods considered here. The \%SI in  
			$[91\%;93\%]$ for all states and methods considered.}
		\begin{tabular}{l r rrr}
			\toprule
			& & \multicolumn{3}{c}{$\Delta E_{\ip}$} \\
   \cmidrule(l{3pt}r{3pt}){3-5}
			T&	\ip &  \pip &  \MBPTip{} &  \pMBPTip{}   \\
			\midrule
$ \text{\ip}_{1}^{1/2g} $&	{4.46} &   {-0.05} &     {0.11} 
&      {0.06}        
\\
$ \text{\ip}_{2}^{3/2g} $	&	{5.00} &   {-0.14} &     
{0.01} &     
{-0.10}  \\
$ \text{\ip}_{3}^{1/2u} $&	{4.91} &   {-0.10} &     {0.04} 
&     {-0.03} 
\\
$ \text{\ip}_{4}^{3/2u} $&	{4.28} &   {-0.10} &     {0.04} 
&     {-0.04} 
 \\
			\bottomrule
		\end{tabular}
		\label{tab:I3m1_IP_Gaunt_tab}
	\end{table}
	
 We can see that overall the ionization energies obtained with the different approximate methods (collectively referred to in what follows as \IPA) are very close to the reference \ip{} value. That said, we can identify some trends for the individual approaches : we see that that \pip{} generally underestimates the \ip{} values, \MBPTip{} overestimates them, and \pMBPTip{} falls in between. The latter can be attributed to a systematic error compensation, though we also observe that there are some non-additive effects since adding up the differences with respect to the reference for \pip{} and \MBPTip{} yield results which are close, but exactly those for \pMBPTip{}. Furthermore, we see that these four electronic states are predominantly singly ionized states (all \%SI are between 91\% and 93\%), with the characteristic that the states are dominated by a single ionized configuration (in other words, ionizations occur from essential a single spinor).

The second class of compounds for which valence ionizations were investigated is the halogen monoxides XO$^-$ ($X \in \left\lbrace Cl;\ Br;\ I ;\ At;\ Ts \right\rbrace$). These compounds have the halogen in the (+I) oxidation state, and are for instance involved in ozone degradation~\cite{Burkholder2015} or have an interest in nuclear medicine in the case of astatine\cite{liu2020_these_At,Vaidyanathan2008}. The four \ip{} states under consideration correspond in effect to the spin-orbit split doublet ground (${}^{2}X_{1/2}, {}^{2}X_{3/2}$) and first excited (${}^{2}A_{1/2}, {}^{2}A_{3/2}$) states of the corresponding halogen monoxide radicals. As expected, as the atomic number of the halogen increases the spin-orbit splitting will be more important; \citet{Shee2018} have analyzed in detail these states, and showed that the ground states are dominated by contributions from the halogen, whereas the reverse is true for the first excited state. As shown below, this will have significant implications on the performance of the approximate methods.

\begin{figure}
	\noindent
	\includegraphics[width=\linewidth]{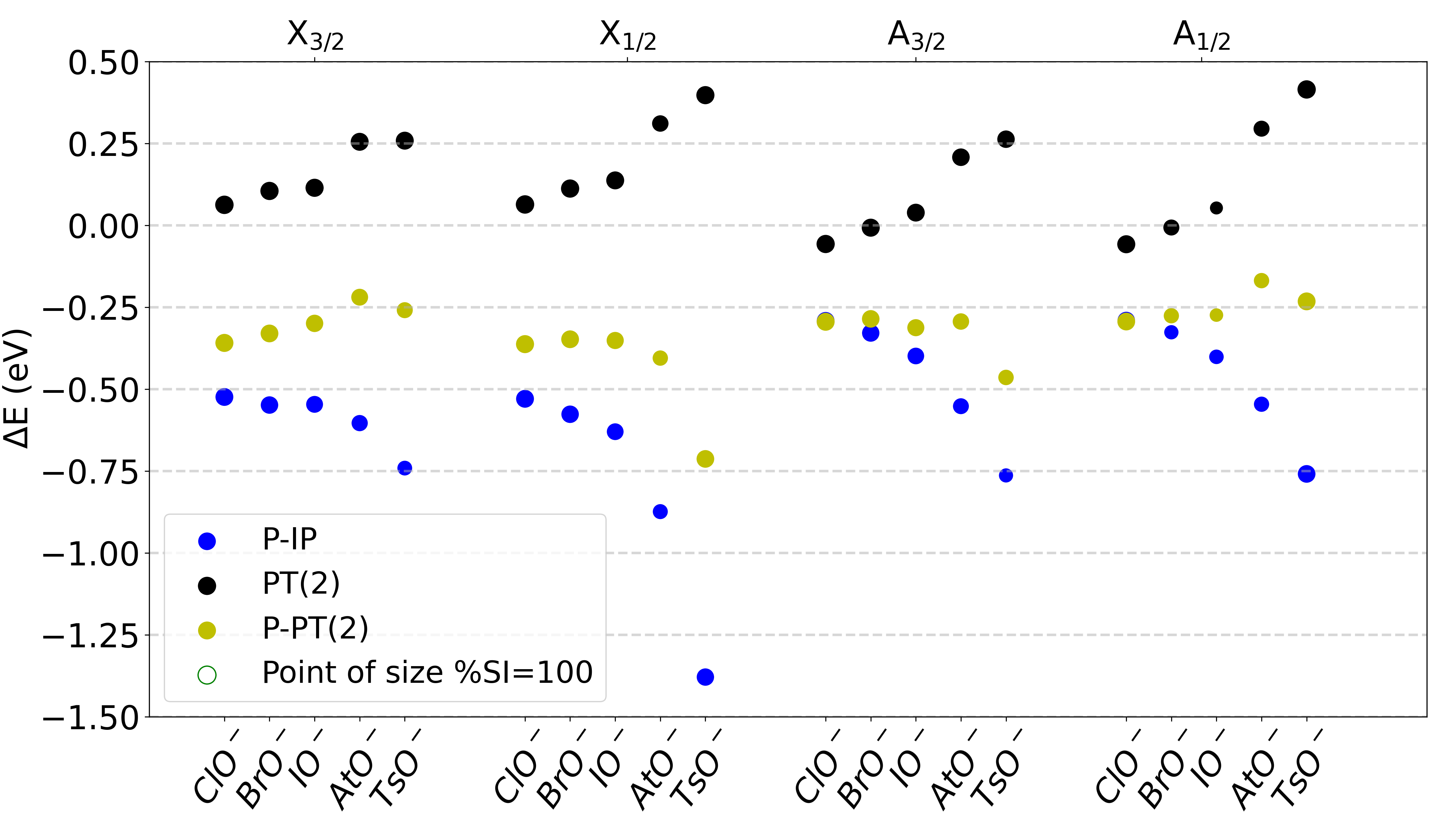}
	\caption{$\Delta E_{\ip}$ values (in \SI{}{\electronvolt}) for the different approximate electronic structure methods considered here as a function of electronic state and halogen. Points have a 
	size proportional to 
		\%SI (see \Supplemental{} for the list of individual ionization energy values). \label{fig:XOm1_All_Diff_IPA_wrt_IP_sym}
	 }
\end{figure}

In figure~\ref{fig:XOm1_All_Diff_IPA_wrt_IP_sym} we show how $\Delta E_{\ip}$ values for the different methods vary with respect to the change in halogen for each of the spin-orbit split ground and excited states, with the size of each point being proportional of the singly ionized character for each of the electronic states (smaller sizes denoting higher 2h1p character). In general, the evolution of $\Delta E_{\ip}$ is the same for the first four solutions as we go through the excited states from the lowest to the highest: \MBPTip{} is increasing, \pip{} decreasing and \pMBPTip{} remains relatively constant across the XO$^-$ series.

Looking closely at the different methods, we see first that for \MBPTip{} we obtain overall the smallest deviations with \ip{} in absolute value, though we see a relatively small but clear break between chlorine, bromine and iodine on one side (which show very similar $\Delta E_{\ip}$ values for all four states), and astatine and tennessine on the other (for which $\Delta E_{\ip}$ tend to increase for the $\Omega = 1/2$ components with respect to the $\Omega = 3/2$ one). 

For chlorine, energies are slightly overestimated for the ground state (\SI{0.06}{\electronvolt}) and similarly underestimated (\SI{-0.06}{\electronvolt}) for the $A$ excited state. As spin-orbit coupling is very weak there are no noticeable differences in errors between the $\Omega = 3/2$ and $\Omega = 1/2$ components. For bromine and iodine, $\Delta E_{\ip}$ values are close to \SI{0.11}{\electronvolt}, though for $A$ these are smaller (less than \SI{0.05}{\electronvolt }). 

There are rather small differences in $\Delta E_{\ip}$ values for the spin-orbit splitting of the ground state (\SI{0.01}{\electronvolt} for bromine and \SI{0.02}{\electronvolt} for iodine), and these become much smaller for the spin-orbit splitting of the $A$ state (less than \SI{1}{\milli\electronvolt} for bromine and \SI{0.02 }{\electronvolt} for iodine). 

Thus, for these three compounds, the average of the absolute errors $\bar{\delta}$ for the ground state is $\bar{\delta}=\SI{0.1}{\electronvolt}$ with a standard deviation of $\sigma=\SI{0.03}{\electronvolt}$, whereas for the $A$ state, these values are $\bar{\delta}=\SI{0.04}{\electronvolt}$ and $\sigma=\SI{0.02}{\electronvolt}$, strongly suggesting a systematic error. For astatine and tenessine, the error becomes larger but remains lower than \SI{0.5}{\electronvolt}.

We observe that \pip{} presents the worst results in terms of the absolute value of $\Delta E_{IP}$ among all three approximations considered, and the errors being all negative means that IPs are underestimated. For the $X_{3/2}$ state we see that the error is somewhat constant (around \SI{-0.5}{\electronvolt}) from chlorine to iodine, and slowly increases for astatine and tennessine to reach around \SI{-0.75}{\electronvolt}. 

In qualitative terms, this is similar to what one sees for the \MBPTip{} case. For both spin-orbit split components of the excited state, we see a similar trend, but it differs slightly from \MBPTip{} case, since for the  $A_{3/2}$ state there is a larger difference between astatine and tennessine. What is remarkable in \pip{} is that for $X_{1/2}$ state there is a strong change in 
$\Delta E_{IP}$ when going from astatine to tennessine, with the error reaching \SI{-1.38}{\electronvolt}. 

One element that may help understand the root of this behavior is the nature of this state : as discussed by~\citet{Shee2018}, in the EOM-IP-CCSD description of TsO, the $X_{1/2}$ state is predominantly made up by a configuration with 1h character, in which  $\pi^*_{1/2}$ spinor ends up being singly occupied. However, there is also a contribution from a configuration with 2h1p character (on top of removing one electron from $\pi^*_{1/2}$, an electron from the doubly occupied $\sigma_{1/2}$ is placed on the empty $\sigma^*_{1/2}$ spinor) that is sufficiently large to affect energies by a few tenths of an \SI{}{\electronvolt} in comparison to Fock-space coupled cluster results~\cite{Shee2018}. 

In such a case \pip{} turns out to be a poor approximation since it suppresses the ability of $\bar{H}_{DD}$ to account for the energetics of relaxation through the 2h1p configurations, though the nature of the $X_{1/2}$ state (spinors centered on tennessine, for which relaxation effects are potentially more important) may also play a role. 

Case in point, the  $\Delta E_{IP}$ values for the $A$ states of TsO are similar, even though EOM-IP-CCSD calculations indicate the $A_{1/2}$ state also has a non-negligible contribution with 2h1p character (the $\pi^*_{1/2}$ remains doubly occupied and two electrons are removed from the $\sigma_{1/2}$ and one place on $\sigma^*_{1/2}$) whereas the $A_{3/2}$ does not. But in constrast to the $X$ states, $A$ states are centered on oxygen.

Finally, it turns out that the approximation of the $\bar{H}_{DD}$ block in the \pMBPTip{} method is compensated to some extent by the errors introduced by the approximate treatment of electron correlation. From chlorine to iodine, the differences are included in the limits: $ \left[- 0.36;\ -0.27 \right] $\SI{}{\electronvolt }, in this case $\bar{\delta}=\SI{0.34}{\electronvolt} $ and $\sigma=\SI{0.02}{\electronvolt}$ for $X$ states and $\bar{\delta}=\SI{0.29}{\electronvolt}$, $\sigma=\SI{0.01}{\electronvolt}$ for $A$ states. 

The standard deviations show us that the error is almost constant, with only the results for the $\Omega = 1/2$ component of the ground state and to a lesser extend, the $\Omega = 3/2$ component of the excited state of TsO deviating significantly from this trend. Paradoxically the most approximate method becomes, in this case, the most stable across the halogen monoxide series.
	
A final remark concerns the change in \%SI for the different complexes and electronic states, shown in the \Supplemental{}. We observe that there is no evident correlation between a particular value (i.e.\ whether a state is mono or multiconfigurational) and the magnitude of $\Delta E_{IP}$. 

We do note, however, that \MBPTip{} are typically very close to that of the reference calculations, whereas \pip{} values are usually smaller than the reference and \MBPTip{} ones, and \pMBPTip{} values as expected fall in between \MBPTip{} and \pip{} (though tend to be closer to the latter). 

Taken together, these suggest that the approximation of $\bar{H}_{DD}$ does have a non-negligible impact on the nature of the wavefunction by favoring an increase in multiconfigurational character for the 1h contributions.

\subsubsection{Valence ionizations, potential energy curves\label{title:subsection_PES} }

\begin{figure}
	\noindent
	\includegraphics[width=\linewidth]{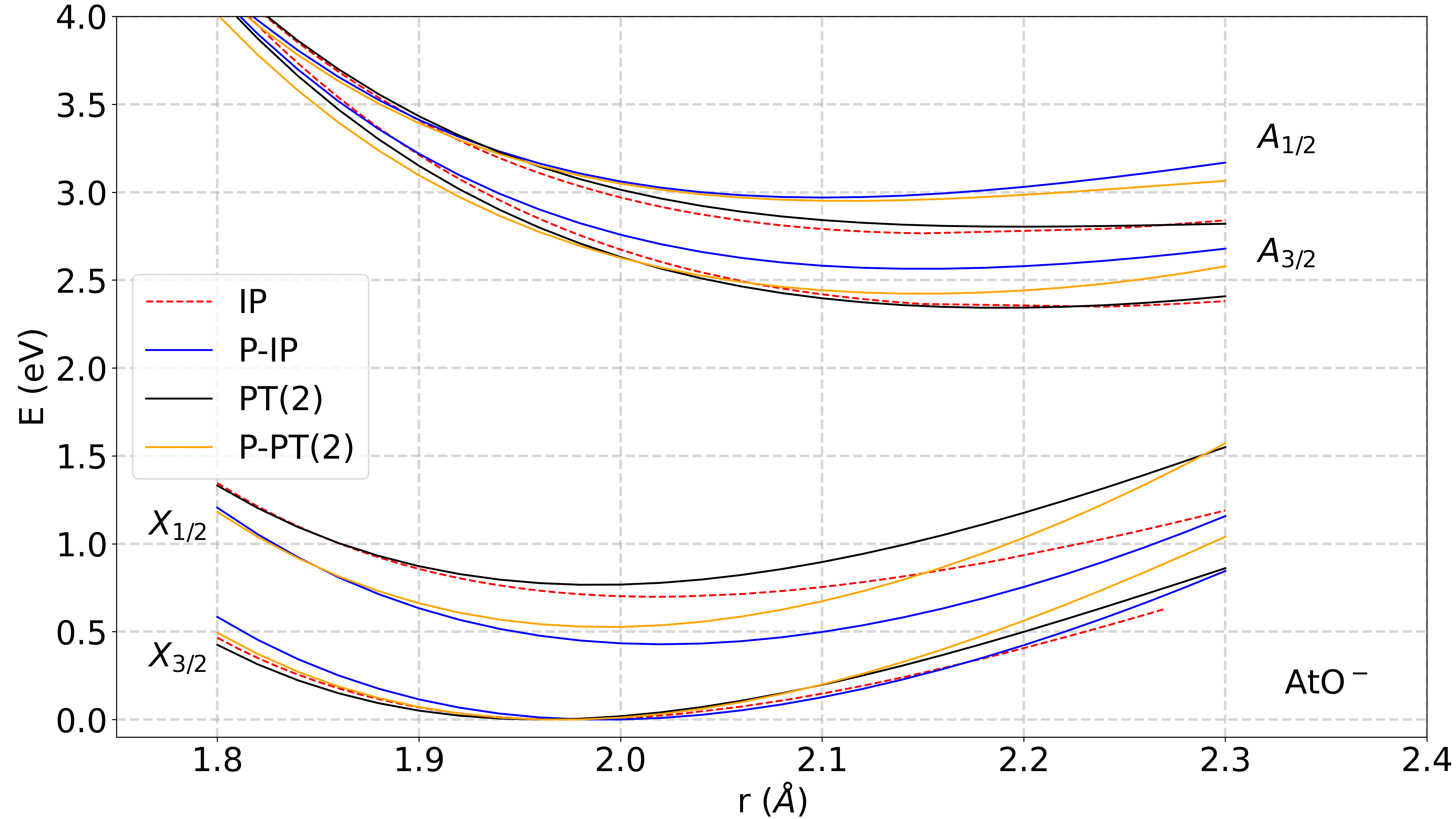}
	\caption{Potential energy curve for AtO (in \SI{}{\electronvolt}, shifted by the energy at the minimum of the $X_{3/2}$ state) as function of the  X-O distance  (r, in \AA), for the reference method and the different approximations.\label{fig:PES_AtOm1_AllSym}}
\end{figure}

Apart from the investigation for ground-state equilibrium structures, we have also investigated the effect of the approximation on the potential energy curves for $X$ and $A$ states of the halogen monoxide radicals. As an example, we shown in figure~\ref{fig:PES_AtOm1_AllSym} the results for the AtO species, as a representative of the overall trends. The curves for the other species as well as their equilibrium distances can be found in the \Supplemental{}.

The first information from figure~\ref{fig:PES_AtOm1_AllSym} is that \MBPTip{} (in  black) shows very good agreement to the reference values (in red) for the $A$ states across the studied geometries. It also reproduces rather well the curves for the $X$ states up to the $X_{3/2}$ equilibrium structure; for larger internuclear distances, it starts to shift towards higher energies for both spin-orbit spit components, but the deviation with respect to the reference becomes more significant for the $X_{1/2}$ state than for the $X_{3/2}$ state.

Second, we see that the curves for the partition-based approaches (\pip{} in blue and \pMBPTip{} in yellow) are quite close to each other for internuclear distances shorther than the $X_{3/2}$ equilibrium structure, and  start to separate out for larger distances; interestingly, \pMBPTip{} energies tend to be higher than \pip{} ones for the $X$ states, while the reverse is true for the $A$ states.

Taken together, these results hint that for halogen monoxides \pMBPTip{} should not necessarily be more reliable than \MBPTip{} for obtaining spectroscopic constants, or investigating excited states for geometries away from the ground-state equilibrium. But whatever the case, they provide further evidence that \pip{} is the least balanced of the three approximate methods under consideration. 

From the graphs it is also interesting to notice that for the $X_{1/2}$ the \pMBPTip{} curve goes to being closer to the \pip{} curve for small internuclear distances to being closer to the \MBPTip{} for longer distances. However, this behavior is not observed for the other three states, indicating that the error cancellation that benefits the \pMBPTip{} approach near the equilibrium structure of the ground state is not consistent across different structures.

\begin{figure*} 
		\noindent
		\includegraphics[width=1.0\textwidth]{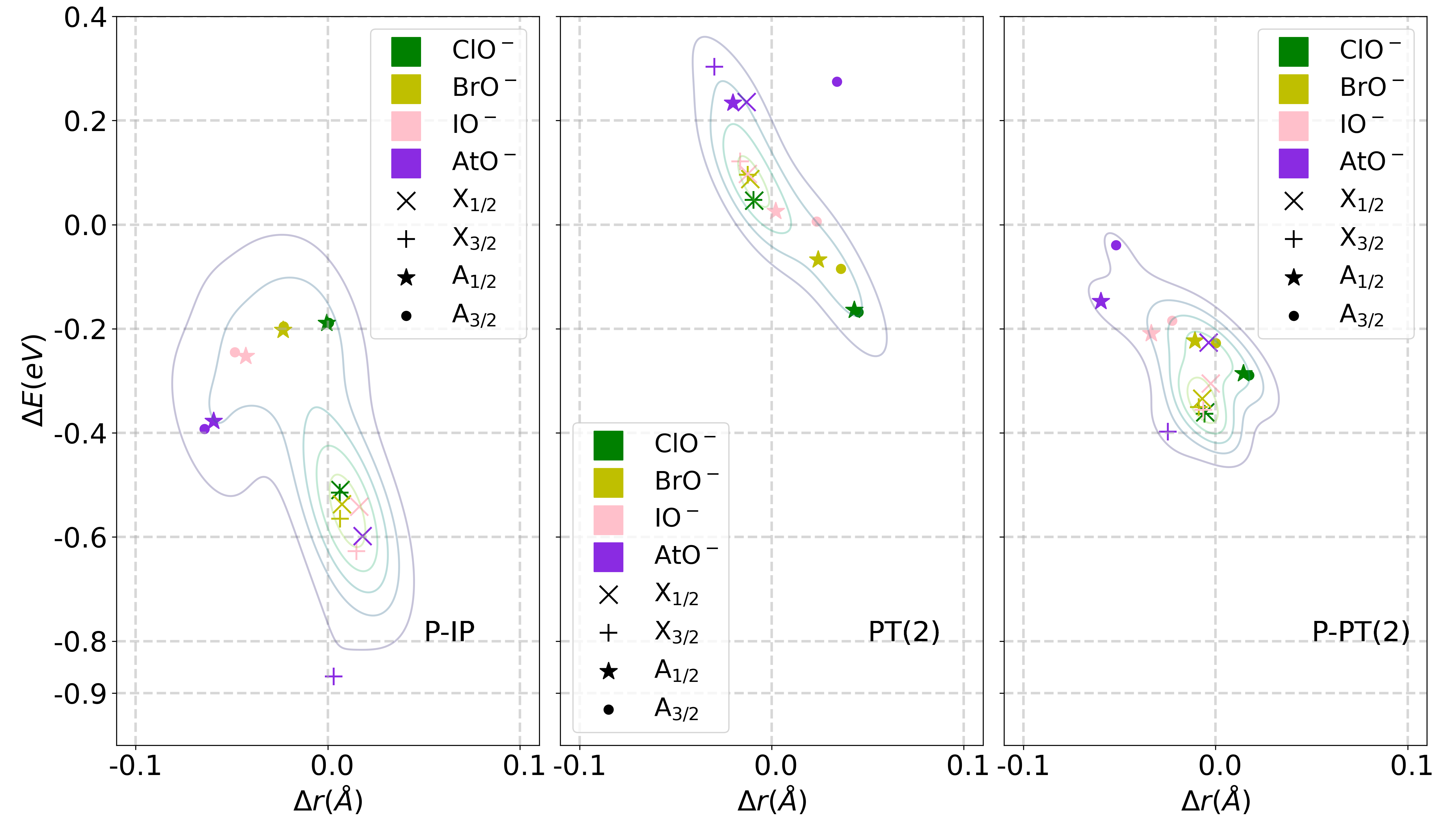}
\caption{Difference in the \emph{PES} minimums positions and energies  $ \Delta E $ (in \SI{}{\electronvolt}) w.r.t. $ \Delta r $ (\AA) 	for the 3 \IPA{} and the 4 compounds. For reasons of readability : colors for compounds and symbols for states. The 4 different isodensities represented were calculated by a Gaussian kde algorithm\cite{ScipyKDE}, resp. in [-0.1; 0.1]~($ \mathring{A} $), [-1.0; 0.4]~(\SI{}{\electronvolt}) limits for $ \Delta r $ and $ \Delta E $.\label{fig:Error_wrt_ip_App}}
\end{figure*}

To complement this comparison we provide in figure~\ref{fig:Error_wrt_ip_App} a closer look at how both energetics and equilibrium structures compare to the reference values ($\Delta E$ and $\Delta_r$, respectively) for the different species. From this figure we clearly see that results for \pip{} are rather dispersed both in energetics and geometries. 

The results for \MBPTip{} on the other hand cluster around the reference values, but not in a very homogeneous way for different species. A similar clustering is seen for \pMBPTip{}, with the systematic error for energies which has been discussed above  comparison to \MBPTip{} being clearly seen. We also observe that for chorine, bromine and iodine, \pMBPTip{} results are more compactly clustered together than for \MBPTip{} (all values are included in a zone of $\pm$\SI{ 0.1}{\electronvolt} and $\pm$\SI{0.025}{\angstrom}).

The difference between \MBPTip{} and \ip{} can be explained by a poorer description of the ground state by this approximation~\cite{Gwaltney1996}. A correction can be made by comparing the CCSD correlation energies with respect to MP2. We corrected our results by the difference between these two energies (relative to their respective minima) along the dissociation curve; this sometimes seems to  improve the results but without a general conclusion appearing. 

This compensation method can be interesting in the explanation but not very advantageous for obtaining the energies, the purpose of \MBPTip{} and  \pMBPTip{} being to avoid doing CCSD calculations. This correction could also have been made in the previous study on XO$^-$ but a change in the baseline only leads to a change in the mean and not in the standard deviation (as the distance is the same in both \ip{} methods).

So far we have discussed the errors for the individual excited state energies, however, it is also crucial to examine how the methods represent the energy differences between electronic states. If such energy differences are correctly described it becomes possible to investigage electronic transitions. Since in our discussion of the potential energy curves we have established that the errors for the different approximate methods are not constant, we restrict ourselves to a comparison around  the ground-state equilibrium structure (the Franck-Condon zone).

\begin{figure} 
	\noindent
	\includegraphics[width=0.5\textwidth]{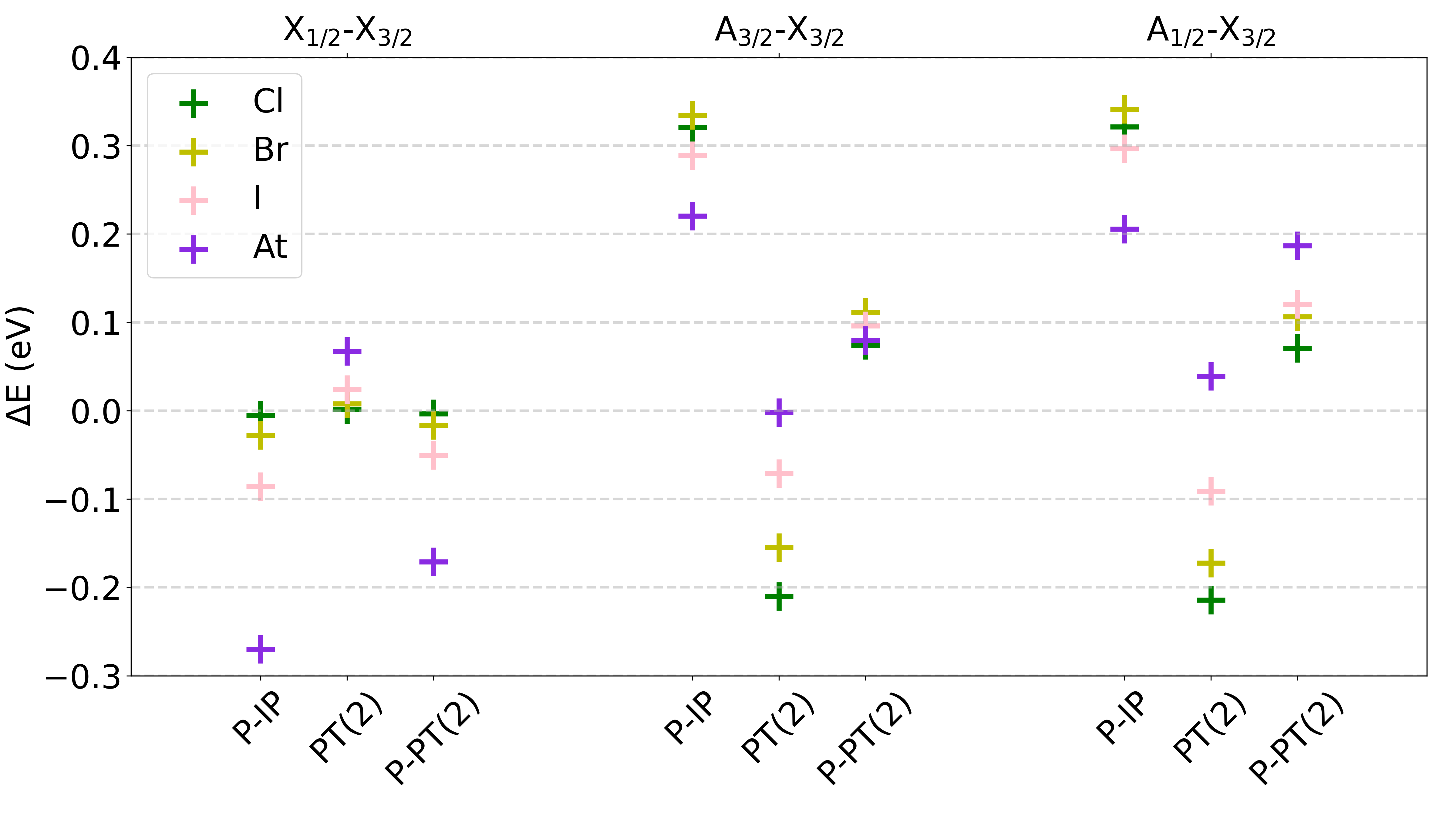}
		\caption{Error in transition energies : $\Delta$\IPA-$\Delta$\ip{} (in \SI{}{\electronvolt}), for all \XOm{X} and \IPA s (see \Supplemental{} for values).\label{fig:Error_in_Transition_Energies}}
\end{figure}

Concerning the errors in transition energies, these are shown in figure~\ref{fig:Error_in_Transition_Energies}; there, we take the lowest state (\trenteetun{}) as reference. As the first transition is between the spin-orbit components of the ground state, for chlorine and bromine results are essentially the same. For the two other compounds, the error increases for \pip{}, for \MBPTip{}, the energies are correct with deviation smaller than \SI{0.1}{\electronvolt}. For \pMBPTip{}, the error remains small up to iodine for which the error is \SI{-0.05}{\electronvolt}.

For the following two transitions (\trentedeux-\trenteetun and \douze-\trenteetun), for the same IP$_{App}$ and the same compound, the errors are essentially the same. \pip{} overestimates the energies, \MBPTip{} underestimates them but becomes more reliable for as systems become heavier (e.g.\  from ClO$^-$ to AtO$^-$).  \pMBPTip{} shows a more systematic behavior for the transitions between $\Omega = 3/2$ than between $\Omega = 3/2$ and $\Omega = 1/2$ states.

\subsubsection{Core ionizations \label{title:subsection_hx}}

On order to avoid complexifying the notation in what follows, we will continue to employ the same acronyms for ther reference and approximate methods as done for valence ionizations, but for the results all approximated methods (\IPA{}) and the reference EOM-CCSD results (\ip{}), it will be implicit that they were obtained the CVS approximation~\cite{Halbert2021}.

Apart from those, since for highly symmetric systems such as HCl, HBr and I$^{-}$ the \Dirac{}  implementation can be used to carry out a complete diagonalization (that is, obtain all possible states from the EOM-CCSD matrix without invoking the CVS approximation), we take the opportunity to present these results (denoted by the shorthand \fullip{}) and compare them to \ip{} ones, thus extending our previous comparison~\cite{Halbert2021}.

Our results are shown in Table~\ref{tab:All_IP_CVS} for K- and L-edges of HCl, the M-edges of HBr, and the K-, L- and M-edges of I$^-$. In it, we present the \fullip{} and \ip{} energies along their respective  differences (denoted by $ \Delta_f $), as well as the $\Delta E_{IP}$ values gauging the difference between approximate methods and the reference \ip{} results.

Before addressing the results for approximate methods, we see from table~\ref{tab:All_IP_CVS} that the effect of the CVS approximation is rather small, with $ \Delta_f $ values being usually of a few tenths of an~\SI{}{\electronvolt} for most edges--including highly energetic ones such as the I$^-$ K-edge--with a few exceptions that we had identified previously~\cite{Halbert2021}. We also see that for molecular systems, the hydrogen-halogen bond slightly lifts the degeneracies that are expected for each of the edges if the systems were spherically symmetric, but for these systems these splittings are not particuarly important. Taken together, these results attest to the reliability of the CVS approximation, for the discussion that follows.

\begin{table}[h!]
	\fontsize{8}{6}\selectfont
	\centering
	\caption{ 
		 Ionization energies (in \SI{}{\electronvolt}) for HCl (K- and L-edges), HBr (M-edges) and I$^-$ (K-, L- and M-edges) for the complete diagonalization of the EOM-IP-CCSD Hamiltonian (\fullip{}) and for the CVS approximation (\ip{}), as well as their difference ($ \Delta_f $, in \SI{}{\electronvolt}). Energy differences ($\Delta E_{\ip}$, in \SI{}{\electronvolt}) between approximate methods and \ip{}. We note that \%SI \IPA{} are included in the range [87\%; 88\%], [89\%; 91\%] and [88\%; 92\%] respectively for HCl, HBr and I$^-$. The differences between labels $(a) $, $(b) $ and $(c)$ denote the splitting of states due to the deviation from spherical symmetry in the presence of the hydrogen atom.
	}
	\begin{tabular}{l r rr r r r }
		\toprule
		& \fullip{} &\multicolumn{2}{c}{CVS-\ip{}  
		}
		&  
		  \multicolumn{3}{c}{$\Delta E_{\ip}  $}  \\
  \cmidrule(l{3pt}r{3pt}){3-4}   \cmidrule(l{3pt}r{3pt}){5-7} 
		T&	E$_{f} $ & E$_{IP}$  & $\Delta_{f}  $   &    
		  \pip{}
		&  \MBPTip{}   &  \pMBPTip{} \\
		\midrule
 		\midrule 
 & \multicolumn{6}{c}{HCl} \\
		\midrule
$K$&{2835.46}& {2835.76}& {0.30}&{-0.45}&{0.23}& {-0.20}\\
		$L_1$&{280.25}&{280.27}&{0.02}&{-0.14}&{0.12}&{-0.01}\\
		$L_2$&{209.66}&{209.59}&{-0.07}&{-0.06}&{0.13}&{0.09}\\
		$L_3(a) $&{208.02}&{207.93}&{-0.09}&{-0.15}&{0.13}&{0.09}\\
		$L_3(b)$&{207.94}&{207.84}&{-0.10}&{-0.06}&{0.13}&{0.08}\\
  \midrule
   & \multicolumn{6}{c}{HBr} \\
  \midrule
$M_2$&{200.77}&{200.92}&{0.16}&{-1.15}&{0.15}&{-0.85}\\
		$M_3(a)$
&{194.21}&{194.00}&{-0.21}&{-1.13}&{0.15}&{-0.83} \\

$M_3(b)$&{194.15}&{193.84}&{-0.31}&{-1.13}&{0.14}&{-0.84}\\
$M_4(a)$&{78.03}&{78.13}&{0.10}&{-1.15}&{0.23}&{-0.78}\\
$M_4(b)$&{77.84}&{77.90}&{0.06}&{-1.16}&{0.23}&{-0.80}\\
  $M_5(a)$&{77.00}&{77.05}&{0.06}&{-1.13}&{0.23} &{-0.77}\\
$M_5(b)$&{77.01}&{76.93}&{-0.08}&{-1.13}&{0.23}&{-0.78}\\
$M_5(c)$&{76.88}&{76.74}&{-0.13}&{-1.14}&{0.22}&{-0.79}\\
  \midrule
 & \multicolumn{6}{c}{I$^-$} \\
  \midrule
   $K$   & 33290.63 &  33290.72 & 0.09 &  -5.77 &  1.27 &   -4.25 \\
 $L_1$  & 5209.13 &   5209.29 & 0.16 &  -4.35 &  1.02 &   -3.11 \\
 $L_2$  & 4866.09 &    4867.68 & 1.59 & -4.67 &   1.07 &  -3.37 \\
 $L_3$     & 4567.06 &    4567.67  & 0.61 &  -4.53 &   1.06 &  -3.25 \\
 $M_1 $  &1078.43&   1078.27 & -0.16  &  -2.25 &  0.72&   -1.41 \\
 $M_2$ &     937.25&     936.37 &  0.12 &-2.31&   0.75 &  -1.44 \\
 $M_3$ & 878.64 &     878.57  & -0.06  & -2.21 &   0.74 &  -1.36 \\
 $M_4$ &631.93 &     631.67 & -0.26 &  -2.42 &   0.80 &  -1.50 \\
 $M_5$& 618.89 &  619.81 &  0.91 &  -2.39 &   0.79  &  -1.48 \\
		\bottomrule
	\end{tabular}
	\label{tab:All_IP_CVS}
\end{table}

Turning now to the performance of the approximate methods, we begin with the results HCl. We observe, for \MBPTip{} a very homogeneous error for the L-edge at around \SI{0.13}{\electronvolt}, and a slightly larger error (\SI{0.23}{\electronvolt}) for the K-edge. For \pip{}, we see errors of opposite sign compared to \MBPTip{}, indicating once more that the reference values are underestimated. More interestingly they are, in absolute value, twice as large for the K-edge, comparable for the $L_1$- and $L_3(a)$- edges, and slightly smaller for the other edges. 

In the \pMBPTip{}, we see a pattern of error cancellation which is very similar to that found for the valence ionization, though since there are much more significant differences between \MBPTip{} and \pip{} for the K-edge than for the others, we see a much larger value of $\Delta E_{IP}$. 

Moving on to HBr, we only compare the M-edge which is of similar or lower energy than the K- and L-edges of HCl. Perhaps unsurprisingly, for \MBPTip{} we see $\Delta E_{\ip}$ values with are similar to, or slighlty larger than those for HCl, attesting to the very systematic nature of this approximation. For \pip{} we also have quite systematic errors, but they underestimate the reference more significantly than for HCl, and now in absolute value the $\Delta E_{IP}$ are nearly \SI{1}{\electronvolt} larger than for \MBPTip{}; and consequently \pMBPTip{} are also affected and in absolute value $\Delta E_{IP}$ are sigificantly larger than for \MBPTip{}.

A similar trend emerges for I$^-$, with \MBPTip{} showing a tendency to overestimate the reference (by about \SI{1}{\electronvolt}), with $\Delta E_{IP}$ which are comparable within each edge. The \pip{} and \MBPTip{} approaches also show $\Delta E_{IP}$ values which are very similar within each edge, but both show $\Delta E_{IP}$ values which in absolute terms are significatly (between two to five times) larger than the \MBPTip{} values. 

\begin{figure} 
	\noindent
	\includegraphics[width=0.5\textwidth]{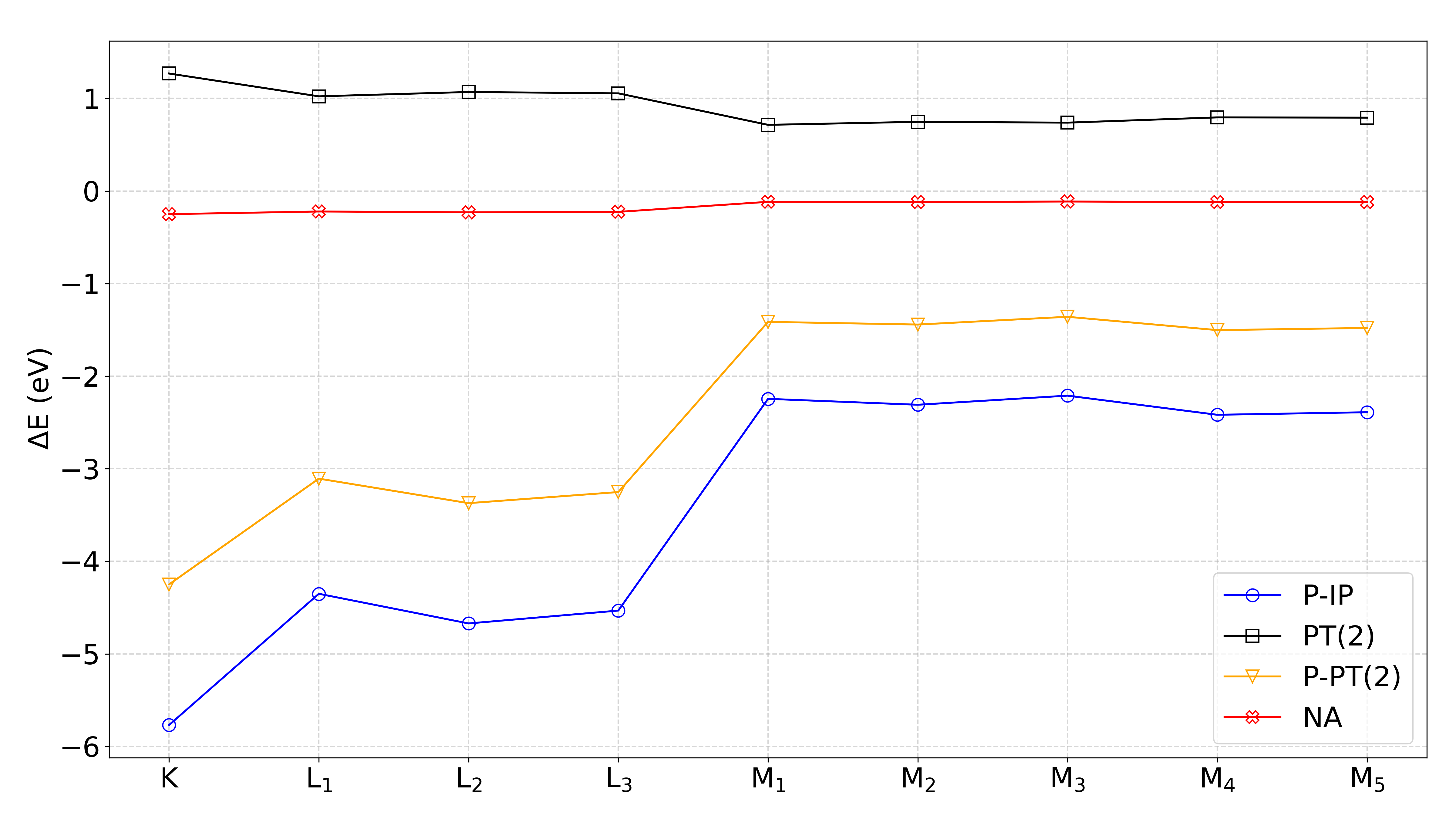}
	 \caption{Values of $\Delta E_{IP}$ (in \SI{}{\electronvolt}) for the K-, L- and M- edges of I$^{-}$ for the \MBPTip{}, \pip{} and \pMBPTip{} approximations, as well as a measure of the non-additivity (NA) of the partition and second-order approximations.\label{fig:Im1_CVS_Diff}}
\end{figure}

In figure~\ref{fig:Im1_CVS_Diff} we provide a graphical representation of $\Delta E_{IP}$ shown in table~\ref{tab:All_IP_CVS} above for each of the approximate methods, adding to it (red) a measure of the non-additivity (NA) of the partition and second-order approximations:
\begin{equation}
\mathrm{NA}=\Delta E_{IP}(\mathrm{\MBPT{}}) - \left[\Delta E_{IP}(\mathrm{\pMBPT{}}) - \Delta E_{IP}(\mathrm{\pip{}})\right],
\end{equation}
from which we see that there is a small non-additivity, that is slightly larger for the K and L edges with respect to the M edges.  

To explain the processes behind these differences, one should first recall that as discussed elsewhere (see e.g.~\citet{Halbert2021} and references therein), for the determination of core ionization energies, the contribution to the electronic states' energies of relaxation effects due to the creation of the core hole can be equally or more important than electron correlation effects as discussed by~\citet{South2016}.

Furthermore, relaxation will be increasingly more important as one probes more energetic edges. With that, electronic structure methods that insufficiently account for relaxation (for example, using Koopmans theorem to approximate ionization energies by the absolute value of orbital energies) will in general largely overestimate core binding energies  and show larger discrepancies from experiment for more energetic edges than for less energetic ones (by several tens of \SI{}{\electronvolt} depending on the case).

Methods such as EOM-CCSD on the other hand have shown to be capable of accounting for a significant part of relaxation, and when combined with corrections for quantum electrodynamics effects and the Breit interaction are taken into account, one can obtain energies in agreement with experiment to within about 1-2~\SI{}{\electronvolt}~\cite{Halbert2021,Opoku2022}, and show fairly constant errors for different edges.

In view of these results from the literature and our results above, in which we see fairly constant $\Delta E_{IP}$ values for \MBPTip{} (but more energetic edges showing slightly larger values than less energetics ones), while at the same time observing a net increase in the magnitude of $\Delta E_{IP}$ for both \pip{} and \pMBPTip{} as one goes towards (a) heavier elements; and (b) more energetic edges, we are led to the conclusion that for core ionizations \MBPTip{} does a better job at accounting for relaxation than the partition-based approaches, due to the approximation to the $\bar{H}_{DD}$ block in the latter two. 

That said, it is important to recall that all $\Delta E_{IP}$ values shown here remain significant smaller than what would be the case if cruder approximations were employed (e.g.\ Koopmans theorem), meaning that these approximate methods (and in particular \MBPTip{} and \pMBPTip{}, due to their more modest computational cost) would remain of interest for the calculation of core ionizations.

Finally, we note that for all ionizations considered, we have \%SI around [87\%; 88\%] for HCl, [89\%; 91\%] for HBr and [88\%; 92\%] for I$^-$, meaning that for all methods and systems we have been able to compare ionized states whose wavefunctions are dominated by the same ionized determinant. It is also interesting to see that for the core spectra of these species we did not see the partition-based approaches increasing the states' multideterminantal character.

\subsection{Electronic Affinities \label{title:EA}}

Addressing now electron affinities, we summarize our results in table~\ref{tab:All_EA}. We see that for all systems considered (I$_3^-$, CH$_2$IBr and CH$_2$I$_2$), we have rather small and uniform errors for all approximations, so it is actually quite difficult to  decide which one shows better performance as done for ionization energies. That said, we still see that \MBPTea{} still appears to be somewhat more systematic than the partition-based approaches \pea{} and \pMBPTea{}.

For these systems most electron affinities are positive and relatively large, corresponding to situation in which we have metastable states. For these cases, we see that the error introduced by the approximations are rather negligible in comparison to the electron affinities themselves.

For CH$_2$IBr and CH$_2$I$_2$, on the other hand, the first electron affinity is negative, denoting bound states that are nevertheless not very stable--\SI{-0.03}{\electronvolt} for CH$_2$IBr and \SI{-0.33}{\electronvolt} for CH$_2$I$_2$ respectively. We see that all of our approximated methods tend to underestimate the energies of these bound states. That makes them yield first electron attachments states which are also bound, and therefore qualitatively in line with EOM-CCSD, though slightly more stable than the latter.

These results underscore the fact that, by introducting approximations one invariably loses in accuracy, and consequently in predictive power when tiny energy differences are involved. This serves as a reminder of the importance of attempting to investigate such subtle effects with a range of methods of verying degrees of accuracy to verify whether computationally less expensive methods are sufficiently accurate.

It is noteworthy that the \%SA are sometimes over 90\%, but also as low as 50\%, indicating the presence of at least one or more configurations, though the values are very similar across all methods for each of the individual electronic states.


\begin{table}
	\fontsize{8}{6}\selectfont
	\centering
	\caption{Electron affinities and $\Delta E_{\ea}$ values (in \SI{}{\electronvolt}) as well as \%SA values for selected low-lying states of the I$_3^-$, CH$_2$IBr and CH$_2$I$_2$ molecules.}
	\begin{tabular}{l rr rr rr rr}
		\toprule
		& \multicolumn{2}{c}{\ea} 
		&\multicolumn{2}{c}{\pea}&\multicolumn{2}{c}{\MBPTea{}} &
		\multicolumn{2}{c}{\pMBPTea{}}\\
   \cmidrule(l{3pt}r{3pt}){2-3}   \cmidrule(l{3pt}r{3pt}){4-5}  \cmidrule(l{3pt}r{3pt}){6-7}   \cmidrule(l{3pt}r{3pt}){8-9} 
		T& 	E &    \%SA &    \%SA &  $\Delta E_{\ea}  $ &   \%SA
		&  $\Delta E_{\ea}   $ & \%SA &  $\Delta E_{\ea}   $ \\
		\midrule
  \midrule
 & \multicolumn{8}{c}{I$_3^{-}$} \\
  \midrule
		EA$_{1}$ & 2.51 & 76 &     76 &    -0.02 &        75 &  -0.01 &         74 &        -0.05 \\
		EA$_{2}$ & 3.65 & 77 &     77 &     0.01 &        76 &        
		0.01 
		&         75 &         0.01 \\
		EA$_{3}$ & 3.88 & 94 &     94 &     0.01 &        94 &        
		0.02 
		&         95 &         0.03 \\
		EA$_{4}$ & 4.39 & 96 &     96 &     0.01 &        96 &        
		0.00 &         96 &         0.01 \\
\midrule
& \multicolumn{8}{c}{CH$_2$IBr} \\
\midrule
  	EA$_1$&{-0.03}&{52}&{53}&{-0.04}&{52}&{-0.06}&{53}&{-0.08}\\
	EA$_2$&{0.45}&{66}&{70}&{0.00}&{67}&{-0.01}&{70}&{-0.00}\\
	EA$_3$&{0.86}&{97}&{97}&{0.01}&{97}&{-0.01}&{97}&{0.00}\\
	EA$_4$&{0.99}&{87}&{85}&{-0.00}&{86}&{-0.01}&{85}&{-0.01}\\
	EA$_5$&{1.62}&{70}&{69}&{0.01}&{68}&{-0.01}&{67}&{0.01}\\
	EA$_6$&{1.92}&{52}&{49}&{0.00}&{55}&{-0.01}&{48}&{-0.00}\\
\midrule
& \multicolumn{8}{c}{CH$_2$I$_2$} \\
\midrule
	EA$_1$&{-0.33}&{48}&{50}&{-0.06}&{48}&{-0.06}&{49}&{-0.10}\\
	EA$_2$&{0.41}&{57}&{60}&{0.01}&{58}&{-0.01}&{60}&{0.00}\\
	EA$_3$&{0.75}&{83}&{81}&{-0.02}&{82}&{-0.01}&{80}&{-0.02}\\
	EA$_4$&{0.86}&{98}&{98}&{0.01}&{98}&{-0.01}&{98}&{0.00}\\
	EA$_5$&{1.49}&{87}&{86}&{0.01}&{86}&{-0.00}&{85}&{0.01}\\
	EA$_6$&{1.70}&{67}&{62}&{-0.01}&{65}&{-0.01}&{61}&{-0.01}\\
		\bottomrule
	\end{tabular}
	\label{tab:All_EA}
\end{table}

\subsection{Excitation Energies \label{title:EE}}

As a conclusion, we focus on the calculation of excitations energies with our approximate methods. Our results are reported on table~\ref{tab:EE_G_tab} for the triiodide I$_3^-$ and dihalomethane  CH$_2$I$_2$ species. 

Starting with the triiodide results, we see that our three approaches work rather well as far as the standard deviation of $\Delta E_{\ee}$ values is concerned-these are all small, meaning that that the errors for \pee{} and \MBPTee{} are shifts of roughly \SI{0.27}{\electronvolt} for the first (meaning it it overestimates the reference values) and nearly zero for the second. Due to this behavior for \pMBPTee{} in this particular example the results are nearly identical to those of \pee{} and there are no significant error cancellations, unlike the case if ionization energies as pointed out in the case of ionization energies.

One thing to note is that in cases in which states are nearly degenerate such as $\left\lbrace 1_g;\ 0^{-}_{u};\ 1_u \right\rbrace$, the approximate methods have a difficult time in placing them in the same order as found in EOM-CCSD (see numbers in bold on table \ref{tab:EE_G_tab}). Furthermore, we note the presence of non-standard values for the first $1_g$ and $1_u$ by \MBPTee{} and \pMBPTee{}. Indeed, while \MBPTee{} presents a difference of at most \SI{0.01}{\electronvolt} in absolute value, the error for this level is \SI{0.11}{\electronvolt}. The same is true for \pMBPTee{} which goes from an error of $\SI{0.27}{\electronvolt}$ to $\SI{0.37}{\electronvolt}$. These differences to EOM-CCSD are in line with limitations pointed out above for electron affinities.

Similar trends are found for CH$_2$I$_2$. For this system, standard deviations are once again satisfactory, of the order of \SI{0.2}{\electronvolt}, with the averages are respectively \SI{0.31}{\electronvolt}, \SI{-0.07}{\electronvolt} and \SI{0.25}{\electronvolt} for \pee{}, \MBPTee{} and \pMBPTee{} respectively. This means that we observe very systematic but nevertheless larger overestimation \pee{}, and now a slight underestimation by  \MBPTee{}, which is actually the opposite trend found for for ionization energies, but that means that \pMBPTee{} again profits from error calcellation between the second-order and partition approximations. As in the case of electron affinities, we find that overall \%EE values are also very similar between different approximate methods, in spite of the fluctuations from state to state.

Finally, for the comparison with CASPT(2) results for I$_3^-$~\cite{Gomes2010} and CH$_2$I$_2$ by~\citet{Liu2007_CH2I2}, we see that in terms of the mean error the \MBPTee{} performs better than CASPT2 for both molecules, whereas \pee{} and \pMBPTee{} shows a simular behavior. However, in contrast to our methods, the $\Delta E_{\ee}$ values for CASPT2 vary much more significantly from state to state, as reflected in its much larger standard deviations. 

This tendency of varying $\Delta E_{\ee}$ for different excited states had already been identified by~\citet{Gomes2010} and~\citet{Shee2018}, but also in other benchmarks for heavy elements such as those by~\citet{Tecmer2011} and~\citet{Tecmer2014}. We consider it a significant advantage of the approximate methods over CASPT2 that former conserve such near constant difference to EOM-CCSD values. This characteristic reduces the risk of incorrectly ordering electronic states, enhancing the reliability of the approximate methods in this regard.

\begin{table}[htbp]
	\fontsize{8}{6}\selectfont
	\centering
	\caption{Excitation energies and $\Delta E_{\ee}$ values (in \SI{}{\electronvolt}) as well as \%SE values for selected low-lying excited states of the I$_3^-$ and CH$_2$I$_2$ molecules. In bold we denote electronic states whose order has been inverted in approximate calculations with respect to the reference EOM-CCSD results. Here $\bar{x}$ and $\sigma $ denote the mean error over all excited states considered for a particular species and the standard deviation, respectively. CASPT2 values from~\citet{Gomes2010} for I$_3^-$ and from~\citet{Liu2007_CH2I2} for CH$_2$I$_2$ are also provided for comparison.}
	\begin{tabular}{l rr rr rr rr rr }
		\toprule
		& \multicolumn{2}{c}{\ee} &\multicolumn{2}{c}{\pee}
		&\multicolumn{2}{c}{\MBPTee{}}&\multicolumn{2}{c}{\pMBPTee{}}    
		&\multicolumn{2}{c}{CASPT2} \\
       \cmidrule(l{3pt}r{3pt}){2-3}   \cmidrule(l{3pt}r{3pt}){4-5}  \cmidrule(l{3pt}r{3pt}){6-7}   \cmidrule(l{3pt}r{3pt}){8-9}   \cmidrule(l{3pt}r{3pt}){10-11}
		T&	EE  &    \%SE & \%SE   &  $\Delta E_{\ee} $   &  \%SE&  
		$\Delta E_{\ee} $  & \%SE & $\Delta E_{\ee}  $  & E & $\Delta E_{\ee} $  \\
		\midrule
  & \multicolumn{10}{c}{I$_3^-$} \\
  \midrule
		2$_g$&{2.25}&{43}&{45}&{0.27}&{43}&{0.00}&{45}&{0.27}&{2.24}&{-0.01}\\
		1$_g$&{2.38}&{23}&{24}&{0.25}&{23}&\textbf{0.11}&{24}&\textbf{0.37}&{2.32}&{-0.06}\\
		0$_u^-$&{2.38}&{41}&{42}&\textbf{0.27}&{41}&\textbf{0.00}&{42}&\textbf{0.27}&{2.47}&{0.09}\\
		1$_u$&{2.38}&{46}&{47}&\textbf{0.25}&{46}&{0.11}&{47}&{0.37}&{2.47}&{0.09}\\
		0$_g^-$&{2.84}&{21}&{22}&{0.27}&{22}&{0.00}&{22}&{0.28}&{2.76}&{-0.08}\\
		0$_g^+$&{2.89}&{21}&{22}&{0.27}&{21}&{0.00}&{22}&{0.27}&{2.82}&{-0.07}\\
		1$_g$&{3.07}&{41}&{43}&{0.28}&{41}&{0.01}&{43}&{0.28}&{2.85}&{-0.22}\\
		2$_u$&{3.33}&{48}&{49}&{0.26}&{48}&{-0.01}&{49}&{0.24}&{3.10}&{-0.23}\\
		1$_u$&{3.42}&{48}&{49}&{0.26}&{48}&{-0.01}&{49}&{0.25}&{3.11}&{-0.31}\\
		$\mathbf{0_u^{+}}$&{3.67}&{18}&{19}&{0.24}&{17}&{-0.01}&{19}&{0.23}&{3.52}&{-0.15}\\
		2$_g$&{4.08}&{24}&{24}&{0.27}&{24}&{-0.01}&{24}&{0.26}&{3.98}&{-0.10}\\
		0$_u^-$&{4.10}&{42}&{43}&{0.29}&{42}&{-0.01}&{43}&{0.28}&{3.79}&{-0.31}\\
		1$_g$&{4.18}&{47}&{48}&{0.28}&{47}&{-0.01}&{49}&{0.27}&{4.06}&{-0.12}\\
		1$_u$&{4.22}&{40}&{41}&{0.30}&{40}&{-0.00}&{40}&{0.29}&{3.80}&{-0.42}\\
		$\mathbf{0_u^{+}}$&{4.49}&{15}&{16}&{0.24}&{15}&{0.00}&{17}&{0.24}&{4.51}&{0.02}\\
		0$_g^-$&{4.69}&{20}&{21}&{0.30}&{20}&{-0.01}&{21}&{0.29}&{4.51}&{-0.18}\\
		0$_g^+$&{4.70}&{20}&{21}&{0.30}&{20}&{-0.01}&{21}&{0.29}&{4.53}&{-0.17}\\
		1$_g$&{4.90}&{40}&{36}&\textbf{0.30}&{40}&{0.01}&{26}&\textbf{0.33}&{4.60}&{-0.30}\\
		\midrule
		$\bar{x}$&&&&{0.27}&&{0.01}&&{0.28}&&{-0.14}\\
		$\sigma$ 
		&&&&{0.02}&&{0.04}&&{0.04}&&{0.14}\\
  \midrule
  & \multicolumn{10}{c}{CH$_2$I$_2$} \\
  \midrule
		\text{a}&{3.60}&{20}&{21}&{0.29}&{20}&{-0.07}&{21}&{0.23}&{3.76}&{0.16}\\
		\text{b}&{3.62}&{20}&{20}&{0.29}&{20}&{-0.07}&{20}&{0.23}&{3.78}&{0.16}\\
		\text{a}&{3.62}&{20}&{21}&{0.29}&{20}&{-0.06}&{21}&{0.23}&{3.78}&{0.15}\\
		\text{b}&{3.84}&{16}&{17}&{0.30}&{15}&{-0.07}&{16}&{0.23}&{4.03}&{0.19}\\
		\text{a}&{3.87}&{19}&{19}&{0.29}&{19}&{-0.07}&{19}&{0.23}&{4.27}&{0.40}\\
		\text{b}&{3.94}&{13}&{13}&{0.31}&{12}&{-0.08}&{13}&{0.24}&{4.27}&{0.33}\\
		\text{a}&{3.99}&{16}&{17}&{0.30}&{16}&{-0.08}&{17}&{0.23}&{4.31}&{0.32}\\
		\text{b}&{4.06}&{16}&{16}&{0.31}&{16}&{-0.07}&{16}&{0.24}&{4.38}&{0.32}\\
		\text{b}&{4.22}&{18}&{19}&{0.30}&{18}&{-0.06}&{19}&{0.25}&{4.50}&{0.28}\\
		\text{a}&{4.32}&{15}&{16}&{0.31}&{15}&{-0.07}&{16}&{0.24}&{4.60}&{0.28}\\
		\text{b}&{4.35}&{15}&{15}&{0.31}&{15}&{-0.06}&{15}&{0.26}&{4.62}&{0.27}\\
		\text{a}&{4.49}&{15}&{15}&{0.31}&{15}&{-0.07}&{16}&{0.25}&      &\\
		\text{b}&{4.63}&{15}&{15}&{0.33}&{15}&{-0.07}&{15}&{0.27}&      &\\
		\text{a}&{4.68}&{17}&{17}&{0.35}&{17}&{-0.06}&{18}&{0.30}&      &\\
		\text{b}&{4.74}&{15}&{14}&{0.34}&{15}&{-0.07}&{15}&{0.28}&      &\\
		\text{a}&{4.91}&{14}&{15}&{0.36}&{15}&{-0.06}&{15}&{0.30}&      &\\
		\midrule
		$\bar{x}$ &  & &  &   0.31 & &     -0.07
		&          &       0.25 & & 0.26 \\
		$\sigma$ & &   &       &   0.02 &       &      0.01 &   &       0.02  & & 0.08\\
		\bottomrule
	\end{tabular}
	\label{tab:EE_G_tab}
\end{table}

	\section{Conclusion}
	
In this manuscript we have detailed a pilot impementation of three approximate methods based on the relativistic EOM-CCSD method--the partitioned EOM-CCSD (P-EOM-CCSD), the second-order approximation to EOM-CCSD (EOM-MBPT(2)) and the partitioned \EOMMBPT (\pEOMMBPT)--and applied them to a number of benchmark systems to assess how they compare with respect to the original EOM-CCSD for core and valence ionization, electron affinities and valence excitation energies.

These approximated methods have been shown to provide, for the most part, sufficiently accurate results with respect to the reference EOM-CCSD calculations. As a general rule, \EOMMBPT{} has shown to overestimate reference values in a very systematic manner for ionizations and electron affinities, often by no more than a few tenths of \SI{}{\electronvolt} for valence ionizations and somewhat less for electron affinities, and at most around 1-2 \SI{}{\electronvolt} for core ionizations (employing the core-valence separation method). For excitation energies on the other hand it tends to slighly underestimate the EOM-CCSD results.

Conversely, the \pEOM{} method tends to systematically underestimate EOM-CCSD results for ionizations and electron affinities, while overestimating excitation energies. For ionizations and excitations, magnitude of the errors is in general larger than those for the \EOMMBPT{} method, but rather similar for electron affinities. We have found however that for ionization processes in which relaxation effects brought about by 2h1p configurations are important (here, the valence ionization of tennessine monoxide, and the core ionizations of halogen-containing systems), the approximation underlying the \pEOM{} method (discarding all off-diagonal elements of $\bar{H}_{DD}$ and approximating the diagonal by orbital energy differences) degrades the reliability of the method, and results differ from the reference EOM-CCSD the more such relaxation effects are important. 

Because it combines both approximations, the \pEOMMBPT{} shows a behavior that is intermediate between \EOMMBPT{} and \pEOM{} due to error cancellation. It should be said however that the method does show interesting and stable systematic errors, and will suffer less of a performance degradation that \pEOM{} even for cases which are particularly difficult for the latter such as the valence ionization of tennesine oxide. 

Another finding is that for ionizations \pEOM{}, and to a lesser extent \pEOMMBPT{}, seem to somewhat reduce the monodeterminantal nature of ionized states, compared to the EOM-CCSD and \EOMMBPT{} calculations, though we consider these to be a consequence of the partitioned approximation rather than an underlying cause for differences in performance. This point requires investigations on a broader class of complexes, that goes beyond the scope of this work.

With respect to their behavior across different structures, our investigation of the potential energy curves for the spin-orbit split ground and first excited states of halogen monoxides shows that all of these methods have different error cancellation patterns in different regions of the potential energy curves, and therefore should be used with caution. That said, the \EOMMBPT{} and \pEOMMBPT{} methods have shown to yield relatively small errors in equilibrium bond lenghts for ClO, BrO and IO, with a small advantage for \pEOMMBPT{}, though for AtO the reverse appears to be true.

With that, \EOMMBPT{} seems to be a good first alternative to EOM-CCSD among the methods considered for carrying out exploratory calculations for valence and core properties on systems across the periodic table, followed by \pEOMMBPT{}. On our view \pEOM{} does not seem to show a very interesting performance to cost ratio, especially for heavy and superheavy elements and should therefore be avoided, unless additional evidence for more molecular systems demonstrates other situations in which  \pEOM{} fares as well or better than \EOMMBPT{} or \pEOMMBPT{}. 

Furthermore, we observe that in the limited comparisons to CASPT2 for excitation energies, \EOMMBPT{} shows overall better (and more systematic) agreement to EOM-CCSD while \pEOMMBPT{} results deviate from EOM-CCSD also in a more systematic way than CASPT2, though in absolute values their errors are comparable to CASPT2. This suggests these approximate EOM methods presented here can be viable alternatives to CASPT2 for investigating the spectroscopy of molecules with a single-reference ground states.

As a perspective for this work we have the implementation and subsequent exploration of additional approximate methods, such as the CC2 and EOM-CC2, in the simulation of  heavy elements systems. There is a need for economical approaches for exploring (core) excited states of heavy element systems, which by denfinition contain many more inner electrons than first or second-row systems, but at the same time little if any knowledge on how these approaches will behave in such cases. These efforts are to be carried out as part of our further development of the ExaCorr code in \Dirac{}, which lifts many of the limitations of the RELCCSD module employed here for the treatment of larger molecular systems.

\begin{acknowledgments}
We acknowledge support from PIA ANR project CaPPA (ANR-11-LABX-0005-01), I-SITE ULNE projects OVERSEE and MESONM International Associated Laboratory (LAI) (ANR-16-IDEX-0004), the French Ministry of Higher Education and Research, region Hauts de France council and European Regional Development Fund (ERDF), project CPER WAVETECH, and the French national supercomputing facilities (grants DARI A0090801859, A0110801859). ASPG acknowledges support from the Franco-German project CompRIXS (Agence nationale de la recherche ANR-19-CE29-0019, Deutsche Forschungsgemeinschaft JA 2329/6-1).
\end{acknowledgments}

\appendix

\section{ Equations for approximated methods\label{title:appendix_1}}

Note that for the following equations : Internal sommations have been 
omitted and $\Pm pq$ is a permutation operator~:  $ \Pm pq 
g(\dots,p,q,\dots)=  g(\dots,p,q,\dots) -g(\dots,q,p,\dots)$.

As an exemple of the link between the following equations and the $\sigma$ vector used in the Davidson procedure, and \emph{Core-Valence-Separation} (eq.\ref{eq:CVS}) we have (see~\citet{Shee2018}):
\begin{align}
    \sigma_i^a &= [\bar H_{SS}\Rr]^{a}_{i} + [\bar H_{SD}\Rr]^{a}_{i} \\
    \sigma_{ij}^{ab} &= [\bar H_{DS}\Rr]_{ij}^{ab} + [\bar H_{DD}\Rr]_{ij}^{ab} 
\end{align}

\subsection{MBPT(2) - EOM-EE}
\begin{align}
[\bar H_{SS}\Rr]^{a}_{i}=&+ \textcolor{black}{\Fb ae} \rs ei 
-\textcolor{black}{\Fb mi} \rs am +\textcolor{black}{\W maei} \rs em
 \\
[\bar H_{SD}\Rr]^{a}_{i}=&  +\Fb me \rd eami  + \undemi \W amef \rd efim 
- 
\undemi \W mnie\rd 
eamn \\
[\bar H_{DS}\Rr]_{ij}^{ab} =& -\Pm ab \textcolor{black}{\W amij} \rs bm 
+\Pm ij 
\textcolor{black}{\W abej} \rs ei \nonumber \\
&+ 
\Pm ab \left( \textcolor{black}{\V bmfe} \rs em \right) \td afij 
\textcolor{black}{-} \Pm ij \left( 
\textcolor{black}{\V nmje} \rs em 
\right) \td abin \\
[\bar H_{DD}\Rr]_{ij}^{ab}=&+ \Pm ab \textcolor{black}{\Fb be} \rd aeij - 
\Pm ij 
\textcolor{black}{\Fb mj} \rd abim \nonumber \\
&+ \undemi \textcolor{black}{\W mnij} 
\rd 
abmn + \Pm ab \Pm ij \textcolor{black}{\W mbej} \rd ae im \nonumber \\
&-\undemi \Pm ab \left( \V nmfe \rd eamn \right) \td fbij - \undemi \Pm 
ij 
\left( \V nmfe \rd fejm \right) \td abin \nonumber\\
& +\undemi \textcolor{black}{\W 
	abef} 
\rd efij
\end{align}

\subsection{MBPT(2) - EOM-IP}

\begin{align}
[\bar H_{SS}\Rr]_i&= - \textcolor{black}{\Fb mi} \rs {}m \\
[\bar H_{SD}\Rr]_i&= +\Fb me \rd e{}mi  - 
\undemi 
\W mnie 
\rd e{}nm \\
[\bar H_{DS}\Rr]_{ij}^{a}&=\textcolor{black}{-} \textcolor{black}{\W amij} \rs 
{}m \\
[\bar H_{DD}\Rr]_{ij}^{a}&=- \Pm ij \textcolor{black}{\Fb mj} \rd 
a{}im +   \textcolor{black}{\Fb ae}\rd 
e{}ij + \textcolor{black}{\W mnij} \rd a{}mn  \nonumber \\
&  +  
\Pm ij  \textcolor{black}{\W amie} 
\rd e{}mj - \undemi \left( \V mnef \rd f{}nm \right) 
\td aeij  
\end{align}

\subsection{MBPT(2) - EOM-EA}

\begin{align}
[\bar H_{SS}\Rr]^a&=+\textcolor{black}{\Fb ae} \rs e{} 
\\
[\bar H_{SD}\Rr]^a&=+\Fb me \rd ea{}m + 
\undemi  \W 
amef  \rd fem{} \\
[\bar H_{DS}\Rr]_{i}^{ab} &= + \textcolor{black}{\W abie} \rs e{} \\
[\bar H_{DD}\Rr]_{i}^{ab}&=+\Pm ab  \textcolor{black}{\Fb ae} \rd 
ebi{}  
-  \textcolor{black}{\Fb  mi}\rd 
abm{} + \Pm ab  \textcolor{black}{\W amie} 
\rd 
ebm{}  
\nonumber \\
&+\undemi \textcolor{black}{\W abef} \rd efi{}  
-\undemi 
\left( \V mnfe \rd 
efn{}  \right) \td abim 
\end{align}

\subsection{MBPT(2) - EOM - Intermediates }

\begin{align}
\allowdisplaybreaks[1]
 \Fb vv : \Fb ea & = \f ea - \f ma \ts em 
\textcolor{black}{+} \V mefa \ts 
fm - 
\undemi \V mnaf \td efmn \\
\Fb oo : \Fb im &= \f im + \f ie \ts em + \V inme \ts en 
+ \undemi \V inef \td efmn \\
 \W oooo : \W ijmn &= \V ijmn +\Pm mn \V ijen \ts em 
+ 
\undemi \V ijef \td efmn \\
\W vvvv : \W  abef &= \V  abef  - \Pm ab \ts bm \V amef  
+ 
\undemi  \td abmn \V mnef \\
 \W ovvo : \W mbej &= \V mbej + \V mbef  \ts fj - \V 
mnej \ts bn \textcolor{black}{-} \V mnef \left( \td fbjn \right)\\
 \W vvvo : \W efam & = \V efam +\Pm ef \V enag \td 
gfmn + 
\V efag \ts gm + \nonumber \\
& \f na \td efmn + \undemi \V noam \td efno - \Pm ef \V nfam \ts en \\
 \W ovoo : \W iemn &= \V iemn + \f if \td efmn - \V 
iomn \ts eo + \undemi \V iefg \td fgmn \nonumber \\
&+ \Pm mn \V iefn \ts fn + \Pm mn \V iomf \td efno 
\end{align}

\begin{align}
\Wb ovvo &:\Wb mbej = \V mbej  - \V menf 
\td bfnj  \\
{}_{\text{MBPT(2)}} \left\lbrace \Wb ovvo \otimes \{t_1;t_2;\hat{T}^{[1]} \}
\right\rbrace &: \V mbej \otimes \{t_1;t_2;\hat{T}^{[1]}\}
\end{align}

\subsection{P-EOM}
\begin{align}
EE&: _{ }[\bar H^{[0]}_{DD}\Rr]_{ij}^{ab}=\Pm ab  \f be \rd aeij 
- \Pm ij 
\f mj\rd abim\\
IP&: [\bar H^{[0]}_{DD}\Rr]_{ij}^{a}=- \Pm ij \f mj\rd a{}im 
+   \f ae \rd 
e{}ij   \\
EA&: [\bar H^{[0]}_{DD}\Rr]_{i}^{ab}=\Pm ab  \f ae \rd ebi{}  
-  \f mi\rd 
abm{}  
\end{align}

\bibliography{Biblio_partitioned_EOM}

\begin{thebibliography}{101}%
\makeatletter
\providecommand \@ifxundefined [1]{%
 \@ifx{#1\undefined}
}%
\providecommand \@ifnum [1]{%
 \ifnum #1\expandafter \@firstoftwo
 \else \expandafter \@secondoftwo
 \fi
}%
\providecommand \@ifx [1]{%
 \ifx #1\expandafter \@firstoftwo
 \else \expandafter \@secondoftwo
 \fi
}%
\providecommand \natexlab [1]{#1}%
\providecommand \enquote  [1]{``#1''}%
\providecommand \bibnamefont  [1]{#1}%
\providecommand \bibfnamefont [1]{#1}%
\providecommand \citenamefont [1]{#1}%
\providecommand \href@noop [0]{\@secondoftwo}%
\providecommand \href [0]{\begingroup \@sanitize@url \@href}%
\providecommand \@href[1]{\@@startlink{#1}\@@href}%
\providecommand \@@href[1]{\endgroup#1\@@endlink}%
\providecommand \@sanitize@url [0]{\catcode `\\12\catcode `\$12\catcode
  `\&12\catcode `\#12\catcode `\^12\catcode `\_12\catcode `\%12\relax}%
\providecommand \@@startlink[1]{}%
\providecommand \@@endlink[0]{}%
\providecommand \url  [0]{\begingroup\@sanitize@url \@url }%
\providecommand \@url [1]{\endgroup\@href {#1}{\urlprefix }}%
\providecommand \urlprefix  [0]{URL }%
\providecommand \Eprint [0]{\href }%
\providecommand \doibase [0]{http://dx.doi.org/}%
\providecommand \selectlanguage [0]{\@gobble}%
\providecommand \bibinfo  [0]{\@secondoftwo}%
\providecommand \bibfield  [0]{\@secondoftwo}%
\providecommand \translation [1]{[#1]}%
\providecommand \BibitemOpen [0]{}%
\providecommand \bibitemStop [0]{}%
\providecommand \bibitemNoStop [0]{.\EOS\space}%
\providecommand \EOS [0]{\spacefactor3000\relax}%
\providecommand \BibitemShut  [1]{\csname bibitem#1\endcsname}%
\let\auto@bib@innerbib\@empty
\bibitem [{\citenamefont {Loos}, \citenamefont {Scemama},\ and\ \citenamefont
  {Jacquemin}(2020)}]{Loos2020}%
  \BibitemOpen
  \bibfield  {author} {\bibinfo {author} {\bibfnamefont {P.-F.}\ \bibnamefont
  {Loos}}, \bibinfo {author} {\bibfnamefont {A.}~\bibnamefont {Scemama}}, \
  and\ \bibinfo {author} {\bibfnamefont {D.}~\bibnamefont {Jacquemin}},\
  }\bibfield  {title} {\enquote {\bibinfo {title} {The quest for highly
  accurate excitation energies: A computational perspective},}\ }\href
  {\doibase 10.1021/acs.jpclett.0c00014} {\bibfield  {journal} {\bibinfo
  {journal} {The Journal of Physical Chemistry Letters}\ }\textbf {\bibinfo
  {volume} {11}},\ \bibinfo {pages} {2374--2383} (\bibinfo {year}
  {2020})}\BibitemShut {NoStop}%
\bibitem [{\citenamefont {Bokarev}\ and\ \citenamefont
  {K\"{u}hn}(2019)}]{Bokarev2019}%
  \BibitemOpen
  \bibfield  {author} {\bibinfo {author} {\bibfnamefont {S.~I.}\ \bibnamefont
  {Bokarev}}\ and\ \bibinfo {author} {\bibfnamefont {O.}~\bibnamefont
  {K\"{u}hn}},\ }\bibfield  {title} {\enquote {\bibinfo {title} {{Theoretical
  {X-Ray} spectroscopy of transition metal compounds}},}\ }\href {\doibase
  10.1002/wcms.1433} {\bibfield  {journal} {\bibinfo  {journal} {{WIREs}
  Computational Molecular Science}\ }\textbf {\bibinfo {volume} {10}},\
  \bibinfo {pages} {--} (\bibinfo {year} {2019})}\BibitemShut {NoStop}%
\bibitem [{\citenamefont {Izs{\'{a}}k}(2019)}]{Izsk2019}%
  \BibitemOpen
  \bibfield  {author} {\bibinfo {author} {\bibfnamefont {R.}~\bibnamefont
  {Izs{\'{a}}k}},\ }\bibfield  {title} {\enquote {\bibinfo {title}
  {Single-reference coupled cluster methods for computing excitation energies
  in large molecules: The efficiency and accuracy of approximations},}\ }\href
  {\doibase 10.1002/wcms.1445} {\bibfield  {journal} {\bibinfo  {journal}
  {{WIREs} Computational Molecular Science}\ }\textbf {\bibinfo {volume}
  {10}},\ \bibinfo {pages} {--} (\bibinfo {year} {2019})}\BibitemShut {NoStop}%
\bibitem [{\citenamefont {Alov}(2005)}]{Alov2005}%
  \BibitemOpen
  \bibfield  {author} {\bibinfo {author} {\bibfnamefont {N.~V.}\ \bibnamefont
  {Alov}},\ }\bibfield  {title} {\enquote {\bibinfo {title} {Fifty years of
  {X-Ray} photoelectron spectroscopy},}\ }\href {\doibase
  10.1007/s10809-005-0087-9} {\bibfield  {journal} {\bibinfo  {journal}
  {Journal of Analytical Chemistry}\ }\textbf {\bibinfo {volume} {60}},\
  \bibinfo {pages} {297--300} (\bibinfo {year} {2005})}\BibitemShut {NoStop}%
\bibitem [{\citenamefont {Fadley}(2010)}]{Fadley2010}%
  \BibitemOpen
  \bibfield  {author} {\bibinfo {author} {\bibfnamefont {C.~S.}\ \bibnamefont
  {Fadley}},\ }\bibfield  {title} {\enquote {\bibinfo {title} {{X-Ray}
  photoelectron spectroscopy: Progress and perspectives},}\ }\href {\doibase
  10.1016/j.elspec.2010.01.006} {\bibfield  {journal} {\bibinfo  {journal}
  {Journal of Electron Spectroscopy and Related Phenomena}\ }\textbf {\bibinfo
  {volume} {178-179}},\ \bibinfo {pages} {2--32} (\bibinfo {year}
  {2010})}\BibitemShut {NoStop}%
\bibitem [{\citenamefont {Doucet}\ and\ \citenamefont
  {Baruchel}(2016)}]{Doucet_2011}%
  \BibitemOpen
  \bibfield  {author} {\bibinfo {author} {\bibfnamefont {J.}~\bibnamefont
  {Doucet}}\ and\ \bibinfo {author} {\bibfnamefont {J.}~\bibnamefont
  {Baruchel}},\ }\bibfield  {title} {\enquote {\bibinfo {title} {Rayonnement
  synchrotron et applications},}\ }\href {\doibase 10.51257/a-v3-p2700}
  {\bibfield  {journal} {\bibinfo  {journal} {{CND} : m{\'{e}}thodes globales
  et volumiques}\ } (\bibinfo {year} {2016}),\ 10.51257/a-v3-p2700}\BibitemShut
  {NoStop}%
\bibitem [{\citenamefont {Couprie}(2014)}]{Couprie2014}%
  \BibitemOpen
  \bibfield  {author} {\bibinfo {author} {\bibfnamefont {M.-E.}\ \bibnamefont
  {Couprie}},\ }\bibfield  {title} {\enquote {\bibinfo {title} {New generation
  of light sources: Present and future},}\ }\href {\doibase
  10.1016/j.elspec.2013.12.007} {\bibfield  {journal} {\bibinfo  {journal}
  {Journal of Electron Spectroscopy and Related Phenomena}\ }\textbf {\bibinfo
  {volume} {196}},\ \bibinfo {pages} {3--13} (\bibinfo {year}
  {2014})}\BibitemShut {NoStop}%
\bibitem [{\citenamefont {Bergmann}, \citenamefont {Yachandra},\ and\
  \citenamefont {Yano}(2017)}]{Bergmann_2017_rsc_XFEL}%
  \BibitemOpen
  \bibinfo {editor} {\bibfnamefont {U.}~\bibnamefont {Bergmann}}, \bibinfo
  {editor} {\bibfnamefont {V.}~\bibnamefont {Yachandra}}, \ and\ \bibinfo
  {editor} {\bibfnamefont {J.}~\bibnamefont {Yano}},\ eds.,\ \href@noop {}
  {\emph {\bibinfo {title} {{X-Ray} Free Electron Lasers: {Applications} in
  Materials, Chemistry and Biology}}},\ \bibinfo {series} {Energy and
  Environment Series}\ No.~\bibinfo {number} {18}\ (\bibinfo  {publisher}
  {Royal Society of Chemistry},\ \bibinfo {year} {2017})\BibitemShut {NoStop}%
\bibitem [{\citenamefont {Young}\ \emph {et~al.}(2018)\citenamefont {Young},
  \citenamefont {Ueda}, \citenamefont {G\"{u}hr}, \citenamefont {Bucksbaum},
  \citenamefont {Simon}, \citenamefont {Mukamel}, \citenamefont {Rohringer},
  \citenamefont {Prince}, \citenamefont {Masciovecchio}, \citenamefont {Meyer},
  \citenamefont {Rudenko}, \citenamefont {Rolles}, \citenamefont {Bostedt},
  \citenamefont {Fuchs}, \citenamefont {Reis}, \citenamefont {Santra},
  \citenamefont {Kapteyn}, \citenamefont {Murnane}, \citenamefont {Ibrahim},
  \citenamefont {L{\'{e}}gar{\'{e}}}, \citenamefont {Vrakking}, \citenamefont
  {Isinger}, \citenamefont {Kroon}, \citenamefont {Gisselbrecht}, \citenamefont
  {L'Huillier}, \citenamefont {W\"{o}rner},\ and\ \citenamefont
  {Leone}}]{Young2018}%
  \BibitemOpen
  \bibfield  {author} {\bibinfo {author} {\bibfnamefont {L.}~\bibnamefont
  {Young}}, \bibinfo {author} {\bibfnamefont {K.}~\bibnamefont {Ueda}},
  \bibinfo {author} {\bibfnamefont {M.}~\bibnamefont {G\"{u}hr}}, \bibinfo
  {author} {\bibfnamefont {P.~H.}\ \bibnamefont {Bucksbaum}}, \bibinfo {author}
  {\bibfnamefont {M.}~\bibnamefont {Simon}}, \bibinfo {author} {\bibfnamefont
  {S.}~\bibnamefont {Mukamel}}, \bibinfo {author} {\bibfnamefont
  {N.}~\bibnamefont {Rohringer}}, \bibinfo {author} {\bibfnamefont {K.~C.}\
  \bibnamefont {Prince}}, \bibinfo {author} {\bibfnamefont {C.}~\bibnamefont
  {Masciovecchio}}, \bibinfo {author} {\bibfnamefont {M.}~\bibnamefont
  {Meyer}}, \bibinfo {author} {\bibfnamefont {A.}~\bibnamefont {Rudenko}},
  \bibinfo {author} {\bibfnamefont {D.}~\bibnamefont {Rolles}}, \bibinfo
  {author} {\bibfnamefont {C.}~\bibnamefont {Bostedt}}, \bibinfo {author}
  {\bibfnamefont {M.}~\bibnamefont {Fuchs}}, \bibinfo {author} {\bibfnamefont
  {D.~A.}\ \bibnamefont {Reis}}, \bibinfo {author} {\bibfnamefont
  {R.}~\bibnamefont {Santra}}, \bibinfo {author} {\bibfnamefont
  {H.}~\bibnamefont {Kapteyn}}, \bibinfo {author} {\bibfnamefont
  {M.}~\bibnamefont {Murnane}}, \bibinfo {author} {\bibfnamefont
  {H.}~\bibnamefont {Ibrahim}}, \bibinfo {author} {\bibfnamefont
  {F.}~\bibnamefont {L{\'{e}}gar{\'{e}}}}, \bibinfo {author} {\bibfnamefont
  {M.}~\bibnamefont {Vrakking}}, \bibinfo {author} {\bibfnamefont
  {M.}~\bibnamefont {Isinger}}, \bibinfo {author} {\bibfnamefont
  {D.}~\bibnamefont {Kroon}}, \bibinfo {author} {\bibfnamefont
  {M.}~\bibnamefont {Gisselbrecht}}, \bibinfo {author} {\bibfnamefont
  {A.}~\bibnamefont {L'Huillier}}, \bibinfo {author} {\bibfnamefont {H.~J.}\
  \bibnamefont {W\"{o}rner}}, \ and\ \bibinfo {author} {\bibfnamefont {S.~R.}\
  \bibnamefont {Leone}},\ }\bibfield  {title} {\enquote {\bibinfo {title}
  {Roadmap of ultrafast {X-Ray} atomic and molecular physics},}\ }\href
  {\doibase 10.1088/1361-6455/aa9735} {\bibfield  {journal} {\bibinfo
  {journal} {Journal of Physics B: Atomic, Molecular and Optical Physics}\
  }\textbf {\bibinfo {volume} {51}},\ \bibinfo {pages} {032003} (\bibinfo
  {year} {2018})}\BibitemShut {NoStop}%
\bibitem [{\citenamefont {Gunzer}, \citenamefont {Kr\"{u}ger},\ and\
  \citenamefont {Grotemeyer}(2018)}]{Gunzer2018}%
  \BibitemOpen
  \bibfield  {author} {\bibinfo {author} {\bibfnamefont {F.}~\bibnamefont
  {Gunzer}}, \bibinfo {author} {\bibfnamefont {S.}~\bibnamefont {Kr\"{u}ger}},
  \ and\ \bibinfo {author} {\bibfnamefont {J.}~\bibnamefont {Grotemeyer}},\
  }\bibfield  {title} {\enquote {\bibinfo {title} {Photoionization and
  photofragmentation in mass spectrometry with visible and {UV} lasers},}\
  }\href {\doibase 10.1002/mas.21579} {\bibfield  {journal} {\bibinfo
  {journal} {Mass Spectrometry Reviews}\ }\textbf {\bibinfo {volume} {38}},\
  \bibinfo {pages} {202--217} (\bibinfo {year} {2018})}\BibitemShut {NoStop}%
\bibitem [{\citenamefont {Rienstra-Kiracofe}\ \emph {et~al.}(2002)\citenamefont
  {Rienstra-Kiracofe}, \citenamefont {Tschumper}, \citenamefont {Schaefer},
  \citenamefont {Nandi},\ and\ \citenamefont {Ellison}}]{RienstraKiracofe2002}%
  \BibitemOpen
  \bibfield  {author} {\bibinfo {author} {\bibfnamefont {J.~C.}\ \bibnamefont
  {Rienstra-Kiracofe}}, \bibinfo {author} {\bibfnamefont {G.~S.}\ \bibnamefont
  {Tschumper}}, \bibinfo {author} {\bibfnamefont {H.~F.}\ \bibnamefont
  {Schaefer}}, \bibinfo {author} {\bibfnamefont {S.}~\bibnamefont {Nandi}}, \
  and\ \bibinfo {author} {\bibfnamefont {G.~B.}\ \bibnamefont {Ellison}},\
  }\bibfield  {title} {\enquote {\bibinfo {title} {Atomic and molecular
  electron affinities:{\hspace{0.167em}} photoelectron experiments and
  theoretical computations},}\ }\href {\doibase 10.1021/cr990044u} {\bibfield
  {journal} {\bibinfo  {journal} {Chemical Reviews}\ }\textbf {\bibinfo
  {volume} {102}},\ \bibinfo {pages} {231--282} (\bibinfo {year}
  {2002})}\BibitemShut {NoStop}%
\bibitem [{\citenamefont {Richard}\ \emph {et~al.}(2016)\citenamefont
  {Richard}, \citenamefont {Marshall}, \citenamefont {Dolgounitcheva},
  \citenamefont {Ortiz}, \citenamefont {Br{\'{e}}das}, \citenamefont {Marom},\
  and\ \citenamefont {Sherrill}}]{Richard2016}%
  \BibitemOpen
  \bibfield  {author} {\bibinfo {author} {\bibfnamefont {R.~M.}\ \bibnamefont
  {Richard}}, \bibinfo {author} {\bibfnamefont {M.~S.}\ \bibnamefont
  {Marshall}}, \bibinfo {author} {\bibfnamefont {O.}~\bibnamefont
  {Dolgounitcheva}}, \bibinfo {author} {\bibfnamefont {J.~V.}\ \bibnamefont
  {Ortiz}}, \bibinfo {author} {\bibfnamefont {J.-L.}\ \bibnamefont
  {Br{\'{e}}das}}, \bibinfo {author} {\bibfnamefont {N.}~\bibnamefont {Marom}},
  \ and\ \bibinfo {author} {\bibfnamefont {C.~D.}\ \bibnamefont {Sherrill}},\
  }\bibfield  {title} {\enquote {\bibinfo {title} {Accurate ionization
  potentials and electron affinities of acceptor molecules {I}. reference data
  at the {CCSD(T)} complete basis set limit},}\ }\href {\doibase
  10.1021/acs.jctc.5b00875} {\bibfield  {journal} {\bibinfo  {journal} {Journal
  of Chemical Theory and Computation}\ }\textbf {\bibinfo {volume} {12}},\
  \bibinfo {pages} {595--604} (\bibinfo {year} {2016})}\BibitemShut {NoStop}%
\bibitem [{\citenamefont {Chakraborty}\ and\ \citenamefont
  {Nandi}(2020)}]{Chakraborty2020}%
  \BibitemOpen
  \bibfield  {author} {\bibinfo {author} {\bibfnamefont {D.}~\bibnamefont
  {Chakraborty}}\ and\ \bibinfo {author} {\bibfnamefont {D.}~\bibnamefont
  {Nandi}},\ }\bibfield  {title} {\enquote {\bibinfo {title} {Absolute
  dissociative electron attachment cross-section measurement of
  difluoromethane},}\ }\href {\doibase 10.1103/physreva.102.052801} {\bibfield
  {journal} {\bibinfo  {journal} {Physical Review A}\ }\textbf {\bibinfo
  {volume} {102}},\ \bibinfo {pages} {--} (\bibinfo {year} {2020})}\BibitemShut
  {NoStop}%
\bibitem [{\citenamefont {Grell}\ \emph {et~al.}(2015)\citenamefont {Grell},
  \citenamefont {Bokarev}, \citenamefont {Winter}, \citenamefont {Seidel},
  \citenamefont {Aziz}, \citenamefont {Aziz},\ and\ \citenamefont
  {K\"{u}hn}}]{Grell2015}%
  \BibitemOpen
  \bibfield  {author} {\bibinfo {author} {\bibfnamefont {G.}~\bibnamefont
  {Grell}}, \bibinfo {author} {\bibfnamefont {S.~I.}\ \bibnamefont {Bokarev}},
  \bibinfo {author} {\bibfnamefont {B.}~\bibnamefont {Winter}}, \bibinfo
  {author} {\bibfnamefont {R.}~\bibnamefont {Seidel}}, \bibinfo {author}
  {\bibfnamefont {E.~F.}\ \bibnamefont {Aziz}}, \bibinfo {author}
  {\bibfnamefont {S.~G.}\ \bibnamefont {Aziz}}, \ and\ \bibinfo {author}
  {\bibfnamefont {O.}~\bibnamefont {K\"{u}hn}},\ }\bibfield  {title} {\enquote
  {\bibinfo {title} {Multi-reference approach to the calculation of
  photoelectron spectra including spin-orbit coupling},}\ }\href {\doibase
  10.1063/1.4928511} {\bibfield  {journal} {\bibinfo  {journal} {The Journal of
  Chemical Physics}\ }\textbf {\bibinfo {volume} {143}},\ \bibinfo {pages}
  {074104} (\bibinfo {year} {2015})}\BibitemShut {NoStop}%
\bibitem [{\citenamefont {Lundberg}\ and\ \citenamefont
  {Delcey}(2019)}]{Lundberg2019}%
  \BibitemOpen
  \bibfield  {author} {\bibinfo {author} {\bibfnamefont {M.}~\bibnamefont
  {Lundberg}}\ and\ \bibinfo {author} {\bibfnamefont {M.~G.}\ \bibnamefont
  {Delcey}},\ }\bibfield  {title} {\enquote {\bibinfo {title}
  {Multiconfigurational approach to {X-Ray} spectroscopy of transition metal
  complexes},}\ }in\ \href {\doibase 10.1007/978-3-030-11714-6_7} {\emph
  {\bibinfo {booktitle} {Transition Metals in Coordination Environments}}}\
  (\bibinfo  {publisher} {Springer International Publishing},\ \bibinfo {year}
  {2019})\ pp.\ \bibinfo {pages} {185--217}\BibitemShut {NoStop}%
\bibitem [{\citenamefont {Maganas}\ \emph {et~al.}(2019)\citenamefont
  {Maganas}, \citenamefont {Kowalska}, \citenamefont {Nooijen}, \citenamefont
  {DeBeer},\ and\ \citenamefont {Neese}}]{Maganas2019}%
  \BibitemOpen
  \bibfield  {author} {\bibinfo {author} {\bibfnamefont {D.}~\bibnamefont
  {Maganas}}, \bibinfo {author} {\bibfnamefont {J.~K.}\ \bibnamefont
  {Kowalska}}, \bibinfo {author} {\bibfnamefont {M.}~\bibnamefont {Nooijen}},
  \bibinfo {author} {\bibfnamefont {S.}~\bibnamefont {DeBeer}}, \ and\ \bibinfo
  {author} {\bibfnamefont {F.}~\bibnamefont {Neese}},\ }\bibfield  {title}
  {\enquote {\bibinfo {title} {Comparison of multireference ab initio
  wavefunction methodologies for {X-Ray} absorption edges: A case study on
  [{Fe}({II}/{III}){Cl}$_{4}$]$^{2-/1-}$ molecules},}\ }\href {\doibase
  10.1063/1.5051613} {\bibfield  {journal} {\bibinfo  {journal} {The Journal of
  Chemical Physics}\ }\textbf {\bibinfo {volume} {150}},\ \bibinfo {pages}
  {104106} (\bibinfo {year} {2019})}\BibitemShut {NoStop}%
\bibitem [{\citenamefont {Brabec}\ \emph {et~al.}(2012)\citenamefont {Brabec},
  \citenamefont {Bhaskaran-Nair}, \citenamefont {Govind}, \citenamefont
  {Pittner},\ and\ \citenamefont {Kowalski}}]{Brabec2012}%
  \BibitemOpen
  \bibfield  {author} {\bibinfo {author} {\bibfnamefont {J.}~\bibnamefont
  {Brabec}}, \bibinfo {author} {\bibfnamefont {K.}~\bibnamefont
  {Bhaskaran-Nair}}, \bibinfo {author} {\bibfnamefont {N.}~\bibnamefont
  {Govind}}, \bibinfo {author} {\bibfnamefont {J.}~\bibnamefont {Pittner}}, \
  and\ \bibinfo {author} {\bibfnamefont {K.}~\bibnamefont {Kowalski}},\
  }\bibfield  {title} {\enquote {\bibinfo {title} {Communication: Application
  of state-specific multireference coupled cluster methods to core-level
  excitations},}\ }\href {\doibase 10.1063/1.4764355} {\bibfield  {journal}
  {\bibinfo  {journal} {The Journal of Chemical Physics}\ }\textbf {\bibinfo
  {volume} {137}} (\bibinfo {year} {2012}),\ 10.1063/1.4764355}\BibitemShut
  {NoStop}%
\bibitem [{\citenamefont {Sen}, \citenamefont {Shee},\ and\ \citenamefont
  {Mukherjee}(2013)}]{Sen2013}%
  \BibitemOpen
  \bibfield  {author} {\bibinfo {author} {\bibfnamefont {S.}~\bibnamefont
  {Sen}}, \bibinfo {author} {\bibfnamefont {A.}~\bibnamefont {Shee}}, \ and\
  \bibinfo {author} {\bibfnamefont {D.}~\bibnamefont {Mukherjee}},\ }\bibfield
  {title} {\enquote {\bibinfo {title} {A study of the ionisation and excitation
  energies of core electrons using a unitary group adapted state universal
  approach},}\ }\href {\doibase 10.1080/00268976.2013.802384} {\bibfield
  {journal} {\bibinfo  {journal} {Molecular Physics}\ }\textbf {\bibinfo
  {volume} {111}},\ \bibinfo {pages} {2625--2639} (\bibinfo {year}
  {2013})}\BibitemShut {NoStop}%
\bibitem [{\citenamefont {Dutta}\ \emph
  {et~al.}(2014{\natexlab{a}})\citenamefont {Dutta}, \citenamefont {Gupta},
  \citenamefont {Vaval},\ and\ \citenamefont {Pal}}]{Dutta2014fs}%
  \BibitemOpen
  \bibfield  {author} {\bibinfo {author} {\bibfnamefont {A.~K.}\ \bibnamefont
  {Dutta}}, \bibinfo {author} {\bibfnamefont {J.}~\bibnamefont {Gupta}},
  \bibinfo {author} {\bibfnamefont {N.}~\bibnamefont {Vaval}}, \ and\ \bibinfo
  {author} {\bibfnamefont {S.}~\bibnamefont {Pal}},\ }\bibfield  {title}
  {\enquote {\bibinfo {title} {Intermediate hamiltonian fock space
  multireference coupled cluster approach to core excitation spectra},}\ }\href
  {\doibase 10.1021/ct500285e} {\bibfield  {journal} {\bibinfo  {journal}
  {Journal of Chemical Theory and Computation}\ }\textbf {\bibinfo {volume}
  {10}},\ \bibinfo {pages} {3656--3668} (\bibinfo {year}
  {2014}{\natexlab{a}})}\BibitemShut {NoStop}%
\bibitem [{\citenamefont {Dreuw}\ and\ \citenamefont
  {Wormit}(2014)}]{Dreuw2014}%
  \BibitemOpen
  \bibfield  {author} {\bibinfo {author} {\bibfnamefont {A.}~\bibnamefont
  {Dreuw}}\ and\ \bibinfo {author} {\bibfnamefont {M.}~\bibnamefont {Wormit}},\
  }\bibfield  {title} {\enquote {\bibinfo {title} {The algebraic diagrammatic
  construction scheme for the polarization propagator for the calculation of
  excited states},}\ }\href {\doibase 10.1002/wcms.1206} {\bibfield  {journal}
  {\bibinfo  {journal} {Wiley Interdisciplinary Reviews: Computational
  Molecular Science}\ }\textbf {\bibinfo {volume} {5}},\ \bibinfo {pages}
  {82--95} (\bibinfo {year} {2014})}\BibitemShut {NoStop}%
\bibitem [{\citenamefont {Wenzel}, \citenamefont {Wormit},\ and\ \citenamefont
  {Dreuw}(2014)}]{Wenzel2014}%
  \BibitemOpen
  \bibfield  {author} {\bibinfo {author} {\bibfnamefont {J.}~\bibnamefont
  {Wenzel}}, \bibinfo {author} {\bibfnamefont {M.}~\bibnamefont {Wormit}}, \
  and\ \bibinfo {author} {\bibfnamefont {A.}~\bibnamefont {Dreuw}},\ }\bibfield
   {title} {\enquote {\bibinfo {title} {Calculating core-level excitations and
  {X-Ray} absorption spectra of medium-sized closed-shell molecules with the
  algebraic-diagrammatic construction scheme for the polarization
  propagator},}\ }\href {\doibase 10.1002/jcc.23703} {\bibfield  {journal}
  {\bibinfo  {journal} {Journal of Computational Chemistry}\ }\textbf {\bibinfo
  {volume} {35}},\ \bibinfo {pages} {1900--1915} (\bibinfo {year}
  {2014})}\BibitemShut {NoStop}%
\bibitem [{\citenamefont {Mazin}\ and\ \citenamefont
  {Sokolov}(2023)}]{Mazin2023}%
  \BibitemOpen
  \bibfield  {author} {\bibinfo {author} {\bibfnamefont {I.~M.}\ \bibnamefont
  {Mazin}}\ and\ \bibinfo {author} {\bibfnamefont {A.~Y.}\ \bibnamefont
  {Sokolov}},\ }\bibfield  {title} {\enquote {\bibinfo {title} {Core-excited
  states and x-ray absorption spectra from multireference algebraic
  diagrammatic construction theory},}\ }\href {\doibase
  10.1021/acs.jctc.3c00477} {\bibfield  {journal} {\bibinfo  {journal} {Journal
  of Chemical Theory and Computation}\ } (\bibinfo {year} {2023}),\
  10.1021/acs.jctc.3c00477}\BibitemShut {NoStop}%
\bibitem [{\citenamefont {Besley}(2020)}]{Besley2020}%
  \BibitemOpen
  \bibfield  {author} {\bibinfo {author} {\bibfnamefont {N.~A.}\ \bibnamefont
  {Besley}},\ }\bibfield  {title} {\enquote {\bibinfo {title} {{Density
  Functional Theory Based Methods for the Calculation of {X-Ray}
  Spectroscopy}},}\ }\href {\doibase 10.1021/acs.accounts.0c00171} {\bibfield
  {journal} {\bibinfo  {journal} {Accounts of Chemical Research}\ }\textbf
  {\bibinfo {volume} {53}},\ \bibinfo {pages} {1306--1315} (\bibinfo {year}
  {2020})}\BibitemShut {NoStop}%
\bibitem [{\citenamefont {Besley}(2021)}]{Besley2021}%
  \BibitemOpen
  \bibfield  {author} {\bibinfo {author} {\bibfnamefont {N.~A.}\ \bibnamefont
  {Besley}},\ }\bibfield  {title} {\enquote {\bibinfo {title} {Modeling of the
  spectroscopy of core electrons with density functional theory},}\ }\href
  {\doibase 10.1002/wcms.1527} {\bibfield  {journal} {\bibinfo  {journal}
  {{WIREs} Computational Molecular Science}\ ,\ \bibinfo {pages} {--}}
  (\bibinfo {year} {2021})}\BibitemShut {NoStop}%
\bibitem [{\citenamefont {Brena}\ and\ \citenamefont {Luo}(2006)}]{Brena2006}%
  \BibitemOpen
  \bibfield  {author} {\bibinfo {author} {\bibfnamefont {B.}~\bibnamefont
  {Brena}}\ and\ \bibinfo {author} {\bibfnamefont {Y.}~\bibnamefont {Luo}},\
  }\bibfield  {title} {\enquote {\bibinfo {title} {Time-dependent {DFT}
  calculations of core electron shake-up states of
  metal-(free)-phthalocyanines},}\ }\href {\doibase
  10.1016/j.radphyschem.2005.07.017} {\bibfield  {journal} {\bibinfo  {journal}
  {Radiation Physics and Chemistry}\ }\textbf {\bibinfo {volume} {75}},\
  \bibinfo {pages} {1578--1581} (\bibinfo {year} {2006})}\BibitemShut {NoStop}%
\bibitem [{\citenamefont {Fouda}\ and\ \citenamefont
  {Besley}(2017)}]{Fouda2017}%
  \BibitemOpen
  \bibfield  {author} {\bibinfo {author} {\bibfnamefont {A.~E.~A.}\
  \bibnamefont {Fouda}}\ and\ \bibinfo {author} {\bibfnamefont {N.~A.}\
  \bibnamefont {Besley}},\ }\bibfield  {title} {\enquote {\bibinfo {title}
  {Assessment of basis sets for density functional theory-based calculations of
  core-electron spectroscopies},}\ }\href {\doibase 10.1007/s00214-017-2181-0}
  {\bibfield  {journal} {\bibinfo  {journal} {Theoretical Chemistry Accounts}\
  }\textbf {\bibinfo {volume} {137}} (\bibinfo {year} {2017}),\
  10.1007/s00214-017-2181-0}\BibitemShut {NoStop}%
\bibitem [{\citenamefont {Santis}, \citenamefont {Vallet},\ and\ \citenamefont
  {Gomes}(2022)}]{DeSantis2022}%
  \BibitemOpen
  \bibfield  {author} {\bibinfo {author} {\bibfnamefont {M.~D.}\ \bibnamefont
  {Santis}}, \bibinfo {author} {\bibfnamefont {V.}~\bibnamefont {Vallet}}, \
  and\ \bibinfo {author} {\bibfnamefont {A.~S.~P.}\ \bibnamefont {Gomes}},\
  }\bibfield  {title} {\enquote {\bibinfo {title} {Environment effects on x-ray
  absorption spectra with quantum embedded real-time time-dependent density
  functional theory approaches},}\ }\href {\doibase 10.3389/fchem.2022.823246}
  {\bibfield  {journal} {\bibinfo  {journal} {Frontiers in Chemistry}\ }\textbf
  {\bibinfo {volume} {10}},\ \bibinfo {pages} {--} (\bibinfo {year}
  {2022})}\BibitemShut {NoStop}%
\bibitem [{\citenamefont {Koch}\ \emph {et~al.}(1990)\citenamefont {Koch},
  \citenamefont {Jensen}, \citenamefont {Jo/rgensen},\ and\ \citenamefont
  {Helgaker}}]{Koch1990}%
  \BibitemOpen
  \bibfield  {author} {\bibinfo {author} {\bibfnamefont {H.}~\bibnamefont
  {Koch}}, \bibinfo {author} {\bibfnamefont {H.~J.~A.}\ \bibnamefont {Jensen}},
  \bibinfo {author} {\bibfnamefont {P.}~\bibnamefont {Jo/rgensen}}, \ and\
  \bibinfo {author} {\bibfnamefont {T.}~\bibnamefont {Helgaker}},\ }\bibfield
  {title} {\enquote {\bibinfo {title} {Excitation energies from the coupled
  cluster singles and doubles linear response function ({CCSDLR}). applications
  to {Be}, {CH}$^{+}$, {CO}, and {H}$_{2}${O}},}\ }\href {\doibase
  10.1063/1.458815} {\bibfield  {journal} {\bibinfo  {journal} {The Journal of
  Chemical Physics}\ }\textbf {\bibinfo {volume} {93}},\ \bibinfo {pages}
  {3345--3350} (\bibinfo {year} {1990})}\BibitemShut {NoStop}%
\bibitem [{\citenamefont {Coriani}\ and\ \citenamefont
  {Koch}(2015)}]{Coriani2015}%
  \BibitemOpen
  \bibfield  {author} {\bibinfo {author} {\bibfnamefont {S.}~\bibnamefont
  {Coriani}}\ and\ \bibinfo {author} {\bibfnamefont {H.}~\bibnamefont {Koch}},\
  }\bibfield  {title} {\enquote {\bibinfo {title} {Communication: {X-Ray}
  absorption spectra and core-ionization potentials within a core-valence
  separated coupled cluster framework},}\ }\href {\doibase 10.1063/1.4935712}
  {\bibfield  {journal} {\bibinfo  {journal} {The Journal of Chemical Physics}\
  }\textbf {\bibinfo {volume} {143}},\ \bibinfo {pages} {181103} (\bibinfo
  {year} {2015})}\BibitemShut {NoStop}%
\bibitem [{\citenamefont {Bartlett}\ and\ \citenamefont
  {Musia{\l}}(2007)}]{Bartlett2007}%
  \BibitemOpen
  \bibfield  {author} {\bibinfo {author} {\bibfnamefont {R.~J.}\ \bibnamefont
  {Bartlett}}\ and\ \bibinfo {author} {\bibfnamefont {M.}~\bibnamefont
  {Musia{\l}}},\ }\bibfield  {title} {\enquote {\bibinfo {title}
  {Coupled-cluster theory in quantum chemistry},}\ }\href {\doibase
  10.1103/revmodphys.79.291} {\bibfield  {journal} {\bibinfo  {journal}
  {Reviews of Modern Physics}\ }\textbf {\bibinfo {volume} {79}},\ \bibinfo
  {pages} {291--352} (\bibinfo {year} {2007})}\BibitemShut {NoStop}%
\bibitem [{\citenamefont {Coriani}\ \emph {et~al.}(2012)\citenamefont
  {Coriani}, \citenamefont {Christiansen}, \citenamefont {Fransson},\ and\
  \citenamefont {Norman}}]{Coriani2012}%
  \BibitemOpen
  \bibfield  {author} {\bibinfo {author} {\bibfnamefont {S.}~\bibnamefont
  {Coriani}}, \bibinfo {author} {\bibfnamefont {O.}~\bibnamefont
  {Christiansen}}, \bibinfo {author} {\bibfnamefont {T.}~\bibnamefont
  {Fransson}}, \ and\ \bibinfo {author} {\bibfnamefont {P.}~\bibnamefont
  {Norman}},\ }\bibfield  {title} {\enquote {\bibinfo {title} {{Coupled-Cluster
  Response Theory for Near-Edge {X-Ray}-Absorption Fine Structure of Atoms and
  Molecules}},}\ }\href {\doibase 10.1103/PhysRevA.85.022507} {\bibfield
  {journal} {\bibinfo  {journal} {Phys. Rev. A}\ }\textbf {\bibinfo {volume}
  {85}},\ \bibinfo {pages} {022507} (\bibinfo {year} {2012})}\BibitemShut
  {NoStop}%
\bibitem [{\citenamefont {Sadybekov}\ and\ \citenamefont
  {Krylov}(2017)}]{Sadybekov2017}%
  \BibitemOpen
  \bibfield  {author} {\bibinfo {author} {\bibfnamefont {A.}~\bibnamefont
  {Sadybekov}}\ and\ \bibinfo {author} {\bibfnamefont {A.~I.}\ \bibnamefont
  {Krylov}},\ }\bibfield  {title} {\enquote {\bibinfo {title} {Coupled-cluster
  based approach for core-ionized and core-excited states in condensed phase:
  {Theory} and application to different protonated forms of aqueous glycine},}\
  }\href@noop {} {\bibfield  {journal} {\bibinfo  {journal} {J. Chem. Phys.}\
  }\textbf {\bibinfo {volume} {147}},\ \bibinfo {pages} {014107} (\bibinfo
  {year} {2017})}\BibitemShut {NoStop}%
\bibitem [{\citenamefont {Vidal}\ \emph {et~al.}(2019)\citenamefont {Vidal},
  \citenamefont {Feng}, \citenamefont {Epifanovsky}, \citenamefont {Krylov},\
  and\ \citenamefont {Coriani}}]{Vidal2019}%
  \BibitemOpen
  \bibfield  {author} {\bibinfo {author} {\bibfnamefont {M.~L.}\ \bibnamefont
  {Vidal}}, \bibinfo {author} {\bibfnamefont {X.}~\bibnamefont {Feng}},
  \bibinfo {author} {\bibfnamefont {E.}~\bibnamefont {Epifanovsky}}, \bibinfo
  {author} {\bibfnamefont {A.~I.}\ \bibnamefont {Krylov}}, \ and\ \bibinfo
  {author} {\bibfnamefont {S.}~\bibnamefont {Coriani}},\ }\bibfield  {title}
  {\enquote {\bibinfo {title} {New and efficient equation-of-motion
  coupled-cluster framework for core-excited and core-ionized states},}\ }\href
  {\doibase 10.1021/acs.jctc.9b00039} {\bibfield  {journal} {\bibinfo
  {journal} {J. Chem. Theory Comput.}\ }\textbf {\bibinfo {volume} {15}},\
  \bibinfo {pages} {3117--3133} (\bibinfo {year} {2019})}\BibitemShut {NoStop}%
\bibitem [{\citenamefont {Peng}\ \emph {et~al.}(2015)\citenamefont {Peng},
  \citenamefont {Lestrange}, \citenamefont {Goings}, \citenamefont {Caricato},\
  and\ \citenamefont {Li}}]{Peng2015}%
  \BibitemOpen
  \bibfield  {author} {\bibinfo {author} {\bibfnamefont {B.}~\bibnamefont
  {Peng}}, \bibinfo {author} {\bibfnamefont {P.~J.}\ \bibnamefont {Lestrange}},
  \bibinfo {author} {\bibfnamefont {J.~J.}\ \bibnamefont {Goings}}, \bibinfo
  {author} {\bibfnamefont {M.}~\bibnamefont {Caricato}}, \ and\ \bibinfo
  {author} {\bibfnamefont {X.}~\bibnamefont {Li}},\ }\bibfield  {title}
  {\enquote {\bibinfo {title} {{Energy-Specific Equation-of-Motion
  Coupled-Cluster Methods for High-Energy Excited States: Application to
  {K-Edge} {X-Ray} Absorption Spectroscopy}},}\ }\href@noop {} {\bibfield
  {journal} {\bibinfo  {journal} {J. Chem. Theory Comput.}\ }\textbf {\bibinfo
  {volume} {11}},\ \bibinfo {pages} {4146} (\bibinfo {year}
  {2015})}\BibitemShut {NoStop}%
\bibitem [{\citenamefont {Park}, \citenamefont {Perera},\ and\ \citenamefont
  {Bartlett}(2019)}]{Park2019}%
  \BibitemOpen
  \bibfield  {author} {\bibinfo {author} {\bibfnamefont {Y.~C.}\ \bibnamefont
  {Park}}, \bibinfo {author} {\bibfnamefont {A.}~\bibnamefont {Perera}}, \ and\
  \bibinfo {author} {\bibfnamefont {R.~J.}\ \bibnamefont {Bartlett}},\
  }\bibfield  {title} {\enquote {\bibinfo {title} {Equation of motion
  coupled-cluster for core excitation spectra: Two complementary approaches},}\
  }\href {\doibase 10.1063/1.5117841} {\bibfield  {journal} {\bibinfo
  {journal} {J. Chem. Phys.}\ }\textbf {\bibinfo {volume} {151}},\ \bibinfo
  {pages} {164117} (\bibinfo {year} {2019})}\BibitemShut {NoStop}%
\bibitem [{\citenamefont {Matthews}(2020)}]{Matthews2020}%
  \BibitemOpen
  \bibfield  {author} {\bibinfo {author} {\bibfnamefont {D.~A.}\ \bibnamefont
  {Matthews}},\ }\bibfield  {title} {\enquote {\bibinfo {title} {{EOM-CC}
  methods with approximate triple excitations applied to core excitation and
  ionisation energies},}\ }\href {\doibase 10.1080/00268976.2020.1771448}
  {\bibfield  {journal} {\bibinfo  {journal} {Mol. Phys.}\ }\textbf {\bibinfo
  {volume} {118}},\ \bibinfo {pages} {e1771448} (\bibinfo {year}
  {2020})}\BibitemShut {NoStop}%
\bibitem [{\citenamefont {Musial}\ and\ \citenamefont
  {Bartlett}(2008{\natexlab{a}})}]{Musial2008b}%
  \BibitemOpen
  \bibfield  {author} {\bibinfo {author} {\bibfnamefont {M.}~\bibnamefont
  {Musial}}\ and\ \bibinfo {author} {\bibfnamefont {R.~J.}\ \bibnamefont
  {Bartlett}},\ }\bibfield  {title} {\enquote {\bibinfo {title} {Multireference
  fock-space coupled-cluster and equation-of-motion coupled-cluster theories:
  The detailed interconnections},}\ }\href {\doibase 10.1063/1.2982788}
  {\bibfield  {journal} {\bibinfo  {journal} {The Journal of Chemical Physics}\
  }\textbf {\bibinfo {volume} {129}},\ \bibinfo {pages} {134105} (\bibinfo
  {year} {2008}{\natexlab{a}})}\BibitemShut {NoStop}%
\bibitem [{\citenamefont {Shee}\ \emph {et~al.}(2018)\citenamefont {Shee},
  \citenamefont {Saue}, \citenamefont {Visscher},\ and\ \citenamefont
  {Gomes}}]{Shee2018}%
  \BibitemOpen
  \bibfield  {author} {\bibinfo {author} {\bibfnamefont {A.}~\bibnamefont
  {Shee}}, \bibinfo {author} {\bibfnamefont {T.}~\bibnamefont {Saue}}, \bibinfo
  {author} {\bibfnamefont {L.}~\bibnamefont {Visscher}}, \ and\ \bibinfo
  {author} {\bibfnamefont {A.~S.~P.}\ \bibnamefont {Gomes}},\ }\bibfield
  {title} {\enquote {\bibinfo {title} {Equation-of-motion coupled-cluster
  theory based on the 4-component dirac{\textendash}coulomb({\textendash}gaunt)
  hamiltonian. energies for single electron detachment, attachment, and
  electronically excited states},}\ }\href {\doibase 10.1063/1.5053846}
  {\bibfield  {journal} {\bibinfo  {journal} {The Journal of Chemical Physics}\
  }\textbf {\bibinfo {volume} {149}},\ \bibinfo {pages} {174113} (\bibinfo
  {year} {2018})}\BibitemShut {NoStop}%
\bibitem [{\citenamefont {Musial}\ and\ \citenamefont
  {Bartlett}(2008{\natexlab{b}})}]{Musial2008a}%
  \BibitemOpen
  \bibfield  {author} {\bibinfo {author} {\bibfnamefont {M.}~\bibnamefont
  {Musial}}\ and\ \bibinfo {author} {\bibfnamefont {R.~J.}\ \bibnamefont
  {Bartlett}},\ }\bibfield  {title} {\enquote {\bibinfo {title} {Benchmark
  calculations of the fock-space coupled cluster single, double, triple
  excitation method in the intermediate hamiltonian formulation for electronic
  excitation energies},}\ }\href {\doibase 10.1016/j.cplett.2008.04.004}
  {\bibfield  {journal} {\bibinfo  {journal} {Chemical Physics Letters}\
  }\textbf {\bibinfo {volume} {457}},\ \bibinfo {pages} {267--270} (\bibinfo
  {year} {2008}{\natexlab{b}})}\BibitemShut {NoStop}%
\bibitem [{\citenamefont {R{\'{e}}al}\ \emph {et~al.}(2009)\citenamefont
  {R{\'{e}}al}, \citenamefont {Gomes}, \citenamefont {Visscher}, \citenamefont
  {Vallet},\ and\ \citenamefont {Eliav}}]{Real2009}%
  \BibitemOpen
  \bibfield  {author} {\bibinfo {author} {\bibfnamefont {F.}~\bibnamefont
  {R{\'{e}}al}}, \bibinfo {author} {\bibfnamefont {A.~S.~P.}\ \bibnamefont
  {Gomes}}, \bibinfo {author} {\bibfnamefont {L.}~\bibnamefont {Visscher}},
  \bibinfo {author} {\bibfnamefont {V.}~\bibnamefont {Vallet}}, \ and\ \bibinfo
  {author} {\bibfnamefont {E.}~\bibnamefont {Eliav}},\ }\bibfield  {title}
  {\enquote {\bibinfo {title} {Benchmarking electronic structure calculations
  on the bare {UO}$_2^{2+}$ ion: How different are single and multireference
  electron correlation methods?}}\ }\href {\doibase 10.1021/jp903758c}
  {\bibfield  {journal} {\bibinfo  {journal} {The Journal of Physical Chemistry
  A}\ }\textbf {\bibinfo {volume} {113}},\ \bibinfo {pages} {12504--12511}
  (\bibinfo {year} {2009})}\BibitemShut {NoStop}%
\bibitem [{\citenamefont {Bagus}(1965)}]{Bagus1965}%
  \BibitemOpen
  \bibfield  {author} {\bibinfo {author} {\bibfnamefont {P.~S.}\ \bibnamefont
  {Bagus}},\ }\bibfield  {title} {\enquote {\bibinfo {title}
  {Self-consistent-field wave functions for hole states of some {Ne}-like and
  {Ar}-like ions},}\ }\href {\doibase 10.1103/physrev.139.a619} {\bibfield
  {journal} {\bibinfo  {journal} {Physical Review}\ }\textbf {\bibinfo {volume}
  {139}},\ \bibinfo {pages} {A619--A634} (\bibinfo {year} {1965})}\BibitemShut
  {NoStop}%
\bibitem [{\citenamefont {Bagus}\ \emph {et~al.}(1999)\citenamefont {Bagus},
  \citenamefont {Illas}, \citenamefont {Pacchioni},\ and\ \citenamefont
  {Parmigiani}}]{Bagus1999}%
  \BibitemOpen
  \bibfield  {author} {\bibinfo {author} {\bibfnamefont {P.~S.}\ \bibnamefont
  {Bagus}}, \bibinfo {author} {\bibfnamefont {F.}~\bibnamefont {Illas}},
  \bibinfo {author} {\bibfnamefont {G.}~\bibnamefont {Pacchioni}}, \ and\
  \bibinfo {author} {\bibfnamefont {F.}~\bibnamefont {Parmigiani}},\ }\bibfield
   {title} {\enquote {\bibinfo {title} {Mechanisms responsible for chemical
  shifts of core-level binding energies and their relationship to chemical
  bonding},}\ }\href {\doibase 10.1016/s0368-2048(99)00048-1} {\bibfield
  {journal} {\bibinfo  {journal} {Journal of Electron Spectroscopy and Related
  Phenomena}\ }\textbf {\bibinfo {volume} {100}},\ \bibinfo {pages} {215--236}
  (\bibinfo {year} {1999})}\BibitemShut {NoStop}%
\bibitem [{\citenamefont {{Naves de Brito}}\ \emph {et~al.}(1991)\citenamefont
  {{Naves de Brito}}, \citenamefont {Correia}, \citenamefont {Svensson},\ and\
  \citenamefont {{\AA}gren}}]{NavesdeBrito1991}%
  \BibitemOpen
  \bibfield  {author} {\bibinfo {author} {\bibfnamefont {A.}~\bibnamefont
  {{Naves de Brito}}}, \bibinfo {author} {\bibfnamefont {N.}~\bibnamefont
  {Correia}}, \bibinfo {author} {\bibfnamefont {S.}~\bibnamefont {Svensson}}, \
  and\ \bibinfo {author} {\bibfnamefont {H.}~\bibnamefont {{\AA}gren}},\
  }\bibfield  {title} {\enquote {\bibinfo {title} {A theoretical study of
  {X-Ray} photoelectron spectra of model molecules for
  polymethylmethacrylate},}\ }\href {\doibase 10.1063/1.460898} {\bibfield
  {journal} {\bibinfo  {journal} {The Journal of Chemical Physics}\ }\textbf
  {\bibinfo {volume} {95}},\ \bibinfo {pages} {2965--2974} (\bibinfo {year}
  {1991})}\BibitemShut {NoStop}%
\bibitem [{\citenamefont {Shim}\ \emph {et~al.}(2011)\citenamefont {Shim},
  \citenamefont {Klobukowski}, \citenamefont {Barysz},\ and\ \citenamefont
  {Leszczynski}}]{Shim2011}%
  \BibitemOpen
  \bibfield  {author} {\bibinfo {author} {\bibfnamefont {J.}~\bibnamefont
  {Shim}}, \bibinfo {author} {\bibfnamefont {M.}~\bibnamefont {Klobukowski}},
  \bibinfo {author} {\bibfnamefont {M.}~\bibnamefont {Barysz}}, \ and\ \bibinfo
  {author} {\bibfnamefont {J.}~\bibnamefont {Leszczynski}},\ }\bibfield
  {title} {\enquote {\bibinfo {title} {Calibration and applications of the
  {$\Delta$}mp2 method for calculating core electron binding energies},}\
  }\href {\doibase 10.1039/c0cp01591a} {\bibfield  {journal} {\bibinfo
  {journal} {Physical Chemistry Chemical Physics}\ }\textbf {\bibinfo {volume}
  {13}},\ \bibinfo {pages} {5703} (\bibinfo {year} {2011})}\BibitemShut
  {NoStop}%
\bibitem [{\citenamefont {South}\ \emph {et~al.}(2016)\citenamefont {South},
  \citenamefont {Shee}, \citenamefont {Mukherjee}, \citenamefont {Wilson},\
  and\ \citenamefont {Saue}}]{South2016}%
  \BibitemOpen
  \bibfield  {author} {\bibinfo {author} {\bibfnamefont {C.}~\bibnamefont
  {South}}, \bibinfo {author} {\bibfnamefont {A.}~\bibnamefont {Shee}},
  \bibinfo {author} {\bibfnamefont {D.}~\bibnamefont {Mukherjee}}, \bibinfo
  {author} {\bibfnamefont {A.~K.}\ \bibnamefont {Wilson}}, \ and\ \bibinfo
  {author} {\bibfnamefont {T.}~\bibnamefont {Saue}},\ }\bibfield  {title}
  {\enquote {\bibinfo {title} {4-component relativistic calculations of
  {L}$_{3}$ ionization and excitations for the isoelectronic species
  {UO}$_{2}^{2+}$, {OUN}$^{+}$ and {UN}$_{2}$},}\ }\href {\doibase
  10.1039/c6cp00262e} {\bibfield  {journal} {\bibinfo  {journal} {Physical
  Chemistry Chemical Physics}\ }\textbf {\bibinfo {volume} {18}},\ \bibinfo
  {pages} {21010--21023} (\bibinfo {year} {2016})}\BibitemShut {NoStop}%
\bibitem [{\citenamefont {Besley}, \citenamefont {Gilbert},\ and\ \citenamefont
  {Gill}(2009)}]{Besley2009}%
  \BibitemOpen
  \bibfield  {author} {\bibinfo {author} {\bibfnamefont {N.~A.}\ \bibnamefont
  {Besley}}, \bibinfo {author} {\bibfnamefont {A.~T.~B.}\ \bibnamefont
  {Gilbert}}, \ and\ \bibinfo {author} {\bibfnamefont {P.~M.~W.}\ \bibnamefont
  {Gill}},\ }\bibfield  {title} {\enquote {\bibinfo {title}
  {Self-consistent-field calculations of core excited states},}\ }\href
  {\doibase 10.1063/1.3092928} {\bibfield  {journal} {\bibinfo  {journal} {The
  Journal of Chemical Physics}\ }\textbf {\bibinfo {volume} {130}},\ \bibinfo
  {pages} {124308} (\bibinfo {year} {2009})}\BibitemShut {NoStop}%
\bibitem [{\citenamefont {{Pueyo Bellafont}}, \citenamefont {Bagus},\ and\
  \citenamefont {Illas}(2015)}]{PueyoBellafont2015}%
  \BibitemOpen
  \bibfield  {author} {\bibinfo {author} {\bibfnamefont {N.}~\bibnamefont
  {{Pueyo Bellafont}}}, \bibinfo {author} {\bibfnamefont {P.~S.}\ \bibnamefont
  {Bagus}}, \ and\ \bibinfo {author} {\bibfnamefont {F.}~\bibnamefont
  {Illas}},\ }\bibfield  {title} {\enquote {\bibinfo {title} {Prediction of
  core level binding energies in density functional theory: Rigorous definition
  of initial and final state contributions and implications on the physical
  meaning of kohn-sham energies},}\ }\href {\doibase 10.1063/1.4921823}
  {\bibfield  {journal} {\bibinfo  {journal} {The Journal of Chemical Physics}\
  }\textbf {\bibinfo {volume} {142}},\ \bibinfo {pages} {214102} (\bibinfo
  {year} {2015})}\BibitemShut {NoStop}%
\bibitem [{\citenamefont {Takahata}\ and\ \citenamefont
  {Chong}(2012)}]{Takahata2012}%
  \BibitemOpen
  \bibfield  {author} {\bibinfo {author} {\bibfnamefont {Y.}~\bibnamefont
  {Takahata}}\ and\ \bibinfo {author} {\bibfnamefont {D.~P.}\ \bibnamefont
  {Chong}},\ }\bibfield  {title} {\enquote {\bibinfo {title} {{DFT} calculation
  of core{\textendash} and valence{\textendash}shell electron excitation and
  ionization energies of 2, 1, 3-benzo thiadiazole {C}$_6${H}$_4${SN}$_2$,$\ $
  1, 3, 2, 4-benzodithiadiazine {C}$_6${H}$_4${S}$_2${N}$_2$, and 1, 3, 5, 2,
  4-benzotrithiadiazepine {C}$_6${H}$_4${S}$_3${N}$_2$},}\ }\href {\doibase
  10.1016/j.elspec.2012.09.015} {\bibfield  {journal} {\bibinfo  {journal}
  {Journal of Electron Spectroscopy and Related Phenomena}\ }\textbf {\bibinfo
  {volume} {185}},\ \bibinfo {pages} {475--485} (\bibinfo {year}
  {2012})}\BibitemShut {NoStop}%
\bibitem [{\citenamefont {Watts}\ and\ \citenamefont
  {Bartlett}(1990)}]{Watts1990}%
  \BibitemOpen
  \bibfield  {author} {\bibinfo {author} {\bibfnamefont {J.~D.}\ \bibnamefont
  {Watts}}\ and\ \bibinfo {author} {\bibfnamefont {R.~J.}\ \bibnamefont
  {Bartlett}},\ }\bibfield  {title} {\enquote {\bibinfo {title} {The
  coupled-cluster single, double, and triple excitation model for open-shell
  single reference functions},}\ }\href {\doibase 10.1063/1.459002} {\bibfield
  {journal} {\bibinfo  {journal} {The Journal of Chemical Physics}\ }\textbf
  {\bibinfo {volume} {93}},\ \bibinfo {pages} {6104--6105} (\bibinfo {year}
  {1990})}\BibitemShut {NoStop}%
\bibitem [{\citenamefont {Zheng}\ and\ \citenamefont
  {Cheng}(2019)}]{Zheng2019}%
  \BibitemOpen
  \bibfield  {author} {\bibinfo {author} {\bibfnamefont {X.}~\bibnamefont
  {Zheng}}\ and\ \bibinfo {author} {\bibfnamefont {L.}~\bibnamefont {Cheng}},\
  }\bibfield  {title} {\enquote {\bibinfo {title} {Performance of
  delta-coupled-cluster methods for calculations of core-ionization energies of
  first-row elements},}\ }\href {\doibase 10.1021/acs.jctc.9b00568} {\bibfield
  {journal} {\bibinfo  {journal} {Journal of Chemical Theory and Computation}\
  }\textbf {\bibinfo {volume} {15}},\ \bibinfo {pages} {4945--4955} (\bibinfo
  {year} {2019})}\BibitemShut {NoStop}%
\bibitem [{\citenamefont {Nooijen}\ and\ \citenamefont
  {Bartlett}(1995)}]{Nooijen1995_EA}%
  \BibitemOpen
  \bibfield  {author} {\bibinfo {author} {\bibfnamefont {M.}~\bibnamefont
  {Nooijen}}\ and\ \bibinfo {author} {\bibfnamefont {R.~J.}\ \bibnamefont
  {Bartlett}},\ }\bibfield  {title} {\enquote {\bibinfo {title} {Equation of
  motion coupled cluster method for electron attachment},}\ }\href {\doibase
  10.1063/1.468592} {\bibfield  {journal} {\bibinfo  {journal} {The Journal of
  Chemical Physics}\ }\textbf {\bibinfo {volume} {102}},\ \bibinfo {pages}
  {3629--3647} (\bibinfo {year} {1995})}\BibitemShut {NoStop}%
\bibitem [{\citenamefont {Goings}\ \emph {et~al.}(2014)\citenamefont {Goings},
  \citenamefont {Caricato}, \citenamefont {Frisch},\ and\ \citenamefont
  {Li}}]{Goings2014}%
  \BibitemOpen
  \bibfield  {author} {\bibinfo {author} {\bibfnamefont {J.~J.}\ \bibnamefont
  {Goings}}, \bibinfo {author} {\bibfnamefont {M.}~\bibnamefont {Caricato}},
  \bibinfo {author} {\bibfnamefont {M.~J.}\ \bibnamefont {Frisch}}, \ and\
  \bibinfo {author} {\bibfnamefont {X.}~\bibnamefont {Li}},\ }\bibfield
  {title} {\enquote {\bibinfo {title} {Assessment of low-scaling approximations
  to the equation of motion coupled-cluster singles and doubles equations},}\
  }\href {\doibase 10.1063/1.4898709} {\bibfield  {journal} {\bibinfo
  {journal} {The Journal of Chemical Physics}\ }\textbf {\bibinfo {volume}
  {141}},\ \bibinfo {pages} {164116} (\bibinfo {year} {2014})}\BibitemShut
  {NoStop}%
\bibitem [{\citenamefont {Dutta}\ \emph
  {et~al.}(2014{\natexlab{b}})\citenamefont {Dutta}, \citenamefont {Gupta},
  \citenamefont {Pathak}, \citenamefont {Vaval},\ and\ \citenamefont
  {Pal}}]{Dutta2014}%
  \BibitemOpen
  \bibfield  {author} {\bibinfo {author} {\bibfnamefont {A.~K.}\ \bibnamefont
  {Dutta}}, \bibinfo {author} {\bibfnamefont {J.}~\bibnamefont {Gupta}},
  \bibinfo {author} {\bibfnamefont {H.}~\bibnamefont {Pathak}}, \bibinfo
  {author} {\bibfnamefont {N.}~\bibnamefont {Vaval}}, \ and\ \bibinfo {author}
  {\bibfnamefont {S.}~\bibnamefont {Pal}},\ }\bibfield  {title} {\enquote
  {\bibinfo {title} {Partitioned {EOMEA}-{MBPT}(2): An efficient n5 scaling
  method for calculation of electron affinities},}\ }\href {\doibase
  10.1021/ct4009409} {\bibfield  {journal} {\bibinfo  {journal} {Journal of
  Chemical Theory and Computation}\ }\textbf {\bibinfo {volume} {10}},\
  \bibinfo {pages} {1923--1933} (\bibinfo {year}
  {2014}{\natexlab{b}})}\BibitemShut {NoStop}%
\bibitem [{\citenamefont {Dutta}, \citenamefont {Vaval},\ and\ \citenamefont
  {Pal}(2018)}]{Dutta2018}%
  \BibitemOpen
  \bibfield  {author} {\bibinfo {author} {\bibfnamefont {A.~K.}\ \bibnamefont
  {Dutta}}, \bibinfo {author} {\bibfnamefont {N.}~\bibnamefont {Vaval}}, \ and\
  \bibinfo {author} {\bibfnamefont {S.}~\bibnamefont {Pal}},\ }\bibfield
  {title} {\enquote {\bibinfo {title} {Lower scaling approximation to
  {EOM}-{CCSD}: A critical assessment of the ionization problem},}\ }\href
  {\doibase 10.1002/qua.25594} {\bibfield  {journal} {\bibinfo  {journal}
  {International Journal of Quantum Chemistry}\ }\textbf {\bibinfo {volume}
  {118}},\ \bibinfo {pages} {e25594} (\bibinfo {year} {2018})}\BibitemShut
  {NoStop}%
\bibitem [{\citenamefont {Nooijen}\ and\ \citenamefont
  {Snijders}(1995)}]{Nooijen1995}%
  \BibitemOpen
  \bibfield  {author} {\bibinfo {author} {\bibfnamefont {M.}~\bibnamefont
  {Nooijen}}\ and\ \bibinfo {author} {\bibfnamefont {J.~G.}\ \bibnamefont
  {Snijders}},\ }\bibfield  {title} {\enquote {\bibinfo {title} {Second order
  many-body perturbation approximations to the coupled cluster green's
  function},}\ }\href {\doibase 10.1063/1.468900} {\bibfield  {journal}
  {\bibinfo  {journal} {The Journal of Chemical Physics}\ }\textbf {\bibinfo
  {volume} {102}},\ \bibinfo {pages} {1681--1688} (\bibinfo {year}
  {1995})}\BibitemShut {NoStop}%
\bibitem [{\citenamefont {Stanton}\ and\ \citenamefont
  {Gauss}(1995)}]{Stanton1995}%
  \BibitemOpen
  \bibfield  {author} {\bibinfo {author} {\bibfnamefont {J.~F.}\ \bibnamefont
  {Stanton}}\ and\ \bibinfo {author} {\bibfnamefont {J.}~\bibnamefont
  {Gauss}},\ }\bibfield  {title} {\enquote {\bibinfo {title} {Perturbative
  treatment of the similarity transformed hamiltonian in equation-of-motion
  coupled-cluster approximations},}\ }\href {\doibase 10.1063/1.469817}
  {\bibfield  {journal} {\bibinfo  {journal} {The Journal of Chemical Physics}\
  }\textbf {\bibinfo {volume} {103}},\ \bibinfo {pages} {1064--1076} (\bibinfo
  {year} {1995})}\BibitemShut {NoStop}%
\bibitem [{\citenamefont {Gwaltney}, \citenamefont {Nooijen},\ and\
  \citenamefont {Bartlett}(1996)}]{Gwaltney1996}%
  \BibitemOpen
  \bibfield  {author} {\bibinfo {author} {\bibfnamefont {S.~R.}\ \bibnamefont
  {Gwaltney}}, \bibinfo {author} {\bibfnamefont {M.}~\bibnamefont {Nooijen}}, \
  and\ \bibinfo {author} {\bibfnamefont {R.~J.}\ \bibnamefont {Bartlett}},\
  }\bibfield  {title} {\enquote {\bibinfo {title} {Simplified methods for
  equation-of-motion coupled-cluster excited state calculations},}\ }\href
  {\doibase 10.1016/0009-2614(95)01329-6} {\bibfield  {journal} {\bibinfo
  {journal} {Chemical Physics Letters}\ }\textbf {\bibinfo {volume} {248}},\
  \bibinfo {pages} {189--198} (\bibinfo {year} {1996})}\BibitemShut {NoStop}%
\bibitem [{\citenamefont {Nooijen}, \citenamefont {Perera},\ and\ \citenamefont
  {Bartlett}(1997)}]{Nooijen1997}%
  \BibitemOpen
  \bibfield  {author} {\bibinfo {author} {\bibfnamefont {M.}~\bibnamefont
  {Nooijen}}, \bibinfo {author} {\bibfnamefont {S.~A.}\ \bibnamefont {Perera}},
  \ and\ \bibinfo {author} {\bibfnamefont {R.~J.}\ \bibnamefont {Bartlett}},\
  }\bibfield  {title} {\enquote {\bibinfo {title} {Partitioned
  equation-of-motion coupled cluster approach to indirect nuclear spin-spin
  coupling constants},}\ }\href {\doibase 10.1016/s0009-2614(97)00048-1}
  {\bibfield  {journal} {\bibinfo  {journal} {Chemical Physics Letters}\
  }\textbf {\bibinfo {volume} {266}},\ \bibinfo {pages} {456--464} (\bibinfo
  {year} {1997})}\BibitemShut {NoStop}%
\bibitem [{\citenamefont {Christiansen}, \citenamefont {Koch},\ and\
  \citenamefont {Jo/rgensen}(1995)}]{Christiansen1995b}%
  \BibitemOpen
  \bibfield  {author} {\bibinfo {author} {\bibfnamefont {O.}~\bibnamefont
  {Christiansen}}, \bibinfo {author} {\bibfnamefont {H.}~\bibnamefont {Koch}},
  \ and\ \bibinfo {author} {\bibfnamefont {P.}~\bibnamefont {Jo/rgensen}},\
  }\bibfield  {title} {\enquote {\bibinfo {title} {Response functions in the
  {CC}3 iterative triple excitation model},}\ }\href {\doibase
  10.1063/1.470315} {\bibfield  {journal} {\bibinfo  {journal} {The Journal of
  Chemical Physics}\ }\textbf {\bibinfo {volume} {103}},\ \bibinfo {pages}
  {7429--7441} (\bibinfo {year} {1995})}\BibitemShut {NoStop}%
\bibitem [{\citenamefont {Koch}\ \emph {et~al.}(1997)\citenamefont {Koch},
  \citenamefont {Christiansen}, \citenamefont {Jo/rgensen}, \citenamefont
  {de~Mer{\'{a}}s},\ and\ \citenamefont {Helgaker}}]{Koch1997}%
  \BibitemOpen
  \bibfield  {author} {\bibinfo {author} {\bibfnamefont {H.}~\bibnamefont
  {Koch}}, \bibinfo {author} {\bibfnamefont {O.}~\bibnamefont {Christiansen}},
  \bibinfo {author} {\bibfnamefont {P.}~\bibnamefont {Jo/rgensen}}, \bibinfo
  {author} {\bibfnamefont {A.~M.~S.}\ \bibnamefont {de~Mer{\'{a}}s}}, \ and\
  \bibinfo {author} {\bibfnamefont {T.}~\bibnamefont {Helgaker}},\ }\bibfield
  {title} {\enquote {\bibinfo {title} {The {CC}3 model: An iterative coupled
  cluster approach including connected triples},}\ }\href {\doibase
  10.1063/1.473322} {\bibfield  {journal} {\bibinfo  {journal} {The Journal of
  Chemical Physics}\ }\textbf {\bibinfo {volume} {106}},\ \bibinfo {pages}
  {1808--1818} (\bibinfo {year} {1997})}\BibitemShut {NoStop}%
\bibitem [{\citenamefont {Christiansen}, \citenamefont {Koch},\ and\
  \citenamefont {J{\o}rgensen}(1995)}]{Christiansen1995a}%
  \BibitemOpen
  \bibfield  {author} {\bibinfo {author} {\bibfnamefont {O.}~\bibnamefont
  {Christiansen}}, \bibinfo {author} {\bibfnamefont {H.}~\bibnamefont {Koch}},
  \ and\ \bibinfo {author} {\bibfnamefont {P.}~\bibnamefont {J{\o}rgensen}},\
  }\bibfield  {title} {\enquote {\bibinfo {title} {The second-order approximate
  coupled cluster singles and doubles model {CC}2},}\ }\href {\doibase
  10.1016/0009-2614(95)00841-q} {\bibfield  {journal} {\bibinfo  {journal}
  {Chemical Physics Letters}\ }\textbf {\bibinfo {volume} {243}},\ \bibinfo
  {pages} {409--418} (\bibinfo {year} {1995})}\BibitemShut {NoStop}%
\bibitem [{\citenamefont {Tajti}\ and\ \citenamefont
  {Szalay}(2016)}]{Tajti2016}%
  \BibitemOpen
  \bibfield  {author} {\bibinfo {author} {\bibfnamefont {A.}~\bibnamefont
  {Tajti}}\ and\ \bibinfo {author} {\bibfnamefont {P.~G.}\ \bibnamefont
  {Szalay}},\ }\bibfield  {title} {\enquote {\bibinfo {title} {Investigation of
  the impact of different terms in the second order hamiltonian on excitation
  energies of valence and rydberg states},}\ }\href {\doibase
  10.1021/acs.jctc.6b00723} {\bibfield  {journal} {\bibinfo  {journal} {Journal
  of Chemical Theory and Computation}\ }\textbf {\bibinfo {volume} {12}},\
  \bibinfo {pages} {5477--5482} (\bibinfo {year} {2016})}\BibitemShut {NoStop}%
\bibitem [{\citenamefont {Saue}\ \emph {et~al.}(1997)\citenamefont {Saue},
  \citenamefont {{Faegri}}, \citenamefont {Helgaker},\ and\ \citenamefont
  {Gropen}}]{Saue1997}%
  \BibitemOpen
  \bibfield  {author} {\bibinfo {author} {\bibfnamefont {T.}~\bibnamefont
  {Saue}}, \bibinfo {author} {\bibfnamefont {K.}~\bibnamefont {{Faegri}},
  \bibfnamefont {Jr.}}, \bibinfo {author} {\bibfnamefont {T.}~\bibnamefont
  {Helgaker}}, \ and\ \bibinfo {author} {\bibfnamefont {O.}~\bibnamefont
  {Gropen}},\ }\bibfield  {title} {\enquote {\bibinfo {title} {Principles of
  direct 4-component relativistic {SCF}: application to caesium auride},}\
  }\href {\doibase 10.1080/002689797171058} {\bibfield  {journal} {\bibinfo
  {journal} {Molecular Physics}\ }\textbf {\bibinfo {volume} {91}},\ \bibinfo
  {pages} {937--950} (\bibinfo {year} {1997})}\BibitemShut {NoStop}%
\bibitem [{\citenamefont {Pyykk\"{o}}\ and\ \citenamefont
  {Desclaux}(1979)}]{Pyykko1979}%
  \BibitemOpen
  \bibfield  {author} {\bibinfo {author} {\bibfnamefont {P.}~\bibnamefont
  {Pyykk\"{o}}}\ and\ \bibinfo {author} {\bibfnamefont {J.~P.}\ \bibnamefont
  {Desclaux}},\ }\bibfield  {title} {\enquote {\bibinfo {title} {Relativity and
  the periodic system of elements},}\ }\href {\doibase 10.1021/ar50140a002}
  {\bibfield  {journal} {\bibinfo  {journal} {Accounts of Chemical Research}\
  }\textbf {\bibinfo {volume} {12}},\ \bibinfo {pages} {276--281} (\bibinfo
  {year} {1979})}\BibitemShut {NoStop}%
\bibitem [{\citenamefont {Pyykk\"{o}}(1988)}]{pyykko_relativistic_1988}%
  \BibitemOpen
  \bibfield  {author} {\bibinfo {author} {\bibfnamefont {P.}~\bibnamefont
  {Pyykk\"{o}}},\ }\bibfield  {title} {\enquote {\bibinfo {title} {Relativistic
  effects in structural chemistry},}\ }\href {\doibase 10.1021/cr00085a006}
  {\bibfield  {journal} {\bibinfo  {journal} {Chemical Reviews}\ }\textbf
  {\bibinfo {volume} {88}},\ \bibinfo {pages} {563--594} (\bibinfo {year}
  {1988})}\BibitemShut {NoStop}%
\bibitem [{\citenamefont {Pyykk\"{o}}(2011)}]{Pyykk2011}%
  \BibitemOpen
  \bibfield  {author} {\bibinfo {author} {\bibfnamefont {P.}~\bibnamefont
  {Pyykk\"{o}}},\ }\bibfield  {title} {\enquote {\bibinfo {title} {The physics
  behind chemistry and the periodic table},}\ }\href {\doibase
  10.1021/cr200042e} {\bibfield  {journal} {\bibinfo  {journal} {Chemical
  Reviews}\ }\textbf {\bibinfo {volume} {112}},\ \bibinfo {pages} {371--384}
  (\bibinfo {year} {2011})}\BibitemShut {NoStop}%
\bibitem [{\citenamefont {Opoku}, \citenamefont {Toubin},\ and\ \citenamefont
  {Gomes}(2022)}]{Opoku2022}%
  \BibitemOpen
  \bibfield  {author} {\bibinfo {author} {\bibfnamefont {R.~A.}\ \bibnamefont
  {Opoku}}, \bibinfo {author} {\bibfnamefont {C.}~\bibnamefont {Toubin}}, \
  and\ \bibinfo {author} {\bibfnamefont {A.~S.~P.}\ \bibnamefont {Gomes}},\
  }\bibfield  {title} {\enquote {\bibinfo {title} {Simulating core electron
  binding energies of halogenated species adsorbed on ice surfaces and in
  solution {$via$} relativistic quantum embedding calculations},}\ }\href
  {\doibase 10.1039/d1cp05836c} {\bibfield  {journal} {\bibinfo  {journal}
  {Physical Chemistry Chemical Physics}\ }\textbf {\bibinfo {volume} {24}},\
  \bibinfo {pages} {14390--14407} (\bibinfo {year} {2022})}\BibitemShut
  {NoStop}%
\bibitem [{\citenamefont {Halbert}\ \emph {et~al.}(2021)\citenamefont
  {Halbert}, \citenamefont {Vidal}, \citenamefont {Shee}, \citenamefont
  {Coriani},\ and\ \citenamefont {Gomes}}]{Halbert2021}%
  \BibitemOpen
  \bibfield  {author} {\bibinfo {author} {\bibfnamefont {L.}~\bibnamefont
  {Halbert}}, \bibinfo {author} {\bibfnamefont {M.~L.}\ \bibnamefont {Vidal}},
  \bibinfo {author} {\bibfnamefont {A.}~\bibnamefont {Shee}}, \bibinfo {author}
  {\bibfnamefont {S.}~\bibnamefont {Coriani}}, \ and\ \bibinfo {author}
  {\bibfnamefont {A.~S.~P.}\ \bibnamefont {Gomes}},\ }\bibfield  {title}
  {\enquote {\bibinfo {title} {Relativistic {EOM}-{CCSD} for core-excited and
  core-ionized state energies based on the four-component
  dirac{\textendash}coulomb(-gaunt) hamiltonian},}\ }\href {\doibase
  10.1021/acs.jctc.0c01203} {\bibfield  {journal} {\bibinfo  {journal} {Journal
  of Chemical Theory and Computation}\ }\textbf {\bibinfo {volume} {17}},\
  \bibinfo {pages} {3583--3598} (\bibinfo {year} {2021})}\BibitemShut {NoStop}%
\bibitem [{\citenamefont {Saiz-Lopez}\ \emph {et~al.}(2011)\citenamefont
  {Saiz-Lopez}, \citenamefont {Plane}, \citenamefont {Baker}, \citenamefont
  {Carpenter}, \citenamefont {von Glasow}, \citenamefont {Mart{\'{\i}}n},
  \citenamefont {McFiggans},\ and\ \citenamefont {Saunders}}]{SaizLopez2011}%
  \BibitemOpen
  \bibfield  {author} {\bibinfo {author} {\bibfnamefont {A.}~\bibnamefont
  {Saiz-Lopez}}, \bibinfo {author} {\bibfnamefont {J.~M.~C.}\ \bibnamefont
  {Plane}}, \bibinfo {author} {\bibfnamefont {A.~R.}\ \bibnamefont {Baker}},
  \bibinfo {author} {\bibfnamefont {L.~J.}\ \bibnamefont {Carpenter}}, \bibinfo
  {author} {\bibfnamefont {R.}~\bibnamefont {von Glasow}}, \bibinfo {author}
  {\bibfnamefont {J.~C.~G.}\ \bibnamefont {Mart{\'{\i}}n}}, \bibinfo {author}
  {\bibfnamefont {G.}~\bibnamefont {McFiggans}}, \ and\ \bibinfo {author}
  {\bibfnamefont {R.~W.}\ \bibnamefont {Saunders}},\ }\bibfield  {title}
  {\enquote {\bibinfo {title} {Atmospheric chemistry of iodine},}\ }\href
  {\doibase 10.1021/cr200029u} {\bibfield  {journal} {\bibinfo  {journal}
  {Chemical Reviews}\ }\textbf {\bibinfo {volume} {112}},\ \bibinfo {pages}
  {1773--1804} (\bibinfo {year} {2011})}\BibitemShut {NoStop}%
\bibitem [{\citenamefont {Steinhauser}, \citenamefont {Brandl},\ and\
  \citenamefont {Johnson}(2014)}]{Steinhauser2014}%
  \BibitemOpen
  \bibfield  {author} {\bibinfo {author} {\bibfnamefont {G.}~\bibnamefont
  {Steinhauser}}, \bibinfo {author} {\bibfnamefont {A.}~\bibnamefont {Brandl}},
  \ and\ \bibinfo {author} {\bibfnamefont {T.~E.}\ \bibnamefont {Johnson}},\
  }\bibfield  {title} {\enquote {\bibinfo {title} {Comparison of the chernobyl
  and fukushima nuclear accidents: A review of the environmental impacts},}\
  }\href {\doibase 10.1016/j.scitotenv.2013.10.029} {\bibfield  {journal}
  {\bibinfo  {journal} {Science of The Total Environment}\ }\textbf {\bibinfo
  {volume} {470-471}},\ \bibinfo {pages} {800--817} (\bibinfo {year}
  {2014})}\BibitemShut {NoStop}%
\bibitem [{\citenamefont {Yuan}\ \emph {et~al.}(2023)\citenamefont {Yuan},
  \citenamefont {Halbert}, \citenamefont {Pototschnig}, \citenamefont
  {Papadopoulos}, \citenamefont {Coriani}, \citenamefont {Visscher},\ and\
  \citenamefont {Gomes}}]{Yuan2023a}%
  \BibitemOpen
  \bibfield  {author} {\bibinfo {author} {\bibfnamefont {X.}~\bibnamefont
  {Yuan}}, \bibinfo {author} {\bibfnamefont {L.}~\bibnamefont {Halbert}},
  \bibinfo {author} {\bibfnamefont {J.}~\bibnamefont {Pototschnig}}, \bibinfo
  {author} {\bibfnamefont {A.}~\bibnamefont {Papadopoulos}}, \bibinfo {author}
  {\bibfnamefont {S.}~\bibnamefont {Coriani}}, \bibinfo {author} {\bibfnamefont
  {L.}~\bibnamefont {Visscher}}, \ and\ \bibinfo {author} {\bibfnamefont
  {A.~S.~P.}\ \bibnamefont {Gomes}},\ }\href {\doibase
  10.48550/ARXIV.2307.14296} {\enquote {\bibinfo {title} {Formulation and
  implementation of frequency-dependent linear response properties with
  relativistic coupled cluster theory for gpu-accelerated computer
  architectures},}\ } (\bibinfo {year} {2023}),\ \Eprint
  {http://arxiv.org/abs/2307.14296} {arXiv:2307.14296 [physics.chem-ph]}
  \BibitemShut {NoStop}%
\bibitem [{\citenamefont {Pototschnig}\ \emph {et~al.}(2021)\citenamefont
  {Pototschnig}, \citenamefont {Papadopoulos}, \citenamefont {Lyakh},
  \citenamefont {Repisky}, \citenamefont {Halbert}, \citenamefont {Gomes},
  \citenamefont {Jensen},\ and\ \citenamefont {Visscher}}]{Pototschnig2021}%
  \BibitemOpen
  \bibfield  {author} {\bibinfo {author} {\bibfnamefont {J.~V.}\ \bibnamefont
  {Pototschnig}}, \bibinfo {author} {\bibfnamefont {A.}~\bibnamefont
  {Papadopoulos}}, \bibinfo {author} {\bibfnamefont {D.~I.}\ \bibnamefont
  {Lyakh}}, \bibinfo {author} {\bibfnamefont {M.}~\bibnamefont {Repisky}},
  \bibinfo {author} {\bibfnamefont {L.}~\bibnamefont {Halbert}}, \bibinfo
  {author} {\bibfnamefont {A.~S.~P.}\ \bibnamefont {Gomes}}, \bibinfo {author}
  {\bibfnamefont {H.~J.~A.}\ \bibnamefont {Jensen}}, \ and\ \bibinfo {author}
  {\bibfnamefont {L.}~\bibnamefont {Visscher}},\ }\bibfield  {title} {\enquote
  {\bibinfo {title} {Implementation of relativistic coupled cluster theory for
  massively parallel {GPU}-accelerated computing architectures},}\ }\href
  {\doibase 10.1021/acs.jctc.1c00260} {\bibfield  {journal} {\bibinfo
  {journal} {Journal of Chemical Theory and Computation}\ }\textbf {\bibinfo
  {volume} {17}},\ \bibinfo {pages} {5509--5529} (\bibinfo {year}
  {2021})}\BibitemShut {NoStop}%
\bibitem [{\citenamefont {Saue}\ \emph {et~al.}(2020)\citenamefont {Saue},
  \citenamefont {Bast}, \citenamefont {Gomes}, \citenamefont {Jensen},
  \citenamefont {Visscher}, \citenamefont {Aucar}, \citenamefont {Remigio},
  \citenamefont {Dyall}, \citenamefont {Eliav}, \citenamefont {Fasshauer},
  \citenamefont {Fleig}, \citenamefont {Halbert}, \citenamefont {Hedeg{\aa}rd},
  \citenamefont {Helmich-Paris}, \citenamefont {Ilia{\v{s}}}, \citenamefont
  {Jacob}, \citenamefont {Knecht}, \citenamefont {Laerdahl}, \citenamefont
  {Vidal}, \citenamefont {Nayak}, \citenamefont {Olejniczak}, \citenamefont
  {Olsen}, \citenamefont {Pernpointner}, \citenamefont {Senjean}, \citenamefont
  {Shee}, \citenamefont {Sunaga},\ and\ \citenamefont {van
  Stralen}}]{DiracSaue2020}%
  \BibitemOpen
  \bibfield  {author} {\bibinfo {author} {\bibfnamefont {T.}~\bibnamefont
  {Saue}}, \bibinfo {author} {\bibfnamefont {R.}~\bibnamefont {Bast}}, \bibinfo
  {author} {\bibfnamefont {A.~S.~P.}\ \bibnamefont {Gomes}}, \bibinfo {author}
  {\bibfnamefont {H.~J.~A.}\ \bibnamefont {Jensen}}, \bibinfo {author}
  {\bibfnamefont {L.}~\bibnamefont {Visscher}}, \bibinfo {author}
  {\bibfnamefont {I.~A.}\ \bibnamefont {Aucar}}, \bibinfo {author}
  {\bibfnamefont {R.~D.}\ \bibnamefont {Remigio}}, \bibinfo {author}
  {\bibfnamefont {K.~G.}\ \bibnamefont {Dyall}}, \bibinfo {author}
  {\bibfnamefont {E.}~\bibnamefont {Eliav}}, \bibinfo {author} {\bibfnamefont
  {E.}~\bibnamefont {Fasshauer}}, \bibinfo {author} {\bibfnamefont
  {T.}~\bibnamefont {Fleig}}, \bibinfo {author} {\bibfnamefont
  {L.}~\bibnamefont {Halbert}}, \bibinfo {author} {\bibfnamefont {E.~D.}\
  \bibnamefont {Hedeg{\aa}rd}}, \bibinfo {author} {\bibfnamefont
  {B.}~\bibnamefont {Helmich-Paris}}, \bibinfo {author} {\bibfnamefont
  {M.}~\bibnamefont {Ilia{\v{s}}}}, \bibinfo {author} {\bibfnamefont {C.~R.}\
  \bibnamefont {Jacob}}, \bibinfo {author} {\bibfnamefont {S.}~\bibnamefont
  {Knecht}}, \bibinfo {author} {\bibfnamefont {J.~K.}\ \bibnamefont
  {Laerdahl}}, \bibinfo {author} {\bibfnamefont {M.~L.}\ \bibnamefont {Vidal}},
  \bibinfo {author} {\bibfnamefont {M.~K.}\ \bibnamefont {Nayak}}, \bibinfo
  {author} {\bibfnamefont {M.}~\bibnamefont {Olejniczak}}, \bibinfo {author}
  {\bibfnamefont {J.~M.~H.}\ \bibnamefont {Olsen}}, \bibinfo {author}
  {\bibfnamefont {M.}~\bibnamefont {Pernpointner}}, \bibinfo {author}
  {\bibfnamefont {B.}~\bibnamefont {Senjean}}, \bibinfo {author} {\bibfnamefont
  {A.}~\bibnamefont {Shee}}, \bibinfo {author} {\bibfnamefont {A.}~\bibnamefont
  {Sunaga}}, \ and\ \bibinfo {author} {\bibfnamefont {J.~N.~P.}\ \bibnamefont
  {van Stralen}},\ }\bibfield  {title} {\enquote {\bibinfo {title} {The {DIRAC}
  code for relativistic molecular calculations},}\ }\href {\doibase
  10.1063/5.0004844} {\bibfield  {journal} {\bibinfo  {journal} {The Journal of
  Chemical Physics}\ }\textbf {\bibinfo {volume} {152}},\ \bibinfo {pages}
  {204104} (\bibinfo {year} {2020})}\BibitemShut {NoStop}%
\bibitem [{\citenamefont {Crawford}\ and\ \citenamefont
  {Schaefer}(2007)}]{Crawford2007}%
  \BibitemOpen
  \bibfield  {author} {\bibinfo {author} {\bibfnamefont {T.~D.}\ \bibnamefont
  {Crawford}}\ and\ \bibinfo {author} {\bibfnamefont {H.~F.}\ \bibnamefont
  {Schaefer}},\ }\bibfield  {title} {\enquote {\bibinfo {title} {An
  introduction to coupled cluster theory for computational chemists},}\ }in\
  \href {\doibase 10.1002/9780470125915.ch2} {\emph {\bibinfo {booktitle}
  {Reviews in Computational Chemistry}}}\ (\bibinfo  {publisher} {John Wiley
  {\&} Sons, Inc.},\ \bibinfo {year} {2007})\ pp.\ \bibinfo {pages}
  {33--136}\BibitemShut {NoStop}%
\bibitem [{\citenamefont {Cederbaum}, \citenamefont {Domcke},\ and\
  \citenamefont {Schirmer}(1980)}]{cederbaum_many-body_1980}%
  \BibitemOpen
  \bibfield  {author} {\bibinfo {author} {\bibfnamefont {L.~S.}\ \bibnamefont
  {Cederbaum}}, \bibinfo {author} {\bibfnamefont {W.}~\bibnamefont {Domcke}}, \
  and\ \bibinfo {author} {\bibfnamefont {J.}~\bibnamefont {Schirmer}},\
  }\bibfield  {title} {\enquote {\bibinfo {title} {Many-body theory of core
  holes},}\ }\href {\doibase 10.1103/PhysRevA.22.206} {\bibfield  {journal}
  {\bibinfo  {journal} {Physical Review A}\ }\textbf {\bibinfo {volume} {22}},\
  \bibinfo {pages} {206--222} (\bibinfo {year} {1980})}\BibitemShut {NoStop}%
\bibitem [{\citenamefont {L\"{o}wdin}(1963)}]{Lwdin1963}%
  \BibitemOpen
  \bibfield  {author} {\bibinfo {author} {\bibfnamefont {P.-O.}\ \bibnamefont
  {L\"{o}wdin}},\ }\bibfield  {title} {\enquote {\bibinfo {title} {Studies in
  perturbation theory},}\ }\href {\doibase 10.1016/0022-2852(63)90151-6}
  {\bibfield  {journal} {\bibinfo  {journal} {Journal of Molecular
  Spectroscopy}\ }\textbf {\bibinfo {volume} {10}},\ \bibinfo {pages} {12--33}
  (\bibinfo {year} {1963})}\BibitemShut {NoStop}%
\bibitem [{\citenamefont {Lawley}(1987)}]{lawley_ab_1987}%
  \BibitemOpen
  \bibinfo {editor} {\bibfnamefont {K.~P.}\ \bibnamefont {Lawley}},\ ed.,\
  \href@noop {} {\emph {\bibinfo {title} {Ab initio methods in quantum
  chemistry}}},\ \bibinfo {series} {Advances in chemical physics}\ No.\
  \bibinfo {number} {v. 67, 69}\ (\bibinfo  {publisher} {Wiley},\ \bibinfo
  {address} {Chichester [West Sussex] ; New York},\ \bibinfo {year}
  {1987})\BibitemShut {NoStop}%
\bibitem [{\citenamefont {Geertsen}, \citenamefont {Rittby},\ and\
  \citenamefont {Bartlett}(1989)}]{Geertsen1989}%
  \BibitemOpen
  \bibfield  {author} {\bibinfo {author} {\bibfnamefont {J.}~\bibnamefont
  {Geertsen}}, \bibinfo {author} {\bibfnamefont {M.}~\bibnamefont {Rittby}}, \
  and\ \bibinfo {author} {\bibfnamefont {R.~J.}\ \bibnamefont {Bartlett}},\
  }\bibfield  {title} {\enquote {\bibinfo {title} {The equation-of-motion
  coupled-cluster method: Excitation energies of be and {CO}},}\ }\href
  {\doibase 10.1016/0009-2614(89)85202-9} {\bibfield  {journal} {\bibinfo
  {journal} {Chemical Physics Letters}\ }\textbf {\bibinfo {volume} {164}},\
  \bibinfo {pages} {57--62} (\bibinfo {year} {1989})}\BibitemShut {NoStop}%
\bibitem [{\citenamefont {Gauss}\ and\ \citenamefont
  {Stanton}(1995)}]{Gauss1995}%
  \BibitemOpen
  \bibfield  {author} {\bibinfo {author} {\bibfnamefont {J.}~\bibnamefont
  {Gauss}}\ and\ \bibinfo {author} {\bibfnamefont {J.~F.}\ \bibnamefont
  {Stanton}},\ }\bibfield  {title} {\enquote {\bibinfo {title} {Coupled-cluster
  calculations of nuclear magnetic resonance chemical shifts},}\ }\href
  {\doibase 10.1063/1.470240} {\bibfield  {journal} {\bibinfo  {journal} {The
  Journal of Chemical Physics}\ }\textbf {\bibinfo {volume} {103}},\ \bibinfo
  {pages} {3561--3577} (\bibinfo {year} {1995})}\BibitemShut {NoStop}%
\bibitem [{DIR(2019)}]{DIRAC19}%
  \BibitemOpen
  \href@noop {} {} (\bibinfo {year} {2019}),\ \bibinfo {note} {{DIRAC}, a
  relativistic ab initio electronic structure program, Release {DIRAC19}
  (2019), written by A.~S.~P.~Gomes, T.~Saue, L.~Visscher, H.~J.~{\relax
  Aa}.~Jensen, and R.~Bast, with contributions from I.~A.~Aucar, V.~Bakken,
  K.~G.~Dyall, S.~Dubillard, U.~Ekstr{\"o}m, E.~Eliav, T.~Enevoldsen,
  E.~Fa{\ss}hauer, T.~Fleig, O.~Fossgaard, L.~Halbert, E.~D.~Hedeg{\aa}rd,
  B.~Heimlich--Paris, T.~Helgaker, J.~Henriksson, M.~Ilia{\v{s}}, Ch.~R.~Jacob,
  S.~Knecht, S.~Komorovsk{\'y}, O.~Kullie, J.~K.~L{\ae}rdahl, C.~V.~Larsen,
  Y.~S.~Lee, H.~S.~Nataraj, M.~K.~Nayak, P.~Norman, G.~Olejniczak, J.~Olsen,
  J.~M.~H.~Olsen, Y.~C.~Park, J.~K.~Pedersen, M.~Pernpointner, R.~di~Remigio,
  K.~Ruud, P.~Sa{\l}ek, B.~Schimmelpfennig, B.~Senjean, A.~Shee, J.~Sikkema,
  A.~J.~Thorvaldsen, J.~Thyssen, J.~van~Stralen, M.~L.~Vidal, S.~Villaume,
  O.~Visser, T.~Winther, and S.~Yamamoto (available at
  \url{http://dx.doi.org/10.5281/zenodo.3572669}, see also
  \url{http://www.diracprogram.org})}\BibitemShut {NoStop}%
\bibitem [{\citenamefont {Visscher}, \citenamefont {Styszy{\~{n}}ski},\ and\
  \citenamefont {Nieuwpoort}(1996)}]{Visscher1996}%
  \BibitemOpen
  \bibfield  {author} {\bibinfo {author} {\bibfnamefont {L.}~\bibnamefont
  {Visscher}}, \bibinfo {author} {\bibfnamefont {J.}~\bibnamefont
  {Styszy{\~{n}}ski}}, \ and\ \bibinfo {author} {\bibfnamefont {W.~C.}\
  \bibnamefont {Nieuwpoort}},\ }\bibfield  {title} {\enquote {\bibinfo {title}
  {Relativistic and correlation effects on molecular properties. {II}. the
  hydrogen halides {HF}, {HCl}, {HBr}, {HI}, and {HAt}},}\ }\href {\doibase
  10.1063/1.472066} {\bibfield  {journal} {\bibinfo  {journal} {The Journal of
  Chemical Physics}\ }\textbf {\bibinfo {volume} {105}},\ \bibinfo {pages}
  {1987--1994} (\bibinfo {year} {1996})}\BibitemShut {NoStop}%
\bibitem [{\citenamefont {Dyall}(2006)}]{Dyall2006}%
  \BibitemOpen
  \bibfield  {author} {\bibinfo {author} {\bibfnamefont {K.~G.}\ \bibnamefont
  {Dyall}},\ }\bibfield  {title} {\enquote {\bibinfo {title} {Relativistic
  quadruple-zeta and revised triple-zeta and double-zeta basis sets for the 4p,
  5p, and 6p elements},}\ }\href {\doibase 10.1007/s00214-006-0126-0}
  {\bibfield  {journal} {\bibinfo  {journal} {Theoretical Chemistry Accounts}\
  }\textbf {\bibinfo {volume} {115}},\ \bibinfo {pages} {441--447} (\bibinfo
  {year} {2006})}\BibitemShut {NoStop}%
\bibitem [{\citenamefont {Kendall}, \citenamefont {Dunning},\ and\
  \citenamefont {Harrison}(1992)}]{Kendall1992}%
  \BibitemOpen
  \bibfield  {author} {\bibinfo {author} {\bibfnamefont {R.~A.}\ \bibnamefont
  {Kendall}}, \bibinfo {author} {\bibfnamefont {T.~H.}\ \bibnamefont
  {Dunning}}, \ and\ \bibinfo {author} {\bibfnamefont {R.~J.}\ \bibnamefont
  {Harrison}},\ }\bibfield  {title} {\enquote {\bibinfo {title} {Electron
  affinities of the first-row atoms revisited. systematic basis sets and wave
  functions},}\ }\href {\doibase 10.1063/1.462569} {\bibfield  {journal}
  {\bibinfo  {journal} {The Journal of Chemical Physics}\ }\textbf {\bibinfo
  {volume} {96}},\ \bibinfo {pages} {6796--6806} (\bibinfo {year}
  {1992})}\BibitemShut {NoStop}%
\bibitem [{\citenamefont {Sikkema}\ \emph {et~al.}(2009)\citenamefont
  {Sikkema}, \citenamefont {Visscher}, \citenamefont {Saue},\ and\
  \citenamefont {Ilia{\v{s}}}}]{Sikkema2009}%
  \BibitemOpen
  \bibfield  {author} {\bibinfo {author} {\bibfnamefont {J.}~\bibnamefont
  {Sikkema}}, \bibinfo {author} {\bibfnamefont {L.}~\bibnamefont {Visscher}},
  \bibinfo {author} {\bibfnamefont {T.}~\bibnamefont {Saue}}, \ and\ \bibinfo
  {author} {\bibfnamefont {M.}~\bibnamefont {Ilia{\v{s}}}},\ }\bibfield
  {title} {\enquote {\bibinfo {title} {The molecular mean-field approach for
  correlated relativistic calculations},}\ }\href {\doibase 10.1063/1.3239505}
  {\bibfield  {journal} {\bibinfo  {journal} {The Journal of Chemical Physics}\
  }\textbf {\bibinfo {volume} {131}},\ \bibinfo {pages} {124116} (\bibinfo
  {year} {2009})}\BibitemShut {NoStop}%
\bibitem [{\citenamefont {Visscher}(1997)}]{Visscher1997}%
  \BibitemOpen
  \bibfield  {author} {\bibinfo {author} {\bibfnamefont {L.}~\bibnamefont
  {Visscher}},\ }\bibfield  {title} {\enquote {\bibinfo {title} {Approximate
  molecular relativistic dirac-coulomb calculations using a simple coulombic
  correction},}\ }\href {\doibase 10.1007/s002140050280} {\bibfield  {journal}
  {\bibinfo  {journal} {Theoretical Chemistry Accounts: Theory, Computation,
  and Modeling (Theoretica Chimica Acta)}\ }\textbf {\bibinfo {volume} {98}},\
  \bibinfo {pages} {68--70} (\bibinfo {year} {1997})}\BibitemShut {NoStop}%
\bibitem [{\citenamefont {Visscher}\ and\ \citenamefont
  {Dyall}(1997)}]{Visscher1997_nuclearpotential}%
  \BibitemOpen
  \bibfield  {author} {\bibinfo {author} {\bibfnamefont {L.}~\bibnamefont
  {Visscher}}\ and\ \bibinfo {author} {\bibfnamefont {K.~G.}\ \bibnamefont
  {Dyall}},\ }\bibfield  {title} {\enquote {\bibinfo {title} {Dirac-fock atomic
  electronic structure calculations using different nuclear charge
  distributions},}\ }\href {\doibase 10.1006/adnd.1997.0751} {\bibfield
  {journal} {\bibinfo  {journal} {Atomic Data and Nuclear Data Tables}\
  }\textbf {\bibinfo {volume} {67}},\ \bibinfo {pages} {207--224} (\bibinfo
  {year} {1997})}\BibitemShut {NoStop}%
\bibitem [{\citenamefont {Davidson}(1975)}]{Davidson1975}%
  \BibitemOpen
  \bibfield  {author} {\bibinfo {author} {\bibfnamefont {E.~R.}\ \bibnamefont
  {Davidson}},\ }\bibfield  {title} {\enquote {\bibinfo {title} {The iterative
  calculation of a few of the lowest eigenvalues and corresponding eigenvectors
  of large real-symmetric matrices},}\ }\href {\doibase
  10.1016/0021-9991(75)90065-0} {\bibfield  {journal} {\bibinfo  {journal}
  {Journal of Computational Physics}\ }\textbf {\bibinfo {volume} {17}},\
  \bibinfo {pages} {87--94} (\bibinfo {year} {1975})}\BibitemShut {NoStop}%
\bibitem [{\citenamefont {Hirao}\ and\ \citenamefont
  {Nakatsuji}(1982)}]{Hirao1982}%
  \BibitemOpen
  \bibfield  {author} {\bibinfo {author} {\bibfnamefont {K.}~\bibnamefont
  {Hirao}}\ and\ \bibinfo {author} {\bibfnamefont {H.}~\bibnamefont
  {Nakatsuji}},\ }\bibfield  {title} {\enquote {\bibinfo {title} {A
  generalization of the davidson{\textquotesingle}s method to large
  nonsymmetric eigenvalue problems},}\ }\href {\doibase
  10.1016/0021-9991(82)90119-x} {\bibfield  {journal} {\bibinfo  {journal}
  {Journal of Computational Physics}\ }\textbf {\bibinfo {volume} {45}},\
  \bibinfo {pages} {246--254} (\bibinfo {year} {1982})}\BibitemShut {NoStop}%
\bibitem [{\citenamefont {Hunter}(2007)}]{Hunter:2007}%
  \BibitemOpen
  \bibfield  {author} {\bibinfo {author} {\bibfnamefont {J.~D.}\ \bibnamefont
  {Hunter}},\ }\bibfield  {title} {\enquote {\bibinfo {title} {Matplotlib: A 2d
  graphics environment},}\ }\href {\doibase 10.1109/MCSE.2007.55} {\bibfield
  {journal} {\bibinfo  {journal} {Computing in Science \& Engineering}\
  }\textbf {\bibinfo {volume} {9}},\ \bibinfo {pages} {90--95} (\bibinfo {year}
  {2007})}\BibitemShut {NoStop}%
\bibitem [{\citenamefont {Halbert}\ and\ \citenamefont {Severo
  Pereira~Gomes}(2023)}]{halbert2023-dataset}%
  \BibitemOpen
  \bibfield  {author} {\bibinfo {author} {\bibfnamefont {L.}~\bibnamefont
  {Halbert}}\ and\ \bibinfo {author} {\bibfnamefont {A.}~\bibnamefont {Severo
  Pereira~Gomes}},\ }\href {\doibase 10.5281/ZENODO.8094645} {\enquote
  {\bibinfo {title} {Dataset: The performance of approximate equation of motion
  coupled cluster for valence and core states of heavy element systems},}\ }
  (\bibinfo {year} {2023})\BibitemShut {NoStop}%
\bibitem [{\citenamefont {Gomes}\ \emph {et~al.}(2010)\citenamefont {Gomes},
  \citenamefont {Visscher}, \citenamefont {Bolvin}, \citenamefont {Saue},
  \citenamefont {Knecht}, \citenamefont {Fleig},\ and\ \citenamefont
  {Eliav}}]{Gomes2010}%
  \BibitemOpen
  \bibfield  {author} {\bibinfo {author} {\bibfnamefont {A.~S.~P.}\
  \bibnamefont {Gomes}}, \bibinfo {author} {\bibfnamefont {L.}~\bibnamefont
  {Visscher}}, \bibinfo {author} {\bibfnamefont {H.}~\bibnamefont {Bolvin}},
  \bibinfo {author} {\bibfnamefont {T.}~\bibnamefont {Saue}}, \bibinfo {author}
  {\bibfnamefont {S.}~\bibnamefont {Knecht}}, \bibinfo {author} {\bibfnamefont
  {T.}~\bibnamefont {Fleig}}, \ and\ \bibinfo {author} {\bibfnamefont
  {E.}~\bibnamefont {Eliav}},\ }\bibfield  {title} {\enquote {\bibinfo {title}
  {The electronic structure of the triiodide ion from relativistic correlated
  calculations: A comparison of different methodologies},}\ }\href {\doibase
  10.1063/1.3474571} {\bibfield  {journal} {\bibinfo  {journal} {The Journal of
  Chemical Physics}\ }\textbf {\bibinfo {volume} {133}},\ \bibinfo {pages}
  {064305} (\bibinfo {year} {2010})}\BibitemShut {NoStop}%
\bibitem [{\citenamefont {Wang}, \citenamefont {Tu},\ and\ \citenamefont
  {Wang}(2014)}]{Wang2014}%
  \BibitemOpen
  \bibfield  {author} {\bibinfo {author} {\bibfnamefont {Z.}~\bibnamefont
  {Wang}}, \bibinfo {author} {\bibfnamefont {Z.}~\bibnamefont {Tu}}, \ and\
  \bibinfo {author} {\bibfnamefont {F.}~\bibnamefont {Wang}},\ }\bibfield
  {title} {\enquote {\bibinfo {title} {Equation-of-motion coupled-cluster
  theory for excitation energies of closed-shell systems with
  spin{\textendash}orbit coupling},}\ }\href {\doibase 10.1021/ct500854m}
  {\bibfield  {journal} {\bibinfo  {journal} {Journal of Chemical Theory and
  Computation}\ }\textbf {\bibinfo {volume} {10}},\ \bibinfo {pages}
  {5567--5576} (\bibinfo {year} {2014})}\BibitemShut {NoStop}%
\bibitem [{\citenamefont {Zhu}\ \emph {et~al.}(2001)\citenamefont {Zhu},
  \citenamefont {Takahashi}, \citenamefont {Saeki}, \citenamefont {Tsukuda},\
  and\ \citenamefont {Nagata}}]{Zhu2001}%
  \BibitemOpen
  \bibfield  {author} {\bibinfo {author} {\bibfnamefont {L.}~\bibnamefont
  {Zhu}}, \bibinfo {author} {\bibfnamefont {K.}~\bibnamefont {Takahashi}},
  \bibinfo {author} {\bibfnamefont {M.}~\bibnamefont {Saeki}}, \bibinfo
  {author} {\bibfnamefont {T.}~\bibnamefont {Tsukuda}}, \ and\ \bibinfo
  {author} {\bibfnamefont {T.}~\bibnamefont {Nagata}},\ }\bibfield  {title}
  {\enquote {\bibinfo {title} {Photodissociation of gas-phase {I}$_3^-$:
  product branching in the visible and {UV} regions},}\ }\href {\doibase
  10.1016/s0009-2614(01)01288-x} {\bibfield  {journal} {\bibinfo  {journal}
  {Chemical Physics Letters}\ }\textbf {\bibinfo {volume} {350}},\ \bibinfo
  {pages} {233--239} (\bibinfo {year} {2001})}\BibitemShut {NoStop}%
\bibitem [{\citenamefont {Choi}\ \emph {et~al.}(2000)\citenamefont {Choi},
  \citenamefont {Bise}, \citenamefont {Hoops},\ and\ \citenamefont
  {Neumark}}]{Choi2000}%
  \BibitemOpen
  \bibfield  {author} {\bibinfo {author} {\bibfnamefont {H.}~\bibnamefont
  {Choi}}, \bibinfo {author} {\bibfnamefont {R.~T.}\ \bibnamefont {Bise}},
  \bibinfo {author} {\bibfnamefont {A.~A.}\ \bibnamefont {Hoops}}, \ and\
  \bibinfo {author} {\bibfnamefont {D.~M.}\ \bibnamefont {Neumark}},\
  }\bibfield  {title} {\enquote {\bibinfo {title} {Photodissociation dynamics
  of the triiodide anion ({I}$_3^-$)},}\ }\href {\doibase 10.1063/1.482040}
  {\bibfield  {journal} {\bibinfo  {journal} {The Journal of Chemical Physics}\
  }\textbf {\bibinfo {volume} {113}},\ \bibinfo {pages} {2255--2262} (\bibinfo
  {year} {2000})}\BibitemShut {NoStop}%
\bibitem [{\citenamefont {Burkholder}, \citenamefont {Cox},\ and\ \citenamefont
  {Ravishankara}(2015)}]{Burkholder2015}%
  \BibitemOpen
  \bibfield  {author} {\bibinfo {author} {\bibfnamefont {J.~B.}\ \bibnamefont
  {Burkholder}}, \bibinfo {author} {\bibfnamefont {R.~A.}\ \bibnamefont {Cox}},
  \ and\ \bibinfo {author} {\bibfnamefont {A.~R.}\ \bibnamefont
  {Ravishankara}},\ }\bibfield  {title} {\enquote {\bibinfo {title}
  {Atmospheric degradation of ozone depleting substances, their substitutes,
  and related species},}\ }\href {\doibase 10.1021/cr5006759} {\bibfield
  {journal} {\bibinfo  {journal} {Chemical Reviews}\ }\textbf {\bibinfo
  {volume} {115}},\ \bibinfo {pages} {3704--3759} (\bibinfo {year}
  {2015})}\BibitemShut {NoStop}%
\bibitem [{\citenamefont {Liu}(2020)}]{liu2020_these_At}%
  \BibitemOpen
  \bibfield  {author} {\bibinfo {author} {\bibfnamefont {L.}~\bibnamefont
  {Liu}},\ }\emph {\bibinfo {title} {{Exploration of astatine chemistry in
  solution : focus on the Pourbaix diagram in noncomplexing medium and
  characterization of astatine-mediated halogen bonds}}},\ \href
  {https://tel.archives-ouvertes.fr/tel-03123005} {\bibinfo {type} {Theses}},\
  \bibinfo  {school} {{Ecole nationale sup{\'e}rieure Mines-T{\'e}l{\'e}com
  Atlantique}} (\bibinfo {year} {2020})\BibitemShut {NoStop}%
\bibitem [{\citenamefont {Vaidyanathan}\ and\ \citenamefont
  {Zalutsky}(2008)}]{Vaidyanathan2008}%
  \BibitemOpen
  \bibfield  {author} {\bibinfo {author} {\bibfnamefont {G.}~\bibnamefont
  {Vaidyanathan}}\ and\ \bibinfo {author} {\bibfnamefont {M.}~\bibnamefont
  {Zalutsky}},\ }\bibfield  {title} {\enquote {\bibinfo {title} {Astatine
  radiopharmaceuticals: Prospects and problems},}\ }\href {\doibase
  10.2174/1874471010801030177} {\bibfield  {journal} {\bibinfo  {journal}
  {Current Radiopharmaceuticalse}\ }\textbf {\bibinfo {volume} {1}},\ \bibinfo
  {pages} {177--196} (\bibinfo {year} {2008})}\BibitemShut {NoStop}%
\bibitem [{Sci()}]{ScipyKDE}%
  \BibitemOpen
  \href@noop {} {\enquote {\bibinfo {title} {{Scipy} representation of a kernel
  density estimate using gaussian kernels},}\ }\bibinfo {howpublished}
  {\url{https://docs.scipy.org/doc/scipy/reference/generated/scipy.stats.gaussian_kde.html}},\
  \bibinfo {note} {accessed: 2023-07-01}\BibitemShut {NoStop}%
\bibitem [{\citenamefont {Liu}\ \emph {et~al.}(2007)\citenamefont {Liu},
  \citenamefont {Vico}, \citenamefont {Lindh},\ and\ \citenamefont
  {Fang}}]{Liu2007_CH2I2}%
  \BibitemOpen
  \bibfield  {author} {\bibinfo {author} {\bibfnamefont {Y.-J.}\ \bibnamefont
  {Liu}}, \bibinfo {author} {\bibfnamefont {L.~D.}\ \bibnamefont {Vico}},
  \bibinfo {author} {\bibfnamefont {R.}~\bibnamefont {Lindh}}, \ and\ \bibinfo
  {author} {\bibfnamefont {W.-H.}\ \bibnamefont {Fang}},\ }\bibfield  {title}
  {\enquote {\bibinfo {title} {Spin-orbit ab initio investigation of the
  ultraviolet photolysis of diiodomethane},}\ }\href {\doibase
  10.1002/cphc.200600737} {\bibfield  {journal} {\bibinfo  {journal}
  {{ChemPhysChem}}\ }\textbf {\bibinfo {volume} {8}},\ \bibinfo {pages}
  {890--898} (\bibinfo {year} {2007})}\BibitemShut {NoStop}%
\bibitem [{\citenamefont {Tecmer}\ \emph {et~al.}(2011)\citenamefont {Tecmer},
  \citenamefont {Gomes}, \citenamefont {Ekstr\"{o}m},\ and\ \citenamefont
  {Visscher}}]{Tecmer2011}%
  \BibitemOpen
  \bibfield  {author} {\bibinfo {author} {\bibfnamefont {P.}~\bibnamefont
  {Tecmer}}, \bibinfo {author} {\bibfnamefont {A.~S.~P.}\ \bibnamefont
  {Gomes}}, \bibinfo {author} {\bibfnamefont {U.}~\bibnamefont {Ekstr\"{o}m}},
  \ and\ \bibinfo {author} {\bibfnamefont {L.}~\bibnamefont {Visscher}},\
  }\bibfield  {title} {\enquote {\bibinfo {title} {Electronic spectroscopy of
  {UO}${}_2^{2+}$, {NUO}${}^{+}$ and {NUN} : an evaluation of time-dependent
  density functional theory for actinides},}\ }\href {\doibase
  10.1039/c0cp02534h} {\bibfield  {journal} {\bibinfo  {journal} {Physical
  Chemistry Chemical Physics}\ }\textbf {\bibinfo {volume} {13}},\ \bibinfo
  {pages} {6249} (\bibinfo {year} {2011})}\BibitemShut {NoStop}%
\bibitem [{\citenamefont {Tecmer}\ \emph {et~al.}(2014)\citenamefont {Tecmer},
  \citenamefont {Gomes}, \citenamefont {Knecht},\ and\ \citenamefont
  {Visscher}}]{Tecmer2014}%
  \BibitemOpen
  \bibfield  {author} {\bibinfo {author} {\bibfnamefont {P.}~\bibnamefont
  {Tecmer}}, \bibinfo {author} {\bibfnamefont {A.~S.~P.}\ \bibnamefont
  {Gomes}}, \bibinfo {author} {\bibfnamefont {S.}~\bibnamefont {Knecht}}, \
  and\ \bibinfo {author} {\bibfnamefont {L.}~\bibnamefont {Visscher}},\
  }\bibfield  {title} {\enquote {\bibinfo {title} {Communication: Relativistic
  fock-space coupled cluster study of small building blocks of larger uranium
  complexes},}\ }\href {\doibase 10.1063/1.4891801} {\bibfield  {journal}
  {\bibinfo  {journal} {The Journal of Chemical Physics}\ }\textbf {\bibinfo
  {volume} {141}},\ \bibinfo {pages} {041107} (\bibinfo {year}
  {2014})}\BibitemShut {NoStop}%
\end{thebibliography}%

\end{document}


\title{Supplementary information: The performance of approximate equation of motion coupled cluster for valence and core states of heavy element systems}

\author{Loic Halbert}
\email{loic.halbert@univ-
	lille.fr}
\author{André Severo Pereira Gomes}%
 \email{andre.gomes@univ-
 	lille.fr}
\affiliation{ 
Université de Lille, CNRS, UMR 8523—PhLAM—Physique des Lasers 
Atomes et Molécules,
F-59000 Lille, France
}%

\maketitle


\begin{widetext}

\section{Supplementary information}

\subsection{Introduction}

Here again for table :
\begin{itemize}
\item \pee; \pip; \pea : partitioned EOM-\{EE; IP; EA\}
\item \MBPT : Many Body Perturbation Theory(2)
\item \pMBPT : partitioned  Many Body Perturbation Theory(2)
\end{itemize}

Finally, remember that : for the k\upfr{th} solution, with $\left (\rs {\mathbf {}} i \right)_k $ of the determinant 'single ionization' $ \left| \Phi_i^{\mathbf 	{}} \right \rangle $ , we then define~:~$\%SI=100 \left(\left (\rs {\mathbf {}} i \right)_k\right)^{2}$. Likewise, we define : $\%SA=100 \left(\left (r^a \right)_k\right)^{2}$ and $\%SE=100 \left(\left (r^a_i \right)_k\right)^{2}$.

\subsection{\ch{HCl}-\ch{HBr} in CVS}

For HCl and HBr  : as the full diagonalisation was available for these compunds, the comparaison is made, first, CVS w.r.t. the full diagonalisation. Then, \emph{full diagonalisation-Approximated methods} w.r.t. full diagonalisation is shown.

\begin{table}[H]
	\fontsize{8}{6}\selectfont
	\centering
\caption{  Energy 'E' (in \SI{}{\electronvolt}) and the $\%SI=100 \left(\left (\rs {\mathbf {}} i \right)_k\right)^{2}$ of the selected  
	solution for \ch{HCl}. $ \Delta_f $ is the difference w.r.t. \fullip{} and $ \Delta_{\ip} 
	$ is 
	the difference \IPA{}-\ip{}.  'mean' and 'std' 
	for 
	the L 
	edge, respectively : the mean  and the standard deviation of the 
	difference 
	in energie.  
	 For experimental data precision of $ \pm $ 	\SI{0.5}{\electronvolt}.
 }
\begin{tabular}{l rr rr rrr rrr rrr rr rr rr r}
\toprule
& \multicolumn{2}{c}{\fullip} &\multicolumn{2}{c}{\ip}
		&\multicolumn{3}{c}{\pip}&\multicolumn{3}{c}{\MBPT} &
		\multicolumn{3}{c}{\pMBPT} & \multicolumn{2}{c}{full-\pip}&\multicolumn{2}{c}{full-\MBPT} &
		\multicolumn{2}{c}{full-\pMBPT}  & exp\\
  \cmidrule(l{3pt}r{3pt}){2-3}   \cmidrule(l{3pt}r{3pt}){4-5} \cmidrule(l{3pt}r{3pt}){6-8} \cmidrule(l{3pt}r{3pt}){9-11} \cmidrule(l{3pt}r{3pt}){12-14}  \cmidrule(l{3pt}r{3pt}){15-16} \cmidrule(l{3pt}r{3pt}){17-18} \cmidrule(l{3pt}r{3pt}){19-20}
 T &     E$_f$ &  \%SI &   $\Delta_f$ &   \%SI & $\Delta_f$ &  $\Delta_{IP}$  &  \%SI & $\Delta_f $  &  $\Delta_{IP}$  &   \%SI &  $\Delta_f$ & $\Delta_{IP}$  &   \%SI & $\Delta_f$ &   \%SI & $\Delta_f$ &   \%SI & $\Delta_f$ &   \%SI  & \\
\cmidrule{1-20}
$K$ & 2835.464 &   86 &  0.300 & 88 & -0.153 &  -0.453 &  88 &   0.524 &    0.225 &     88 &    0.096 &    -0.203 &      88 &  -0.137 &   88 &    0.223 &    87 &     0.112 &     88  &\\
\cmidrule{1-20}
$L_1$ &  280.254 &   12 &  0.018 & 89 & -0.123 &  -0.140 &  88 &   0.136 &    0.118 &     89 &    0.010 &    -0.008 &      88 &  -0.216 &   19 &    0.007 &    14 &    -0.164 &     34 & \\
$L_2$  &  209.661 &   87 & -0.074 & 88 & -0.132 &  -0.057 &  87 &   0.055 &    0.130 &     88 &    0.012 &     0.086 &      88 &   0.047 &   78 &    0.125 &    87 &     0.172 &     80 & 209.01\upfr{\cite{Aitken1980}} ; 208.6\upfr{\cite{Hayes1972}} \\
$L_3(a)$  &  208.016 &   87 & -0.088 & 88 & -0.148 &  -0.059 &  87 &   0.042 &    0.131 &     88 &   -0.003 &     0.085 &      88 &  -0.257 &    1 &    0.126 &    87 &    -0.146 &     2 & 207.1 \upfr{\cite{Hayes1972}}\\
 $L_3(b)$ &  207.936 &   86 & -0.100 & 88 & -0.162 &  -0.063 &  88 &   0.027 &    0.127 &     88 &   -0.021 &     0.078 &      88 &  -0.417 &   65 &    0.119 &    87 &    -0.314 &     61 & 207.38 \upfr{\cite{Aitken1980}} \\
 \cmidrule{1-20}
 mean(L) & & & -0.0612 & & -0.1412 & -0.800 & & 0.065 & 0.127 & & -0.001 &  0.061 & & -0.211 & & 0.094 & & -0.113 &\\
 std (L) & & & 0.0465  & &  0.0152 & 0.035 & & 0.042 & 0.005 & & 0.013 
 & 0.040 & & 0.149 & & 0.045 & & 0.158 & \\
\bottomrule
\end{tabular}
	\label{tab:HCl_All}
\end{table}

\begin{figure}[H]
	\noindent
	\caption{Energies (in \SI{}{\electronvolt}) for \ch{HCl} - Comparison between CVS \IPA s and \fullip{} for the K 
	and L edges for the 2 upper graph, for the two second it is the comparison between \IPA s full diagonalized and
	\fullip{}, all values in \SI{}{\electronvolt}.  The points have a size proportional to 
	$\%SI=100 \left(\left (\rs {\mathbf {}} i \right)_k\right)^{2}$, to compare, in the first block, the green circle 
	corresponds to a \%SI of 100\%. Experimental data $ 
	{}^{\dagger } $~\cite{Hayes1972}; $ {}^{\dagger 
	\dagger} $~\cite{Aitken1980} are given with error bars.}
   \begin{center}
	\includegraphics[width=1\linewidth]%
	{p_EOM_figure/%
		HCl_dav3z_x2cmmf_CVS_vs_fullIP_eV_N.png}
	\includegraphics[width=1\linewidth]%
	{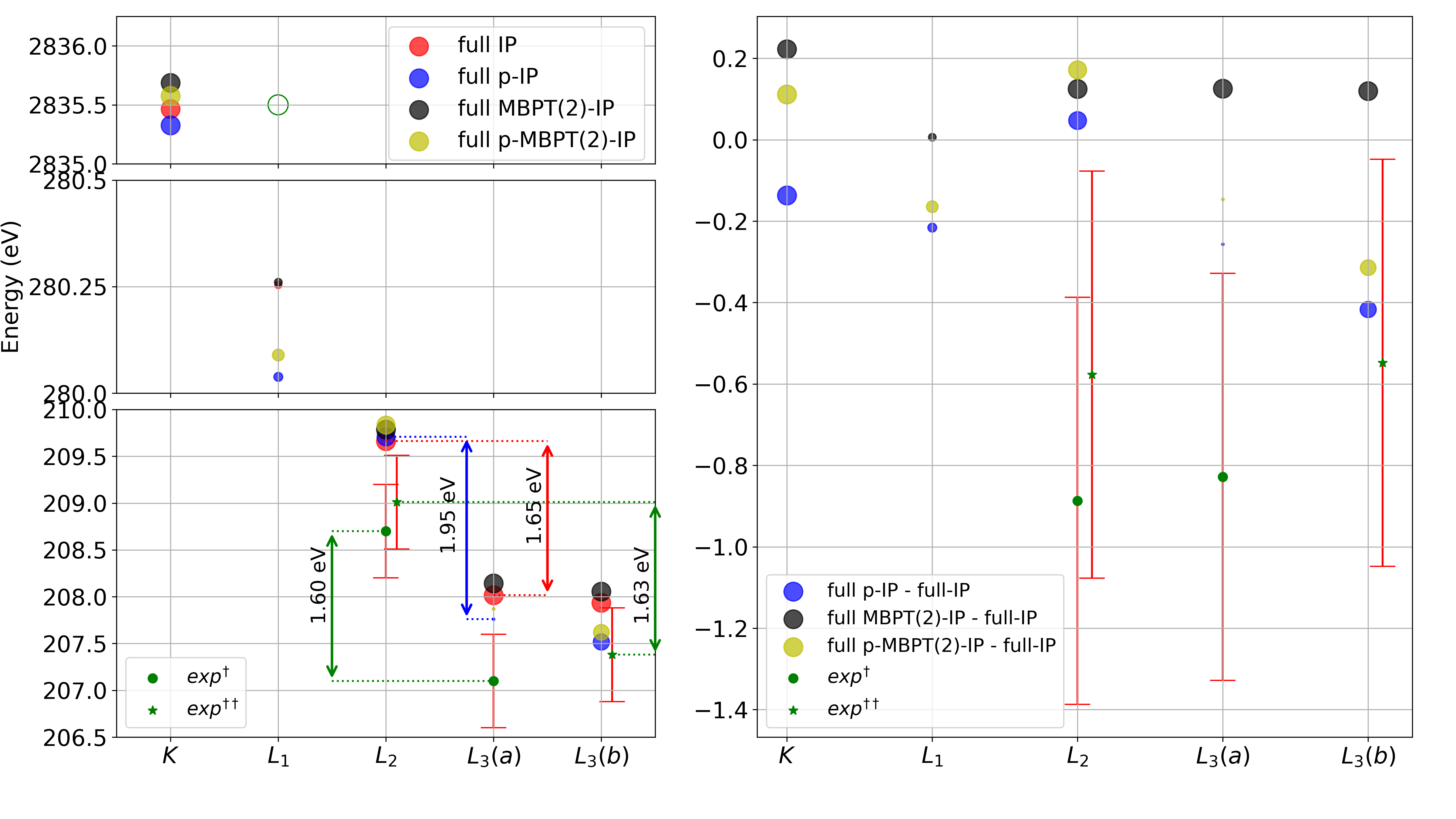}
   \end{center}
	\label{fig:HCl_all}
\end{figure}

\begin{table}[H]
	\fontsize{8}{6}\selectfont
	\centering
	\caption{ Energy 'E' (in \SI{}{\electronvolt}) and the $\%SI=100 \left(\left (\rs {\mathbf {}} i \right)_k\right)^{2}$ of the selected 
	solution for \ch{HBr}. $ \Delta_f $ is the difference w.r.t. \fullip{} and  $ \Delta_{\ip} 
	$ is 
	the difference \IPA{}-\ip{} - 'mean' and 'std' 
	for 
	the M
	edge, respectively : the mean and the standard deviation of the 
	difference 
	in energie . For experimental data precision of $ \pm $ 
	\SI{0.5}{\electronvolt}.}
	\begin{tabular}{l rr rr rrr rrr rrr rr rr rr r}
\toprule
& \multicolumn{2}{c}{\fullip} &\multicolumn{2}{c}{\ip}
		&\multicolumn{3}{c}{\pip}&\multicolumn{3}{c}{\MBPT} &
		\multicolumn{3}{c}{\pMBPT} & \multicolumn{2}{c}{full-\pip}&\multicolumn{2}{c}{full-\MBPT} &
		\multicolumn{2}{c}{full-\pMBPT}  & exp${}^{\dagger}$\\
  \cmidrule(l{3pt}r{3pt}){2-3}   \cmidrule(l{3pt}r{3pt}){4-5} \cmidrule(l{3pt}r{3pt}){6-8} \cmidrule(l{3pt}r{3pt}){9-11} \cmidrule(l{3pt}r{3pt}){12-14}  \cmidrule(l{3pt}r{3pt}){15-16} \cmidrule(l{3pt}r{3pt}){17-18} \cmidrule(l{3pt}r{3pt}){19-20}
 T &     E$_f$ &  \%SI &   $\Delta_f$ &   \%SI & $\Delta_f$ &  $\Delta_{IP}$  &  \%SI & $\Delta_f $  &  $\Delta_{IP}$  &   \%SI &  $\Delta_f$ & $\Delta_{IP}$  &   \%SI & $\Delta_f$ &   \%SI & $\Delta_f$ &   \%SI & $\Delta_f$ &   \%SI  & \\
\cmidrule{1-20}
 $M_{2}$ & 200.765 &   36 &  0.156 & 90 & -0.997 &  -1.153 &  89 &   0.301 &    0.145 &     90 &   -0.695 &    -0.851 &  90 &  -0.830 &   48 &    0.058 &    40 &  -0.668 &  33 &\\
  $M_{3}(a)$ & 194.210 &   50 & -0.213 & 90 & -1.337 &  -1.125 &  89 &  -0.067 &    0.145 &     91 &   -1.039 &    -0.827 & 90 &  -0.873 &   34 &    0.034 &    55 &    -0.691 &   68 &\\
  $M_{3}(b)$ & 194.151 &   33 & -0.307 & 90 & -1.439 &  -1.131 &  89 &  -0.169 &    0.138 &     91 &   -1.147 &    -0.839 &  90 &  -1.141 &   58 &   -0.019 &    39 &   -0.851 &   67 &\\
 \cmidrule{1-20}
 M edge $ 3p $ &&&&&&&&&&&&&&&&&&&& \\
 mean  &  &    & -0.121 & & -1.258 &  -1.136 &   & 0.022 &  0.143 &  &   -0.960 &    -0.839 &   &  -0.948 &    &    0.025 &  &    -0.737 &    &\\
 std &   &    &  0.200 &  &  0.189 &   0.012 &  &   0.202 &    0.003 &      &    0.193 &     0.010 &    &   0.138 &  &    0.032 &  &     0.082 &   &\\
\cmidrule{1-20}
$M_{4}(a)$  &  78.033 &   50 &  0.099 & 90 & -1.046 &  -1.145 &  89 &   0.331 &    0.232 &     91 &   -0.684 &    -0.783 &  89 &  -1.027 &  86 &  0.191 & 38 &    -0.675 &  87 & 78.5\upfr{\cite{Hayes1972}} \\
$M_{5}(a)$ &  76.995 &   35 &  0.059 & 90 & -1.070 &  -1.129 &  89 &   0.289 &    0.230 &     91 &   -0.711 &  -0.770 & 89 &  -1.108 &   61 &    0.327 & 53 &    -0.613 &  47 & 77.4\upfr{\cite{Hayes1972}} \\
$M_{4}(b)$ &  77.839 &   67 &  0.058 & 90 & -1.097 &  -1.155 &  89 &   0.285 &    0.227 &     91 &   -0.739 &    -0.796 &  89 &  -1.215 &   48 &    0.587 &    24 &  -0.724 &  80 &
78.0\upfr{\cite{Hayes1972}} \\
$M_{5}(b)$ &  77.008 &   68 & -0.081 & 90 & -1.214 &  -1.134 &  89 &   0.147 &    0.228 &     91 &   -0.858 &    -0.777 &  89 &  -1.193 &   69 &    0.223 & 57 &    -0.845 &     80 &77.3\upfr{\cite{Hayes1972}} \\
$M_{5}(c)$ &  76.876 &   58 & -0.134 & 90 & -1.277 &  -1.142 &  89 &   0.090 &    0.224 &     91 &   -0.923 &    -0.789 & 89 &  -1.306 & 80 &    0.141 &  73 &    -1.013 & 57 & 77.1\upfr{\cite{Hayes1972}}; 77.36\upfr{\cite{OdlingSmee1999}} \\
\cmidrule{1-20}
M edge $ 3d $ &&&&&&&&&&&&&&&&&&&& \\
mean &  &    & <1e-4 &  & -1.141 &  -1.141 &   &   0.228 &    0.228 &      &   -0.783 &    -0.783 &  &  -1.170 &  &    0.294 & &    -0.774 & &\\
std &   &    &  0.091 &  &  0.089 &   0.009 &  &   0.093 &    0.003 &      &    0.092 &     0.009 &  &   0.095 &  &    0.159 &  &     0.142 &  &\\
\bottomrule
\end{tabular}
	\label{tab:HBr_All}
\end{table}

	\noindent
\begin{figure}[H]
\begin{center}
	\noindent
	\caption{Energies (in \SI{}{\electronvolt}) for \ch{HBr} - Comparison between CVS \IPA s and \fullip{} for the M  edge for the 2 upper graph, for the two second it is the comparison between \IPA s full diagonalized and 
	\fullip{}, all values in \SI{}{\electronvolt}.  The points have a size proportional to 
	$\%SI=100 \left(\left (\rs {\mathbf {}} i \right)_k\right)^{2}$, to compare, in the first block, the green circle 
	corresponds to a \%SI of 100\%. Experimental data  ${}^{\dagger } $~\cite{Hayes1972}; $ {}^{\dagger \dagger} $~\cite{OdlingSmee1999} are given with error bars.}
	\includegraphics[width=0.8\linewidth]%
	{p_EOM_figure/%
		HBr_dav3z_x2cmmf_CVS_vs_fullIP_3p3d_N.png}
	\includegraphics[width=0.8\linewidth]%
	{p_EOM_figure/%
	HBr_dav3z_x2cmmf_full_vs_fullIP_eV_N.png}
	\label{fig:HBr_CVS_full_vs_full_ip}
 \end{center}
\end{figure}


\subsection{Valence Ionization Potential for \XOm{X} ($\text{X}\in \left[\ch{Cl};\ \ch{Br};\ \ch{I};\ \ch{At};\ \ch{Ts}\right]$)}


\begin{table}[H]
	\fontsize{8}{6}\selectfont
	\centering
	\captionsetup{font=small}
	\caption{ Four first IP for \XOm{X} ($\text{X}\in \left[\ch{Cl};\ \ch{Br};\ \ch{I};\ \ch{At};\ \ch{Ts}\right]$) - Energy 'E' (in \SI{}{\electronvolt}) and the $\%SI=100 \left(\left (\rs {\mathbf {}} i \right)_k\right)^{2}$ of the selected 
		solution. $ \Delta_{\ip} $ is the difference \IPA-IP and $ \Delta_{exp} 
		$ is the difference w.r.t. experimental.}
	\begin{tabular}{c cr rr rr rr rr}
		\toprule
	&	&	& \multicolumn{2}{c}{\ip} 
		&\multicolumn{2}{c}{\pip}&\multicolumn{2}{c}{\MBPT} &
		\multicolumn{2}{c}{\pMBPT}\\
        \cmidrule(l{3pt}r{3pt}){4-5}  \cmidrule(l{3pt}r{3pt}){6-7}  \cmidrule(l{3pt}r{3pt}){8-9} \cmidrule(l{3pt}r{3pt}){10-11}
	&	T& &	E &    \%SI &    \%SI &  $\Delta_{\ip}  $ &   \%SI
		&  $\Delta_{\ip}   $ & \%SI &  $\Delta_{\ip}   $ \\
\midrule
\multirow{6}{*}{\XOm{Cl}} & X$_{3/2}$ & &2.040 & 91 & 86 &      -0.524 & 92 &        0.062 & 88 &        
-0.359 \\
&X$_{1/2}$ & &2.083 & 91 & 86 &      -0.530 & 92 &        0.063 & 88 &        
-0.363 \\
&A$_{3/2}$ & &6.855 & 89 & 85 &      -0.292 & 90 &       -0.057 & 87 &        
-0.295 \\
&A$_{1/2}$ & &6.914 & 87 & 82 &      -0.291 & 88 &       -0.058 & 85 &        
-0.294 \\
		\cmidrule{2-11}
&		{} &&$\Delta_{exp}  $& &    &$\Delta_{exp}  $ 
		&& 
		$\Delta_{exp}  
		$&&$\Delta_{exp}  $\\
exp\upfr{\cite{Gilles1992}} &		X$_{1/2}$ & 2.276 &  -0.1933 &    & &  -0.7231  & & -0.1299 & 
		&-0.5563\\
		\midrule
\multirow{6}{*}{\XOm{Br}} &  X$_{3/2}$& & 2.149 & 88 & 82 &      -0.549 & 90 &        0.105 & 85 &        
-0.330 \\
& X$_{1/2}$& & 2.280 & 88 & 82 &      -0.577 & 90 &        0.112 & 85 &        
-0.348 \\
& A$_{3/2}$ && 6.126 & 87 & 81 &      -0.329 & 88 &       -0.007 & 83 &        
-0.286 \\
& A$_{1/2}$ && 6.314 & 64 & 53 &      -0.327 & 68 &       -0.007 & 61 &        
-0.276 \\
		\cmidrule{2-11}
&		{} &&$\Delta_{exp}  $& &    
		&$\Delta_{exp}  $ 
		&& $\Delta_{exp}  
		$&&$\Delta_{exp}  $\\
exp\upfr{\cite{Gilles1992}}&		X$_{1/2}$  & 2.353 &  -0.0732 &    & &  -0.6502 & & 0.0387 & 
		&-0.4213\\
		\midrule
\multirow{6}{*}{\XOm{I}} & X$_{3/2}$ && 2.238 & 86 & 76 &      -0.547 & 89 &        0.114 & 81 &        
-0.300 \\
&X$_{1/2}$ && 2.518 & 85 & 75 &      -0.630 & 87 &        0.137 & 79 &        
-0.352 \\
&A$_{3/2}$ && 5.455 & 84 & 74 &      -0.399 & 86 &        0.038 & 78 &        
-0.313 \\
&A$_{1/2}$ && 5.705 & 49 & 55 &      -0.402 & 43 &        0.053 & 47 &        
-0.274 \\
		\cmidrule{2-11}
&		{} &&$\Delta_{exp}  $& &    
		&$\Delta_{exp}  $ 
		&& $\Delta_{exp}
		$&&$\Delta_{exp} $\\
exp\upfr{\cite{Gilles1992}} &	X$_{1/2}$  & 2.378 &  0.1398 &    & &  -0.4905 & &0.2766& &-0.2121\\
		\midrule
\multirow{6}{*}{\XOm{At}} &X$_{3/2}$ & &1.888 & 84 & 70 &      -0.604 & 87 &        0.255 
& 76 &        -0.220 \\
&X$_{1/2}$ & &2.620 & 71 & 59 &      -0.874 & 72 &        0.310 
& 62 &        -0.406 \\
&A$_{3/2}$ && 4.672 & 81 & 67 &      -0.552 & 83 &        0.208 
& 72 &        -0.294 \\
&A$_{1/2}$ && 4.943 & 68 & 61 &      -0.546 & 68 &        0.295 
& 62 &        -0.169 \\
\midrule
\multirow{6}{*}{\XOm{Ts}} & X$_{3/2}$ && 1.292 & 83 & 57 &      -0.741 & 87 &        0.258 & 68 &        
-0.259 \\
& X$_{1/2}$ && 2.501 & 88 & 81 &      -1.379 & 89 &        0.397 & 84 &        
-0.713 \\
&A$_{3/2}$ && 3.525 & 78 & 52 &      -0.764 & 82 &        0.263 & 63 &        
-0.465 \\
& A$_{1/2}$ && 3.582 & 90 & 84 &      -0.759 & 91 &        0.414 & 86 &        
-0.232 \\
\bottomrule
	\end{tabular}
	\label{tab:XOm1_Ene_SI_Diff_IP_NEW}
\end{table}

\subsection{Potential Energy Curves - \XOm{X} with $\text{X}\in \left[\ch{Cl};\ \ch{Br};\ \ch{I};\ \ch{At}\right]$}


With the curves, we studied position of the minima, the difference w.r.t. EOM-IP and the energies of transition bettween levels.

\subsubsection{Graphics}

\begin{figure}[H]

	\caption{\emph{Potential Energy Curves} for all the compounds with $\text{X}\in \left[\ch{Cl};\ \ch{Br};\ \ch{I};\ \ch{At}\right]$ and all approximated methods \IPA. The energy 
	: E in \SI{}{\electronvolt} and the distance X-O : r in $ \mathring{A} $ 	
	\label{fig:Supplementary_PES_ClOm1}. The overlay graphic is the zoom of the red rectangle.}
	\begin{center}
\begin{minipage}{1\textwidth}
	\includegraphics[width=0.45\textwidth]%
	{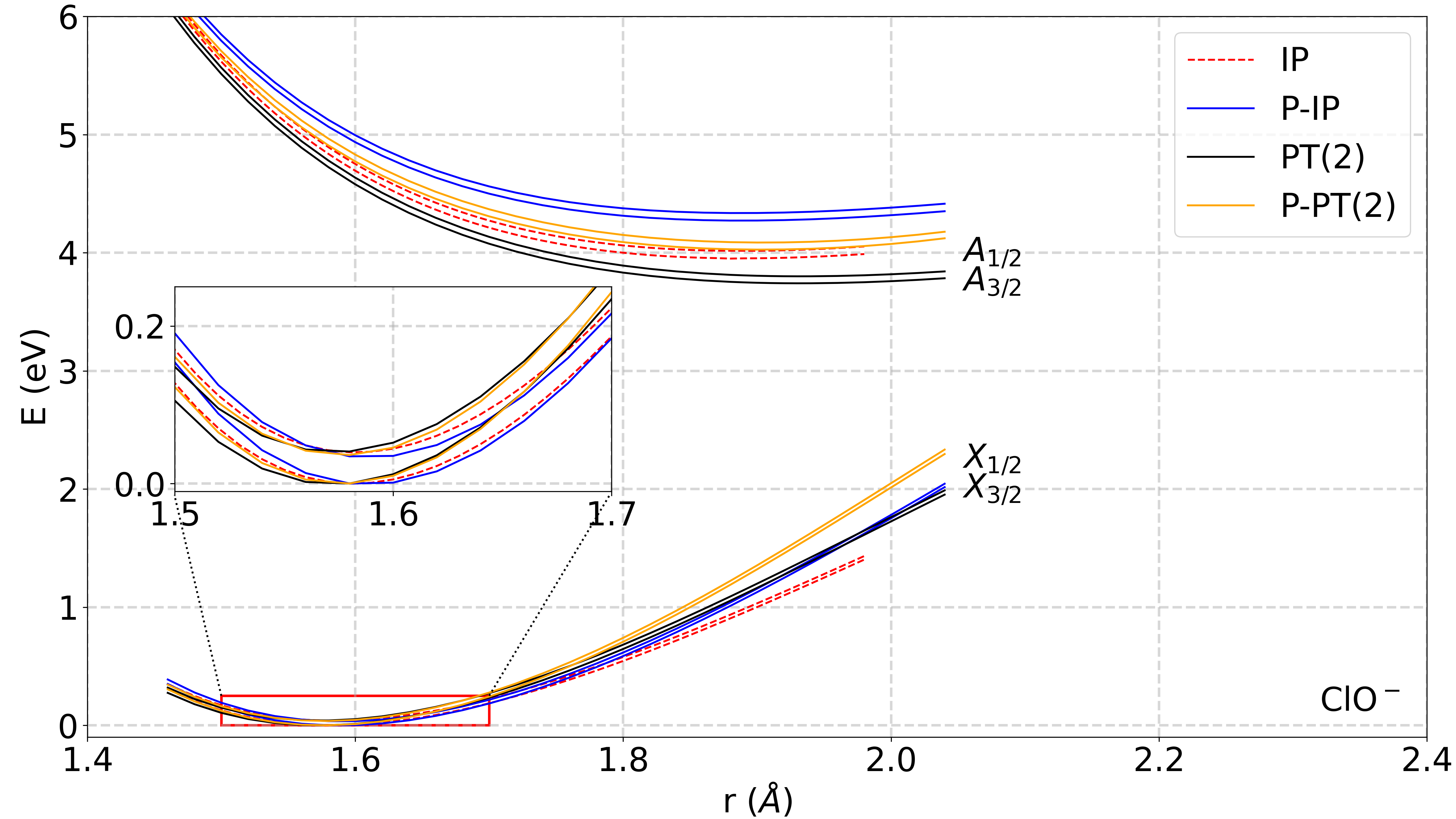}
	\includegraphics[width=0.45\textwidth]%
	{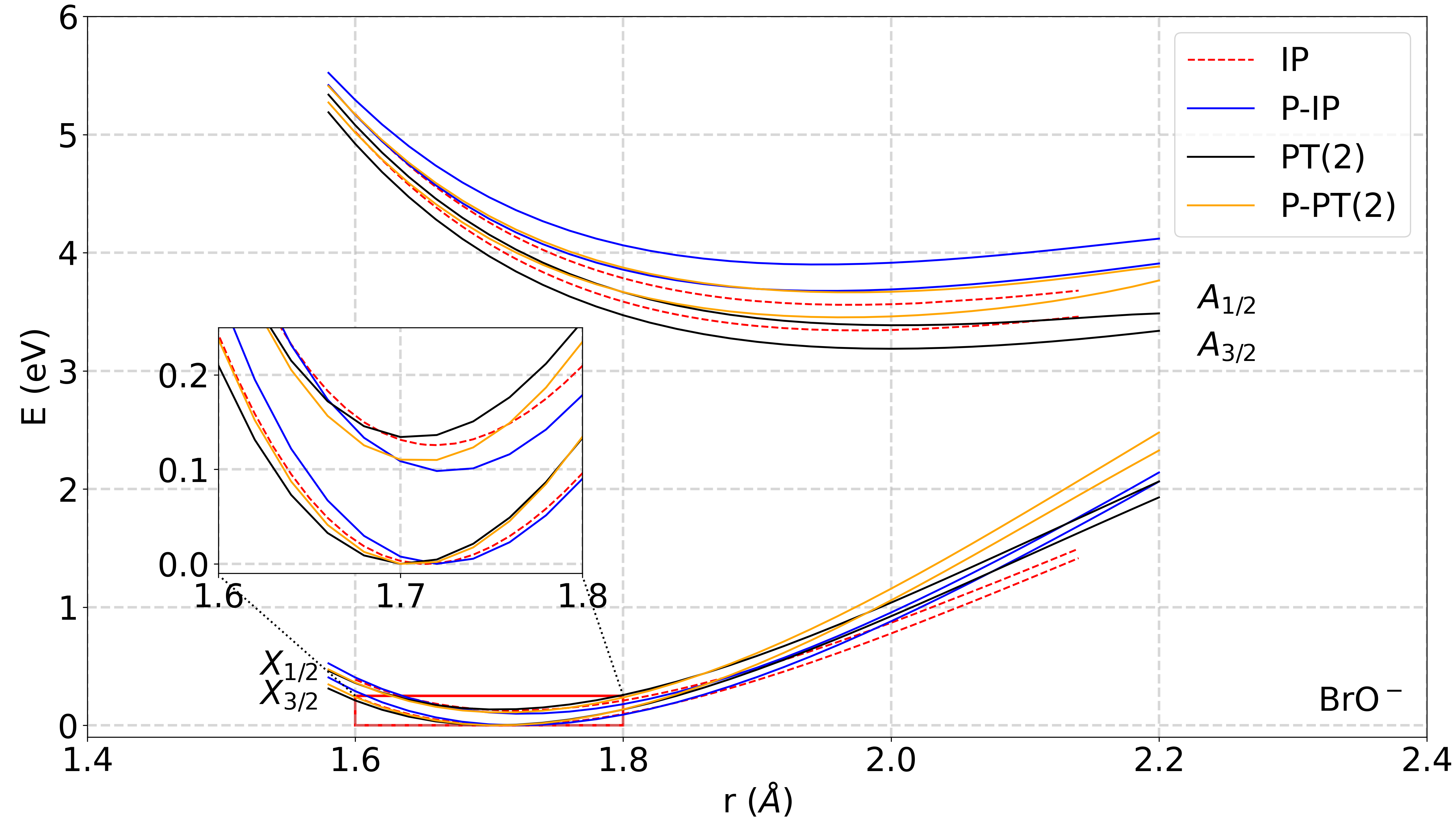}
\end{minipage}\\
\begin{minipage}{1\textwidth}
	\includegraphics[width=0.45\textwidth]%
	{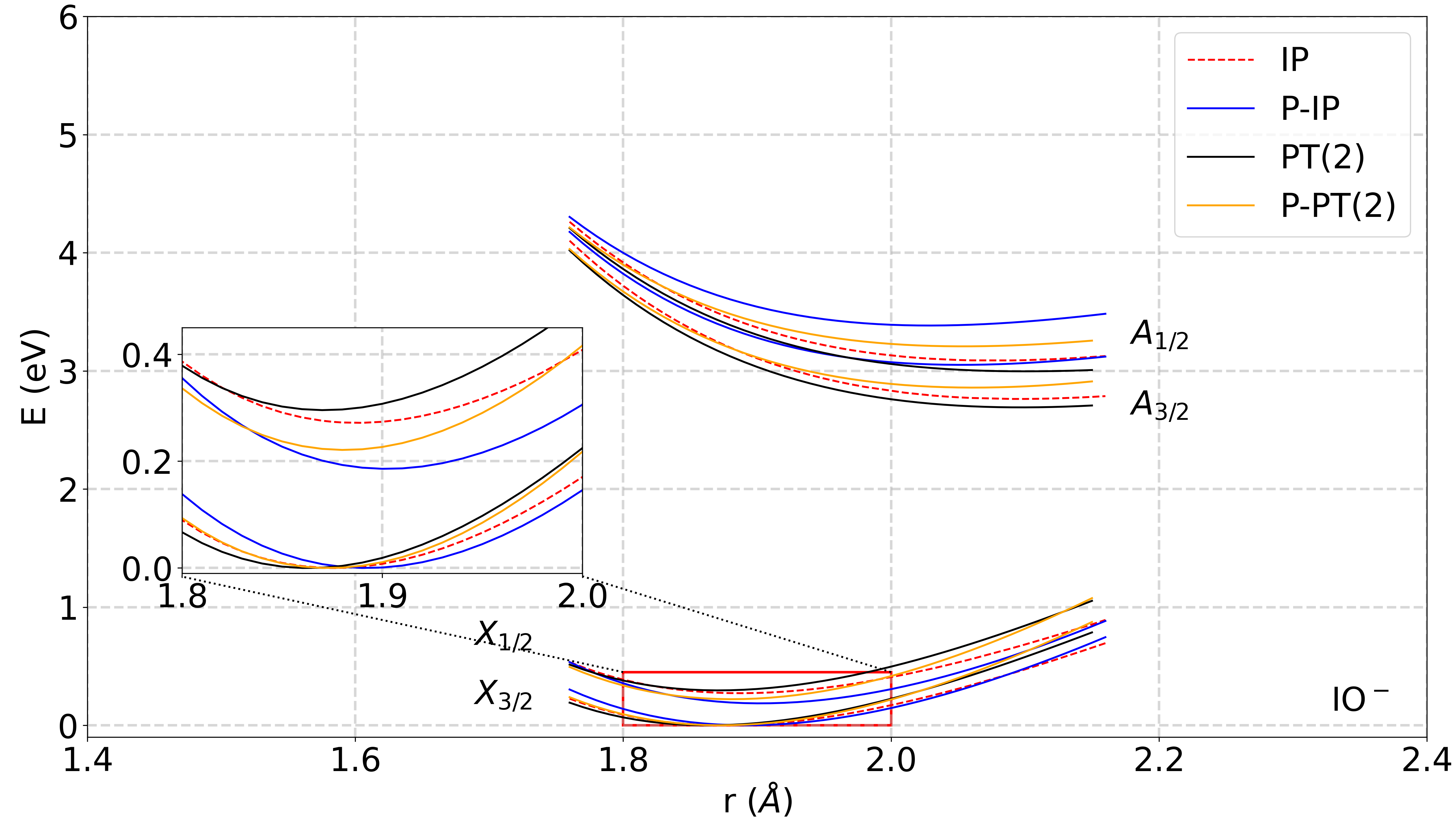}
	\includegraphics[width=0.45\textwidth]%
	{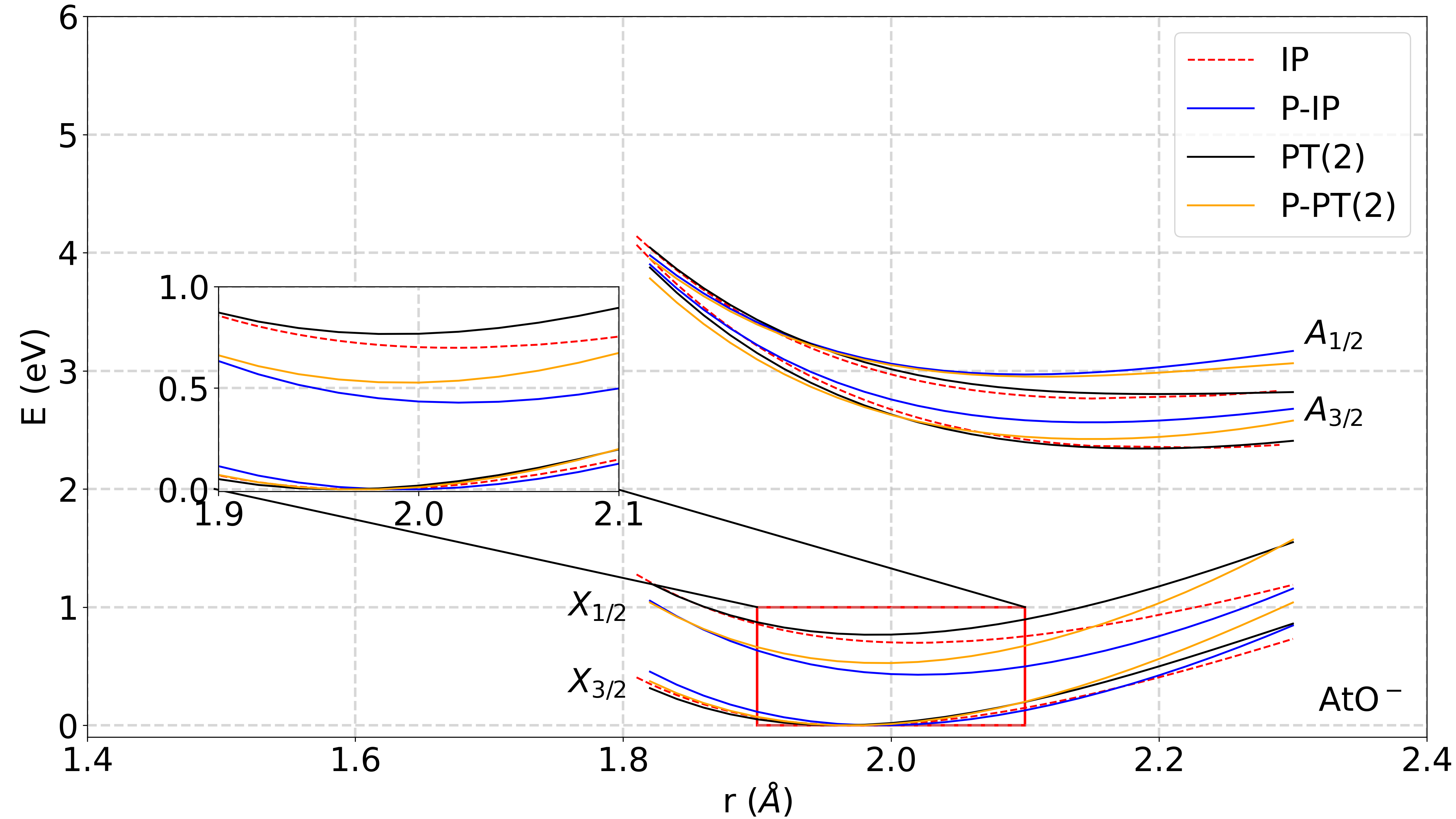}
\end{minipage}
\end{center}
\end{figure}

\subsubsection{Tables : Position of minima}

\begin{table}[H]
	\fontsize{8}{6}\selectfont
	\centering
	\caption{Minimum of the \emph{Potential Energy Curve} for \XOm{X} with $\text{X}\in \left[\ch{Cl};\ \ch{Br};\ \ch{I};\ \ch{At}\right]$.  
		E${}_{MO} $ is the absolute energie (in \SI{}{\electronvolt}), and \emph{r} the distance 
		\ch{X}-\ch{O} 
		of the minimum (in $ \mathring{A} $). $ \Delta_r $  is difference between 
		minimum lenght, and $\Delta_{E} $  is difference 
		between 
		minimum energy.}
	\begin{tabular}{crrrrrrrrr}
		\toprule
	&	& \multicolumn{2}{c}{\ip} 
		&\multicolumn{2}{c}{\pip}&\multicolumn{2}{c}{\MBPT} &
		\multicolumn{2}{c}{\pMBPT}\\
          \cmidrule(l{3pt}r{3pt}){3-4}  \cmidrule(l{3pt}r{3pt}){5-6}  \cmidrule(l{3pt}r{3pt}){7-8} \cmidrule(l{3pt}r{3pt}){9-10}
	&	T& 	r &    E$_{MO}$ &    $ \Delta_r $ &  $\Delta_{E}  $ &   $ 
		\Delta_r $ 
		&  $\Delta_{E}  $ & $ \Delta_{r} $ &  $\Delta_{E}  $ \\
  \midrule
	&	& 	 &  +14.58e3 &   &   &  
		&   & &   \\
	\multirow{4}{*}{\XOm{Cl}} &	\trenteetun & 1.5818 & -10.9061 &       0.0065 &      -0.5095 &         
		-0.0092 
		&          0.0465 &          -0.0054 &          -0.3601 \\
	&	\onze & 1.5828 & -10.8665 &       0.0063 &      -0.5148 &         
		-0.0094 
		&          0.0477 &          -0.0057 &          -0.3634 \\
	&	\trentedeux & 1.8889 &  -6.9541 &      -0.0006 &      -0.1888 &          
		0.0429 
		&         -0.1638 &           0.0145 &          -0.2859 \\
	&	\douze & 1.8875 &  -6.8909 &       0.0004 &      -0.1883 &          
		0.0451 
		&         -0.1678 &           0.0175 &          -0.2893 \\
		\midrule
  	&	& 	 &  +72.90e3 &   &   &  
		&   & &   \\
\multirow{4}{*}{\XOm{Br}} &	\trenteetun& 1.7141 & -7.2208 &       0.0073 &      -0.5370 &         -0.0112 
	&          
	0.0878 &          -0.0070 &          -0.3341 \\
& \onze & 1.7192 & -7.0952 &       0.0064 &      -0.5649 &         -0.0125 &          
	0.0958 &          -0.0088 &          -0.3505 \\
&\trentedeux& 1.9751 & -3.8787 &      -0.0234 &      -0.2027 &          0.0242 
&         
	-0.0673 &          -0.0108 &          -0.2226 \\
& \douze & 1.9692 & -3.6622 &      -0.0229 &      -0.1957 &          0.0360 &         
	-0.0848 &           0.0001 &          -0.2277 \\
	\midrule
 	&	& 	 &  +1.955e5 &   &   &  
		&   & &   \\
\multirow{4}{*}{\XOm{I}}	&	\trenteetun &   1.8758 & -26.7779 &       0.0164 &   -0.5414 &         
		-0.0124 &       0.0974 &          -0.0026 &       -0.3049 \\
&		\onze &   1.8874 & -26.5063 &       0.0149 &   -0.6274 &         
		-0.0165 &       0.1214 &          -0.0063 &       -0.3554 \\
&		\trentedeux &   2.0950 & -24.0159 &      -0.0427 &   -0.2527 
		&          
		0.0020 &       0.0263 &          -0.0334 &       -0.2087 \\
&		\douze &   2.0767 & -23.6909 &      -0.0484 &   -0.2448 &          
		0.0234 &       0.0064 &          -0.0227 &       -0.1845 \\
		\midrule
	&	& 	 &  +6.247e5 &  &   &   &   &  &   \\			
\multirow{4}{*}{\XOm{At}} &			\trenteetun &   1.9734 & -20.3352 &       0.0181 &   -0.5979 &         
		-0.0131 
		&       0.2357 &          -0.0037 &       -0.2263 \\
	&	\onze &   2.0180 & -19.6367 &       0.0031 &   -0.8677 &         
		-0.0298 
		&       0.3031 &          -0.0248 &       -0.3974 \\
	&	\trentedeux &   2.2090 & -17.9910 &      -0.0594 &   -0.3774 &         
		-0.0201 
		&       0.2337 &          -0.0597 &       -0.1468 \\
	&	\douze &   2.1645 & -17.5705 &      -0.0642 &   -0.3925 &          
		0.0339 
		&       0.2746 &          -0.0517 &       -0.0394 \\
		\bottomrule
	\end{tabular}
	\label{tab:PES_XO-_eV}
\end{table}

\subsubsection{Graphic representation of positions and energies minima errors in PES for \XOm{X}}

As reminder of the paper the figure (fig.\ref{fig:Error_wrt_ip_App}) is again reported here. Symbols represent the error in energy and in position of the minimum of the Potential Energy Curves ($\Delta E $ (in \SI{}{\electronvolt})  w.r.t. $ \Delta r $ (\AA)).

	\noindent
	\begin{figure*}[ht!]
		\noindent
			\captionsetup{font=small}
		\caption{Difference in the \emph{PES} minimums positions and energies - $ \Delta E $ (in \SI{}{\electronvolt}) w.r.t. $ \Delta r $ (\AA) 	for the 3 \IPA{} and the 4 compounds.  For reasons of readability : colors for compounds and symbols for states.  The 4 different isodensities represented were calculated by a Gaussian kde algorithm\cite{ScipyKDE}, resp. in [-0.1; 0.1]~($ \mathring{A} $), [-1.0; 0.4]~(\SI{}{\electronvolt}) limits for $ \Delta r $ and $ \Delta E $.}
  \begin{center}
		\includegraphics[width=1.0\textwidth]%
		{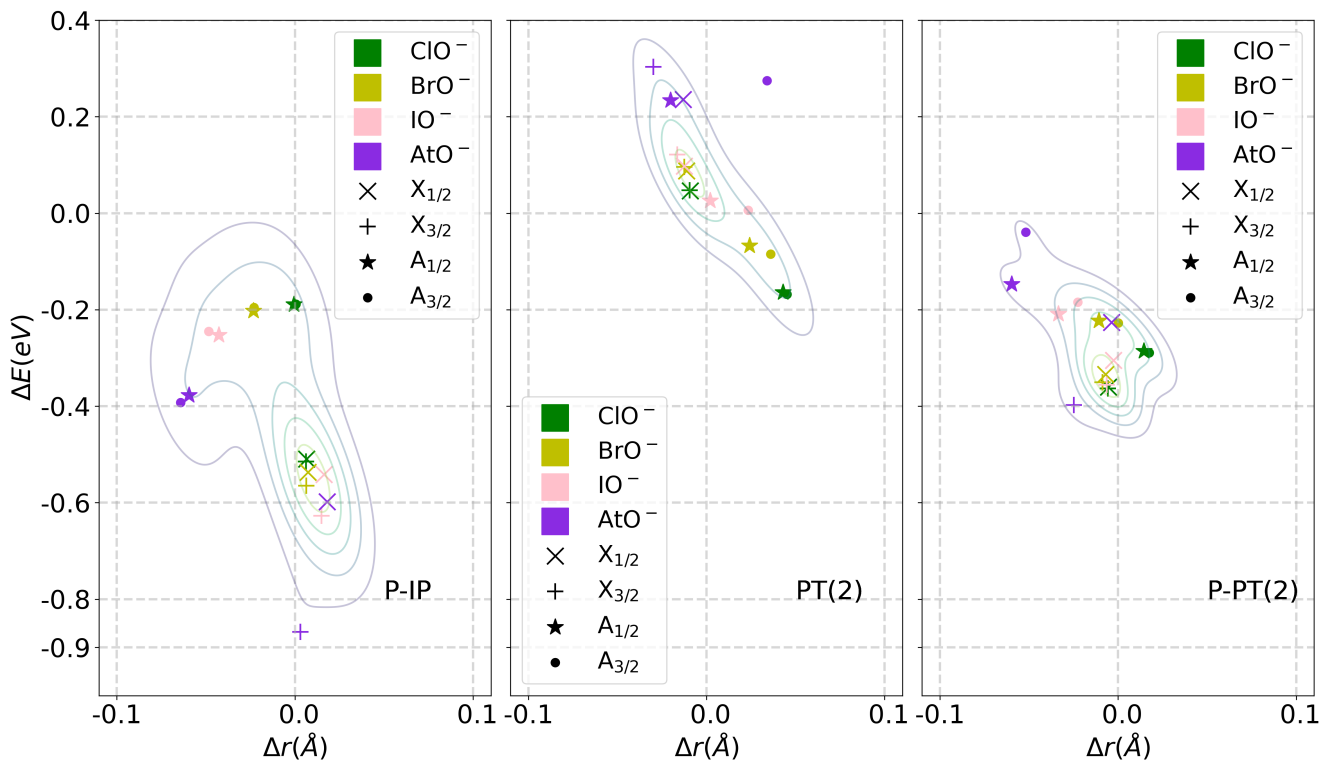}
		\label{fig:Error_wrt_ip_App}
  \end{center}
	\end{figure*}
 
	For the second graph (fig.\ref{fig:Error_wrt_ip_XO}), the information is obviously identical, but on reading we can easily see the increase in the error for \pip{} and the solution \onze{} as we move down the halogen column. We notice that the errors for \trentedeux{} and \douze{} are independent of the magnitude of the spin-orbit coupling.

	\noindent
	\begin{figure*}[!h]
		\noindent
			\captionsetup{font=small}
		\caption{Difference in the \emph{PES} minimums position and energies - $ \Delta E $ (in \SI{}{\electronvolt}) w.r.t. $ \Delta r $ (\AA)	for the 3 \IPA{} and the 4 compounds. For reasons of readability : colors for \IPA s and symbols for states.  The 4 different isodensities represented were calculated by a Gaussian kde algorithm\cite{ScipyKDE}, resp. in [-0.1; 0.1]~($ \mathring{A} $), [-1.0; 0.4]~(\SI{}{\electronvolt}) limits for $ \Delta r $ and $ \Delta E 	$.}
  \begin{center}
		\includegraphics[width=1.0\textwidth]%
		{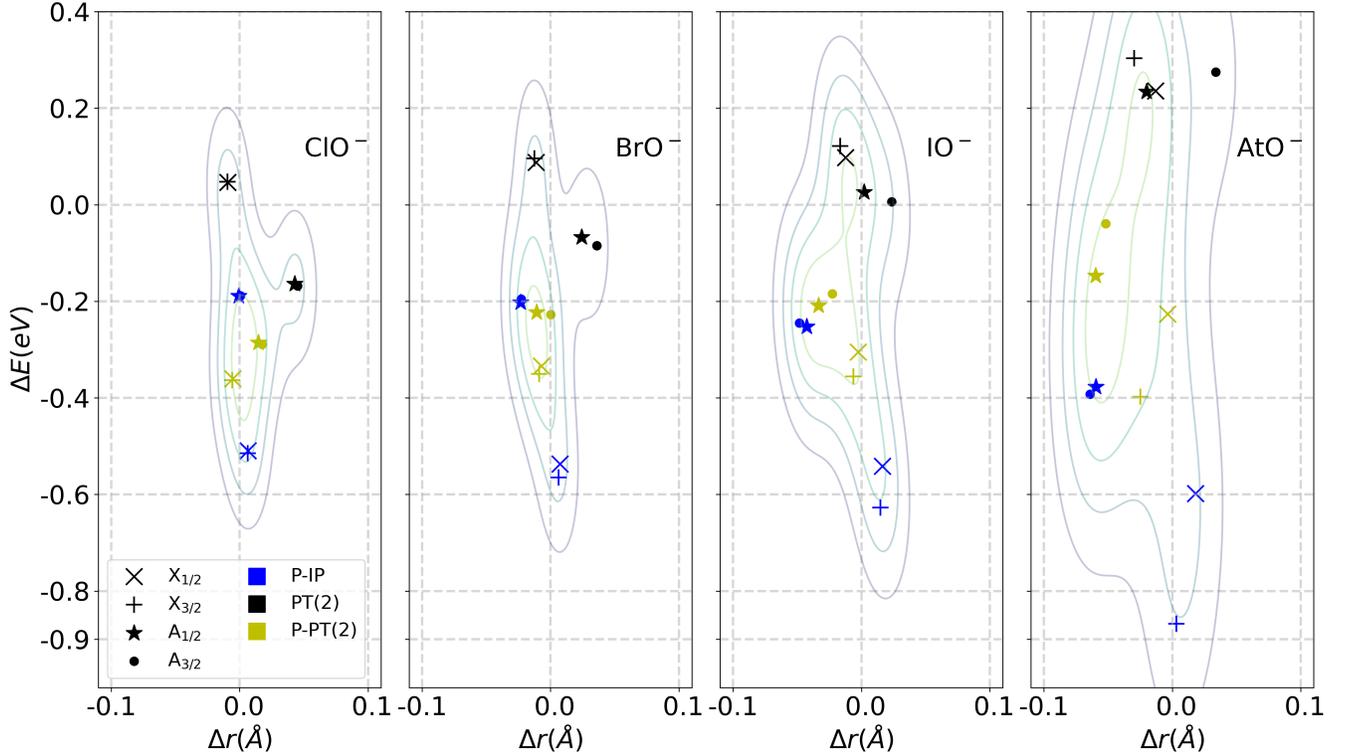}
\end{center}
		\label{fig:Error_wrt_ip_XO}
	\end{figure*}

\subsubsection{ Energies of transitions}

\begin{table}[H]
	\fontsize{8}{6}\selectfont
	\centering
	\caption{ Transition energies  $ \Delta_{E} $ (in \SI{}{\electronvolt}) for \XOm{X} with $\text{X}\in \left[\ch{Cl};\ \ch{Br};\ \ch{I};\ \ch{At}\right]$. and difference 
	with 
	$ \Delta_{\ip} $.}
	\begin{tabular}{c l r rrrr rrr}
		\toprule
& & & \multicolumn{4}{c}{$ \Delta_{E} $} &\multicolumn{3}{c}{$ \Delta_{E} - \Delta_{\ip}  $} \\
  \cmidrule(l{3pt}r{3pt}){4-7} \cmidrule(l{3pt}r{3pt}){8-10}
	&	$ \Delta_T $& &	$ \Delta_{\text{\ip}}$ &  $ 
		\Delta_{\text
			{\pip}}$   & $ \Delta_{\text{\MBPT}}$ 
		&  
		$ \Delta_{\text{\pMBPT}}$ &  $ \Delta_{\text{\pip}}  -  
		\Delta_{\text{\ip}}$  &  $ 
		\Delta_{\text{\MBPT}} - \Delta_{\text{\ip}} $ &  $ 
		\Delta_{\text{\pMBPT}} 
		- \Delta_{\text{\ip}}$\\
		\midrule
	\multirow{4}{*}{\XOm{Cl}}	&\trenteetun-\trenteetun &  &  0.0000 &     0.0000 &        0.0000 		&  	0.0000 &        
		0.0000 &             0.0000 &              0.0000 \\
	&	\onze-\trenteetun &  ~ & 0.0396 &     0.0343 &        0.0408 
		&        	0.0363 	&         
		-0.0054 &             0.0012 &             -0.0034 \\
	&	\trentedeux-\trenteetun & &   3.9520 &     4.2727 &        3.7418 
		&      	4.0262 &          
		0.3207 &            -0.2103 &              0.0741 \\
	&	\douze-\trenteetun &  &  4.0153 &     4.3364 &        3.8009 &         
		4.0860 	&       	0.3212 &            -0.2144 &              0.0708 \\
		\midrule
\multirow{4}{*}{\XOm{Br}}	&  \trenteetun -\trenteetun& &    0.0000 &     0.0000 &        0.0000 &         0.0000 
&          0.0000 &             0.0000 &              0.0000 \\
&\onze -\trenteetun & ~ &   0.1256 &     0.0978 &        0.1336 &         
0.1092 &         -0.0279 &             0.0080 &             -0.0164 \\
&\trentedeux -\trenteetun &   & 3.3421 &     3.6764 &        3.1871 &         3.4536 
&          0.3344 &            -0.1550 &              0.1115 \\
&\douze-\trenteetun  & &   3.5586 &     3.8999 &        3.3860 &         3.6650 
&          0.3414 &            -0.1726 &              0.1065 \\
		\midrule
\multirow{4}{*}{\XOm{I}}	&		\trenteetun-\trenteetun&&	0.0000 & 0.0000 &   0.0000 &    0.0000 		&   	0.0000 &   	0.0000 
		&        0.0000 \\
&		\onze-\trenteetun& ~ &	0.2717 & 0.1856 &   0.2956 &    0.2211 
		&  	-0.0860 	&   	0.0240 	&       -0.0505 \\
&		\trentedeux-\trenteetun& &	2.7620 & 3.0507 &   2.6909 &    2.8583 
		&   	0.2887 &     	-0.0711 	&        0.0962 \\
&		\douze-\trenteetun&	&3.0870 & 3.3835 &   2.9960 &    3.2074 &   
		0.2966 	&     	-0.0910 	&        0.1204 \\
		\midrule
  \multirow{4}{*}{\XOm{At}}	&			\trenteetun-\trenteetun &	& 0.0000 & 0.0000 &   0.0000 &    0.0000 	&   	0.0000 &   	0.0000 
		&        0.0000 \\
	&	\onze-\trenteetun &	& 0.6985 & 0.4287 &   0.7659 &    0.5274 &  
		-0.2698 	&    	0.0674 	&       -0.1711 \\
	&	\trentedeux-\trenteetun &	& 2.3442 & 2.5647 &   2.3422 &    2.4237 
		&   	0.2205 &      	-0.0020 	&        0.0795 \\
	&	\douze-\trenteetun & & 	2.7647 & 2.9701 &   2.8036 &    2.9516 &   
		0.2055 	&      	0.0389 	&        0.1869 \\
		\bottomrule
	\end{tabular}
	\label{tab:PES_XO-_eV_transition_Energie_Diff}
\end{table}



It is therefore possible to use the \IPA{} methods to determine the transition energies between the different levels. We can compare these transition energies with experimental values (tab.~\ref{tab:PES_Experimental}). This table shows the transition energy $\Delta_T = \text{\onze} - \text{\trenteetun} $ for the different approximations. Reading this table, the conclusions remain the same, We can also notice that the method \pMBPTip{} remains correct until iodine with an  error of $\sim$\SI{0.04}{\electronvolt}.

\begin{table}[H]
	\fontsize{8}{6}\selectfont
\centering
\captionsetup{font=small}
	\caption{ Transition energies  $ \Delta T_{\text{method}} $ between \onze-\trenteetun{}  
		for \XOm{Cl}, \XOm{Br} and \XOm{I}, comparaison with experimental 
		data (in \SI{}{\electronvolt}).}
	\renewcommand{\arraystretch}{1.8}
	\begin{tabular}{l r rrrr}
		\toprule
		$ \Delta T $: \onze-\trenteetun & exp~\cite{Gilles1992}&	$ 
		\Delta T_{\text{\ip}}$ &  $ 
		\Delta T_{\text
			{\pip}}$   & $ \Delta T_{\text{\MBPT}}$ 
		&  
		$ \Delta T_{\text{\pMBPT}}$ \\
			\midrule
	\ch{ClO^-}	${}^{\ }$&  \num{0.0397} & \num{0.0396} &     
	\num{0.0343} &        
	\num{0.0408} 
	&   	\num{0.0363} \\
	\ch{BrO^-}  ${}^{\ }$ &  \num{0.1270}  & \num{0.1256} &      
	\num{0.0977} &         
	\num{0.1336} &          
	\num{0.1092}\\
	\ch{IO^-} ${}^{\ }$ &\num{0.2593}&	\num{0.2717} & \num{0.1856} 
	&   
	\num{0.2956} &    \num{0.2211} \\
	\bottomrule
	\end{tabular}
	\label{tab:PES_Experimental}
\end{table}

\subsection{\ch{I3-}}

Paper presents data for \HamilDCG, i.e. with Gaunt, here are  tables for \ch{I3-} values for \ip{}, \ee{}, and \ea{} with \HamilDC.\\
\begin{table}[H]
	\fontsize{8}{6}\selectfont
	\centering
	\caption{Ionization potential (in \SI{}{\electronvolt}) for \ch{I3-}, \HamilDC{} hamiltonian, $\%SI=100 \left(\left (\rs {\mathbf {}} i \right)_k\right)^{2}$ in 
	$ [89\%;93\%] $  in all 
	case}
	\begin{tabular}{l r rrr r rrrr}
		\toprule
		& & \multicolumn{3}{c}{$ \Delta_{\text{\ip}} $} & & 
		\multicolumn{4}{c}{$ \Delta_{\text{exp}} $}\\
        \cmidrule(l{3pt}r{3pt}){3-5}  \cmidrule(l{3pt}r{3pt}){7-10}
		T&	\ip &  \pip &  \MBPT &  \pMBPT &  exp~\cite{Choi2000} &  \ip
		&  	\pip&  \MBPT &  \pMBPT \\
		\midrule
		$ \text{\ip}_{1}^{1/2g} $&	4.467 &   -0.053 &     0.104 &      0.059 
		&       
		4.53 &     -0.063 &      
		-0.116 &        0.041 &        -0.004 \\
		$ \text{\ip}_{2}^{3/2g} $	&	4.995 &   -0.139 &     0.010 &     -0.096 
		&       4.93 &      0.065 &      
		-0.074 &        0.075 &        -0.032 \\
		$ \text{\ip}_{3}^{1/2u} $&	4.915 &   -0.098 &     0.042 &     -0.034 
		&       4.87 &      0.045 &      
		-0.053 &        0.088 &         0.011 \\
		$ \text{\ip}_{4}^{3/2u} $&	4.282 &   -0.100 &     0.041 &     -0.037 
		&       4.25 &      0.032 &      
		-0.067 &        0.073 &        -0.004 \\
		\midrule
		mean  &	 &   -0.097 &     0.049 &     -0.027 &      &      0.020 
		&      -0.078 &        0.069 &        -0.007 \\
		std  &	 &    0.030 &     0.034 &      0.056 &     &      0.049 
		&       0.023 &        0.017 &         0.015 \\
		\bottomrule
	\end{tabular}
	\label{tab:I3m1_IP_tab}
\end{table}

\begin{table}[H]
	\fontsize{8}{6}\selectfont
	\centering
	\caption{Excitation Energies (in \SI{}{\electronvolt}) for \ch{I3-}, \HamilDC{} hamiltonian, $\%SE=100 \left(\left (r^a_i \right)_k\right)^{2}$. In bold
	characters the reversed levels. Experimental data for (0$_u^+$): \num{3.43}
		\num{4.25} \cite{Choi2000} \num{3.45} \num{4.28} \cite{Zhu2001}. }
	\begin{tabular}{l rr rr rr rr rr}
		\toprule
		& \multicolumn{2}{c}{\ee} &\multicolumn{2}{c}{\pee}
		&\multicolumn{2}{c}{\MBPT}&\multicolumn{2}{c}{\pMBPT}    
		&\multicolumn{2}{c}{CASPT2$^{\cite{Gomes2010}}  $} \\
  \cmidrule(l{3pt}r{3pt}){2-3}  \cmidrule(l{3pt}r{3pt}){4-5}  \cmidrule(l{3pt}r{3pt}){6-7}  \cmidrule(l{3pt}r{3pt}){8-9} \cmidrule(l{3pt}r{3pt}){10-11}
		T&	EE  &    \%SE & \%SE   &  $\Delta  $   &  \%SE&  
		$\Delta  $  & \%SE & $\Delta  $  & E & $ \Delta $  \\
		\midrule
		2$_g$ & 2.2422 & 43 &     45 &   0.2627 &      43 &      
		0.0036 &         45 &       0.2655 &  2.24&       -0.0022 \\
		1$_g$ & 2.3703 & 23 &     24 &   0.2527 &      23 &      
		\textbf{0.1139} &         24 &       \textbf{0.3683} &  2.32&       -0.0503 
		\\
		0$_u^-$ & 2.3749 & 41 &     43 &   \textbf{0.2664} &      41 &      
		\textbf{0.0037} &         43 &       \textbf{0.2687} &  2.47&        0.0951 
		\\
		1$_u$ & 2.3770 & 46 &     48 &  \textbf{ 0.2537} &      46 &      
		0.1142 &         48 &       0.3696 &  2.47&        0.0930 \\
		0$_g^-$ & 2.8384 & 22 &     22 &   0.2712 &      22 &      
		0.0044 &         22 &       0.2751 &  2.76&       -0.0784 \\
		0$_g^+$ & 2.8936 & 22 &     22 &   0.2724 &      22 &      
		0.0034 &         22 &       0.2744 &  2.82&       -0.0736 \\
		1$_g$ & 3.0665 & 42 &     43 &   0.2805 &      42 &      
		0.0062 &         43 &       0.2846 &  2.85&       -0.2165 \\
		2$_u$ & 3.3242 & 48 &     50 &   0.2553 &      48 &     
		-0.0135 &         50 &       0.2429 &  3.10&       -0.2242 \\
		1$_u$ & 3.4090 & 48 &     50 &   0.2609 &      48 &     
		-0.0136 &         50 &       0.2470 &  3.11&       -0.2990 \\
		0$_u^+$ & 3.6630 & 18 &     20 &   0.2434 &      18 &     
		-0.0148 &         19 &       0.2248 &  3.52&       -0.1430 \\
		2$_g$ & 4.0797 & 24 &     24 &   0.2692 &      24 &     
		-0.0085 &         24 &       0.2627 &  3.98&       -0.0997 \\
		0$_u^-$ & 4.0888 & 42 &     43 &   0.2871 &      42 &     
		-0.0101 &         43 &       0.2752 &  3.79&       -0.2988 \\
		1$_g$ & 4.1836 & 47 &     49 &   0.2764 &      47 &     
		-0.0092 &         49 &       0.2668 &  4.06&       -0.1236 \\
		1$_u$ & 4.2087 & 40 &     41 &   0.3010 &      40 &     
		-0.0044 &         41 &       0.2934 &  3.80&       -0.4087 \\
		0$_u^+$ & 4.4894 & 16 &     17 &   0.2433 &      16 &      
		0.0007 &         17 &       0.2403 &  4.51&        0.0206 \\
		0$_g^-$ & 4.6890 & 20 &     21 &   0.2987 &      20 &     
		-0.0105 &         21 &       0.2867 &  4.51&       -0.1790 \\
		0$_g^+$ & 4.7004 & 20 &     21 &   0.2996 &      20 &     
		-0.0095 &         21 &       0.2885 &  4.53&       -0.1704 \\
		1$_g$ & 4.9025 & 40 &     36 &   \textbf{0.3025} &      40 &      
		0.0073 &         24 &       \textbf{0.3325} &  4.60&       -0.3025 \\
		\midrule
		mean &  & & &   0.2721 &  & 0.0091 &  &0.2815 & &       -0.1367 \\
		std &  &  &   &   0.0189 &   & 0.0378 &  &0.0386 & &    0.1368  \\
		\bottomrule
	\end{tabular}
	\label{tab:I3m1_EE_tab}
\end{table}

\begin{table}[H]
	\fontsize{8}{6}\selectfont
	\centering
	\caption{Electronic Affinities (in \SI{}{\electronvolt}) for \ch{I3-}, \HamilDC{} hamiltonian, $\%SA=100 \left(\left (r^a \right)_k\right)^{2}$.}
	\begin{tabular}{l rr rr rr rr}
		\toprule
		& \multicolumn{2}{c}{\ea} 
		&\multicolumn{2}{c}{\pea}&\multicolumn{2}{c}{\MBPT} &
		\multicolumn{2}{c}{\pMBPT}\\
   \cmidrule(l{3pt}r{3pt}){2-3}  \cmidrule(l{3pt}r{3pt}){4-5}  \cmidrule(l{3pt}r{3pt}){6-7}  \cmidrule(l{3pt}r{3pt}){8-9}
		T& 	E &    \%SA &    \%SA &  $\Delta_{\ea}  $ &   \%SA
		&  $\Delta_{\ea}   $ & \%SA &  $\Delta_{\ea}   $ \\
		\midrule
		EA$_{1}^{1/2u}$ & 2.5069 & 76 &     76 &    -0.0229 &        76 &       
		-0.0144 &         74 &        -0.0538 \\
		EA$_{2}^{1/2u}$ & 3.6445 & 77 &     77 &     0.0076 &        76 &        
		0.0038 &         75 &         0.0092 \\
		EA$_{3}^{1/2g}$ & 3.8780 & 94 &     94 &     0.0121 &        94 &        
		0.0179 &         95 &         0.0322 \\
		EA$_{4}^{1/2u}$ & 4.3892 & 96 &     96 &     0.0096 &        96 &        
		0.0003 &         96 &         0.0087 \\
		\midrule
		mean &  &  &      &     0.0016 &        &        
		0.0019 &         &        -0.0009 \\
		std &  &  &      &     0.0143 &         &        
		0.0115 &          &         0.0320 \\
		\bottomrule
	\end{tabular}
	\label{tab:I3m1_EA_tab}
\end{table}

\subsection{\ch{CH2I2} - \ch{CH2IBr}}

Paper presents data for \HamilDCG, i.e. with Gaunt, here are  tables  for \ch{CH2I2} and \ch{CH2IBr} \ea{} with \HamilDC.\\

\begin{table}[H]
	\fontsize{8}{6}\selectfont
	\centering
	\caption{Electronic Affinities (in \SI{}{\electronvolt}) for \ch{CH2I2}, \HamilDC{} hamiltonian, $\%SA=100 \left(\left (r^a \right)_k\right)^{2}$.}
	\begin{tabular}{l r rr rr rr rr}
		\toprule
		&	& \multicolumn{2}{c}{\ea} 
		&\multicolumn{2}{c}{\pea}&\multicolumn{2}{c}{\MBPT} &
		\multicolumn{2}{c}{\pMBPT}\\
        \cmidrule(l{3pt}r{3pt}){3-4}  \cmidrule(l{3pt}r{3pt}){5-6}  \cmidrule(l{3pt}r{3pt}){7-8} \cmidrule(l{3pt}r{3pt}){9-10}
		T& &	E &    \%SA &    \%SA &  $\Delta_{\ea}  $ &   \%SA
		&  $\Delta_{\ea}   $ & \%SA &  $\Delta_{\ea}   $ \\
		\midrule
		EA$_1$& & -0.3277 & 48 &     50 &    -0.0617 &        48 
		&       -0.0575 &         49 &        -0.0971 \\
		EA$_2$& &  0.4105 & 57 &     60 &     0.0045 &        58 
		&       -0.0082 &         60 &         0.0003 \\
		EA$_3$& &  0.7549 & 83 &     80 &    -0.0175 &        82 
		&       -0.0095 &         80 &        -0.0167 \\
		EA$_4$&  &  0.8565 & 98 &     98 &     0.0066 &        98 
		&       -0.0086 &         98 &         0.0007 \\
		EA$_5$& &  1.4908 & 87 &     86 &     0.0103 &        86 
		&       -0.0022 &         85 &         0.0116 \\
		EA$_6$& &  1.6984 & 67 &     61 &    -0.0074 &        65 
		&       -0.0125 &         61 &        -0.0102 \\
		\midrule
		mean &&  &  &      &    -0.0109 &        
		&       -0.0164 &         &        -0.0186 \\
		std &&  &  &     &     0.0246 &      &        
		0.0186 &      &         0.0362 \\
		\midrule
		\midrule
		exp$ {}^{\cite{Modelli1992}} $ &&$ \Delta_{exp} $&&&$ \Delta_{exp} 
		$& 
		&$\Delta_{exp} $&&$\Delta_{exp} $\\
		ET${}_2$& 0.68  & -0.2695& &  &  -0.2649 &  
		&-0.2777 &   & -0.2692  \\
		DA${}_2 $ & 0.46  & -0.0495& &  & -0.0449  &  & 
		-0.0577&   & -0.0492  \\
		\bottomrule
	\end{tabular}
	\label{tab:CH2I2_EA_tab}
\end{table}

\begin{table}[H]
	\fontsize{8}{6}\selectfont
	\centering
	\caption{Electronic Affinities (in \SI{}{\electronvolt}) for \ch{CH2IBr}, \HamilDC{} hamiltonian, $\%SA=100 \left(\left (r^a \right)_k\right)^{2}$.}
	\begin{tabular}{l r rr rr rr rr}
		\toprule
		&	& \multicolumn{2}{c}{\ea} 
		&\multicolumn{2}{c}{\pea}&\multicolumn{2}{c}{\MBPT} &
		\multicolumn{2}{c}{\pMBPT}\\
    \cmidrule(l{3pt}r{3pt}){3-4}  \cmidrule(l{3pt}r{3pt}){5-6}  \cmidrule(l{3pt}r{3pt}){7-8} \cmidrule(l{3pt}r{3pt}){9-10}
		T& &	E &    \%SA &    \%SA &  $\Delta_{\ea}  $ &   \%SA
		&  $\Delta_{\ea}   $ & \%SA &  $\Delta_{\ea}   $ \\
		\midrule
		EA$_1$ && -0.0255 & 52 &     53 &    -0.0417 &        52 &       
		-0.0589 &         53 &        -0.0823 \\
		EA$_2$ &&  0.4487 & 66 &     70 &     0.0037 &        67 &       
		-0.0096 &         70 &        -0.0014 \\
		EA$_3$&&  0.8554 & 97 &     97 &     0.0072 &        97 &       
		-0.0086 &         97 &         0.0009 \\
		EA$_4$&&  0.9897 & 87 &     85 &    -0.0025 &        86 &       
		-0.0096 &         85 &        -0.0048 \\
		EA$_5$ &&  1.6154 & 70 &     69 &     0.0080 &        68 &       
		-0.0061 &         67 &         0.0063 \\
		EA$_6$ &&  1.9166 & 51 &     49 &     0.0000 &        55 &       
		-0.0121 &         48 &        -0.0047 \\
		\midrule
		mean& &  &  &  & -0.0042 & &   -0.0175 & & -0.0143 \\
		std & &  &  &  &     0.0171 &  &   0.0186 &  & 0.0306 \\
		\bottomrule
	\end{tabular}
	\label{tab:CH2IBr_EA_tab}
\end{table}

\ch{CH2I2}  \ee{} with \HamilDC.\\

\begin{table}[H]
	\fontsize{8}{6}\selectfont
	\centering
	\captionsetup{font=small}
	\caption{Excitation Energies (in \SI{}{\electronvolt}) for \ch{CH2I2}, \HamilDC{} hamiltonian, $\%SE=100 \left(\left (r^a_i \right)_k\right)^{2}$.}
	\begin{tabular}{l rr rr rr rr }
		\toprule
		& \multicolumn{2}{c}{\ee} &\multicolumn{2}{c}{\pee}
		&\multicolumn{2}{c}{\MBPT}&\multicolumn{2}{c}{\pMBPT}\\
    \cmidrule(l{3pt}r{3pt}){2-3}  \cmidrule(l{3pt}r{3pt}){4-5}  \cmidrule(l{3pt}r{3pt}){6-7}  \cmidrule(l{3pt}r{3pt}){8-9}
		T&	EE  &    \%SE & \%SE   &  $\Delta  $   &  \%SE&  
		$\Delta  $  & \%SE & $\Delta  $  \\
		\midrule
		a & 3.6032 & 20 &     21 &   0.2873 &      20 &     -0.0689 
		&         21 &       0.2274 \\
		b & 3.6168 & 20 &     20 &   0.2879 &      20 &     -0.0670 
		&         20 &       0.2300 \\
		a & 3.6280 & 20 &     21 &   0.2885 &      20 &     -0.0652 
		&         21 &       0.2326 \\
		b & 3.8444 & 16 &     17 &   0.2955 &      15 &     -0.0729 
		&         16 &       0.2302 \\
		a & 3.8684 & 19 &     19 &   0.2946 &      19 &     -0.0683 
		&         19 &       0.2339 \\
		b & 3.9421 & 13 &     13 &   0.3062 &      13 &     -0.0773 
		&         13 &       0.2359 \\
		a & 3.9936 & 16 &     17 &   0.3022 &      16 &     -0.0760 
		&         17 &       0.2336 \\
		b & 4.0597 & 16 &     16 &   0.3084 &      16 &     -0.0714 
		&         16 &       0.2439 \\
		b & 4.2220 & 18 &     19 &   0.2970 &      18 &     -0.0604 
		&         19 &       0.2448 \\
		a & 4.3222 & 15 &     16 &   0.3011 &      15 &     -0.0726 
		&         16 &       0.2365 \\
		b & 4.3522 & 15 &     15 &   0.3056 &      15 &     -0.0597 
		&         15 &       0.2538 \\
		a & 4.4937 & 14 &     15 &   0.3072 &      15 &     -0.0691 
		&         16 &       0.2449 \\
		b & 4.6384 & 15 &     15 &   0.3241 &      15 &     -0.0737 
		&         15 &       0.2584 \\
		a & 4.6829 & 16 &     17 &   0.3489 &      17 &     -0.0647 
		&         18 &       0.2922 \\
		b & 4.7499 & 15 &     14 &   0.3388 &      15 &     -0.0718 
		&         15 &       0.2752 \\
		a & 4.9106 & 14 &     15 &   0.3512 &      15 &     -0.0617 
		&         15 &       0.2962 \\
		\midrule
		mean & &  &  &   0.3090 &   &     -0.0688 &  &  0.2481 \\
		std & &  &   &   0.0201 &  &    0.0052 &  &       0.0212 \\		
		\bottomrule
	\end{tabular}
	\label{tab:CH2I2_EE_tab}
\end{table}
\end{widetext}

\bibliography{Biblio_partitioned_EOM}